\documentclass[12pt,preprint]{aastex}

\usepackage{epsf}

\newcommand{\bR}{\mbox{\boldmath$R$}}
\newcommand{\be}{\mbox{\boldmath$e$}}
\newcommand{\bff}{\mbox{\boldmath$f$}}
\newcommand{\br}{\mbox{\boldmath$r$}}
\newcommand{\bu}{\mbox{\boldmath$u$}}
\newcommand{\bOmega}{\mbox{$\bf\Omega$}}
\newcommand{\bxi}{\mbox{\boldmath$\xi$}}

\begin{document}

\title{Tidal dissipation in rotating giant planets}
\shorttitle{Tidal dissipation in giant planets}
\author{G. I. Ogilvie\altaffilmark{1,2} and D. N. C. Lin\altaffilmark{2,1}}
\altaffiltext{1}{Institute of Astronomy, University of Cambridge,
  Madingley Road, Cambridge CB3 0HA, UK; {\tt gogilvie@ast.cam.ac.uk}}
\altaffiltext{2}{UCO/Lick Observatory,University of California, Santa
  Cruz, CA 95064; {\tt lin@ucolick.org}}
\shortauthors{Ogilvie \& Lin}

\begin{abstract}
  Many extrasolar planets orbit sufficiently close to their host stars
  that significant tidal interactions can be expected, resulting in an
  evolution of the spin and orbital properties of the planets.  The
  accompanying dissipation of energy can also be an important source
  of heat, leading to the inflation of short-period planets and even
  mass loss through Roche-lobe overflow.  Tides may therefore play an
  important role in determining the observed distributions of mass,
  orbital period, and eccentricity of the extrasolar planets.  In
  addition, tidal interactions between gaseous giant planets in the
  solar system and their moons are thought to be responsible for the
  orbital migration of the satellites, leading to their capture into
  resonant configurations.
  
  Traditionally, the efficiency of tidal dissipation is simply
  parametrized by a quality factor $Q$, which depends, in principle,
  in an unknown way on the frequency and amplitude of the tidal
  forcing.  In this paper, we treat the underlying fluid dynamical
  problem with the aim of determining the efficiency of tidal
  dissipation in gaseous giant planets such as Jupiter, Saturn, or the
  short-period extrasolar planets.  Efficient convection enforces a
  nearly adiabatic stratification in these bodies, which may or may
  not contain rocky cores.  With some modifications, our approach can
  also be applied to fully convective low-mass stars.
  
  In cases of interest, the tidal forcing frequencies are typically
  comparable to the spin frequency of the planet but are small
  compared to its dynamical frequency.  We therefore study the
  linearized response of a slowly and possibly differentially rotating
  planet to low-frequency tidal forcing.  Convective regions of the
  planet support inertial waves, which possess a dense or continuous
  frequency spectrum in the absence of viscosity, while any radiative
  regions support generalized Hough waves.  We formulate the relevant
  equations for studying the excitation of these disturbances and
  present a set of illustrative numerical calculations of the tidal
  dissipation rate.
  
  We argue that inertial waves provide a natural avenue for efficient
  tidal dissipation in most cases of interest.  The resulting value of
  $Q$ depends, in principle, in a highly erratic way on the forcing
  frequency, but we provide analytical and numerical evidence that the
  relevant frequency-averaged dissipation rate may be asymptotically
  independent of the viscosity in the limit of small Ekman number.
  For a smaller viscosity, the tidal disturbance has a finer spatial
  structure and individual resonances are more pronounced.
  
  In short-period extrasolar planets, tidal dissipation via inertial
  waves becomes somewhat less efficient once they are spun down to a
  synchronous state.  However, if the stellar irradiation of the
  planet leads to the formation of a radiative outer layer that
  supports generalized Hough modes, the tidal dissipation rate can be
  enhanced, albeit with significant uncertainty, through the
  excitation and damping of these waves.  The dissipative mechanisms
  that we describe offer a promising explanation of the historical
  evolution and current state of the Galilean satellites as well as
  the observed circularization of the orbits of short-period
  extrasolar planets.
\end{abstract}


\section{INTRODUCTION}

\subsection{Tidal interactions in planetary systems}

The discovery of the very first extrasolar planet orbiting a
main-sequence star, 51~Peg~b, brought the surprising revelation that
other planetary worlds can revolve around their host stars in
extraordinarily tight orbits (Mayor \& Queloz 1995). Around
approximately one per cent of all the stars targeted by
radial-velocity searches, similar planets are found with periods of a
few days.  These short-period planets were probably formed within a
protostellar disk several AU away from their host stars through a
sequence of physical processes outlined in the conventional
`core-accretion' theories of planetary formation (Lissauer 1993;
Pollack et al. 1996).  During its final accretion phase, strong
interactions between a protoplanet and its natal disk lead to the
formation of a gap near its orbit, which effectively terminates its
growth.  Angular momentum exchange with the disk causes a protoplanet
formed in the inner region of the disk to migrate towards its host
star (Goldreich \& Tremaine 1980; Lin \& Papaloizou 1986).  Near the
star, the protoplanet's migration may be halted either by its tidal
interaction with a rapidly spinning host star or by entering a cavity
in the disk associated with the stellar magnetosphere (Lin,
Bodenheimer, \& Richardson 1996). Planets may also terminate their
migration with intermediate-period orbits when their natal disks are
rapidly depleted (Trilling, Lunine, \& Benz 2002; Armitage et al.
2002; Ida \& Lin 2003).

Short-period planets may be scattered to the vicinity of (or far away
from) the stellar surface owing to dynamical instabilities associated
with their long-term post-formation gravitational interaction (Rasio
\& Ford 1996; Weidenschilling \& Marzari 1996).  In this case, the
protoplanet's orbit may initially be highly eccentric and subsequently
undergo circularization owing to the tidal dissipation within its
envelope (Rasio et al. 1996).  Scattering by residual planetesimals is
another promising avenue for planetary orbital migration (Murray et
al. 1998).  It has also been suggested that short-period planets may
be formed {\it in situ\/} through a sequence of planetesimal
migration, coagulation, and gas accretion (Papaloizou \& Terquem 1999;
Bodenheimer, Hubickyj, \& Lissauer 2000; Sasselov \& Lecar 2000).
Finally, a totally different theory of protoplanetary formation relies
upon gravitational instability in a massive gaseous protostellar disk
(Kuiper 1951; Cameron 1978; Boss 1997), which produces a
subcondensation with a composition close to that of the central star
and having little, if any, solid core.  Although the inferred presence
of solid cores in Jupiter, Saturn, Uranus, and Neptune (Hubbard 1984)
favors the orderly growth scenario in our solar system, there is as
yet no direct evidence for or against the existence of cores in
extrasolar planets.

In the past eight years, more than 100 planets have been discovered,
around approximately ten per cent of the target stars in various
planet search programs.  A comprehensive distribution of planetary
mass, period, and eccentricity has begun to emerge. These data are
particularly useful for constraining and differentiating some
scenarios for the formation and evolution of short-period planets. For
example, there appears to be a minimum cut-off period at about three
days.  If the short periods of some planets were attained through
orbital migration, their observed period distribution could be
attributable to the stopping mechanism involved.  It is also possible
that the present period distribution is established through
post-formation evolution, e.g., planets with extremely short periods
may have perished subsequent to their formation.  A particularly
interesting characteristic of the planet-bearing stars is that they
tend to be metal-rich with respect to the Sun and the F--G field-star
average in the solar neighborhood (e.g., Gonzalez et al. 2001; Santos,
Israelian, \& Mayor 2001).  This correlation may result from planets
having being consumed following either migration through the
protoplanetary disk (Lin 1997; Laughlin \& Adams 1997; Sandquist et
al. 1998) or gravitational interaction with other planets (Rasio \&
Ford 1996). It may also be interpreted as evidence that an enhanced
metallicity in the planet-forming disk, probably resulting from a
metal-rich parent cloud, is especially conducive to planet formation
(Pollack et al.  1996; Fischer \& Valenti 2003; Ida \& Lin 2003).

Another important set of observational data is the planets'
eccentricities.  While the eccentricities of planets with periods
$P>21$ days are uniformly distributed up to about $0.7$, all planets
with $P<7$ days have nearly circular orbits.  This observed dichotomy
in the eccentricity--period distribution has been attributed to the
circularization of the orbits of short-period planets induced by their
internal tidal dissipation (Rasio et al. 1996; Marcy et al. 1997).
Such a scenario would be necessary to account for the small
eccentricities of the short-period planets if they were scattered
into, or strongly perturbed in, the vicinity of their host stars by
their planetary siblings.  If, instead, the short-period planets
acquired their small semi-major axes through tidal interaction with
their natal disks (Lin et al. 1996), such a process may be able to
damp the eccentricities of planets with masses $M_p\la10\,M_J$ (where
$M_J$ is the mass of Jupiter) (Goldreich \& Tremaine 1980; Artymowicz
1992), although this hypothesis remains uncertain (Papaloizou, Nelson,
\& Masset 2001; Goldreich \& Sari 2003).  However, some of the
potential stopping mechanisms, such as the planets' tidal interaction
with rapidly spinning host stars (Lin et al. 1996), can also excite
their orbital eccentricities (Dobbs-Dixon, Lin, \& Mardling 2003). An
efficient post-formation eccentricity damping mechanism may still be
needed to account for the small eccentricities of the short-period
planets.

Although no information is available on the rotation of extrasolar
planets, tidal interaction with the host star is expected to bring the
rotation of a short-period planet into synchronism with its orbit and
to eliminate any obliquity of its rotation axis.  Since the spin of a
planet accounts for much less angular momentum than its orbit,
synchronization proceeds more rapidly than circularization.  This
assumption neglects the possible effects of thermal tides, which may
tend to drive the planet away from synchronism (Thomson 1882; Gold \&
Soter 1969).

Tidal interaction can also modify the radii and internal structures of
short-period planets.  In principle, the actual sizes of extrasolar
planets may be used, at least as a partial test, to infer whether they
formed through core accretion or through gravitational instability of
massive gaseous protostellar disks. In the fully convective envelopes
of gaseous giant planets, it is difficult for the refractory elements
to condense into droplets and become differentiated from the volatile
elements (cf. Guillot et al. 2003).  Therefore planets formed through
orderly core accretion are more likely to have solid cores and to be
relatively compact, whereas those formed through gravitational
instability are likely to retain a nearly uniform solar composition
throughout their interiors and to be more extended.

The detection of a transiting planet around the star HD~209458
provides an opportunity for us to measure its radius directly and
thereby to constrain its present internal structure.  The observed
radius, $1.35\,R_J$, of this $0.63\,M_J$ planet (Brown et al. 2001) is
larger than that expected for a coreless planet with a similar mass
and age.  A planet with a core would be still more compact, leading to
an even larger discrepancy with the observational measurement.  For
this short-period planet, the presence of a small residual orbital
eccentricity or non-synchronous rotation could lead to internal tidal
dissipation with a heating rate comparable to or larger than that
released by the Kelvin--Helmholtz contraction.  Provided that the
dissipation of the host star's tidal disturbance occurs well below the
planet's photosphere, it increases the planet's internal temperature
and equilibrium radius (Bodenheimer et al. 2000).

Planet--star interactions may also have altered the mass distribution
of the short-period planets.  Above a critical eccentricity, which is
a function of the planet's semi-major axis, tidal dissipation of
energy during the circularization process can cause a planet to
inflate in size before its eccentricity is damped.  For moderate
eccentricities, the planet adjusts through a sequence of stable
thermal equilibria in which the rate of internal tidal dissipation is
balanced by the enhanced radiative flux associated with the planet's
enlarged radius.  For sufficiently large eccentricities, the planet
swells beyond $2\,R_J$ and its internal degeneracy is partially
lifted.  Thereafter, the thermal equilibria become unstable and the
planet undergoes runaway inflation until its radius exceeds the Roche
radius (Gu, Lin, \& Bodenheimer 2003).  The critical eccentricity of
about 0.2 (for a young Jupiter-mass planet with a period less than
three days) may be easily attained, with the result that many
short-period planets may have migrated to the vicinity of their host
stars and perished there.

The above discussion clearly indicates the importance of planet--star
tidal interactions to the origin and destiny of short-period planets.
They may determine (1) the asymptotic semi-major axis of migrating
protoplanets, (2) the absence of planets with periods less than three
days, (3) the structure of those planets that survive in the vicinity
of their host stars, and (4) the small eccentricities and relatively
low masses of the short-period planets.  However, the efficiency of
these processes depends crucially on the ability of the planet to
dissipate tidal disturbances, which is poorly understood.

\subsection{Theory of the equilibrium tide}

An early analysis of the nature of the tidal interaction between Earth
and the Moon was introduced by Darwin (1880) based on the concept of
the equilibrium tide (see Cartwright 2000 for a historical
discussion).  In this model, due originally to Newton, a homogeneous
spherical body continually adjusts to maintain a state of
quasi-hydrostatic equilibrium in the varying gravitational potential
of its orbital companion.  Darwin introduced a phase lag into the
response, the lag being proportional to the tidal forcing frequency
and attributable to the viscosity of the body.  The phase lag gives
rise to a net tidal torque and dissipation of energy.  Subsequent
formulations (Munk \& MacDonald 1960; Goldreich \& Soter 1966)
parametrize the efficiency of the tidal dissipation, whatever its
origin, by a specific dissipation function or $Q$-value (quality
factor).  Although, in principle, $Q$ is an unknown function of the
frequency and amplitude of the tidal forcing, in the planetary science
community it is usually treated as a constant property of each body in
the solar system, corresponding to a constant phase lag of $\arcsin
Q^{-1}$ (e.g., Murray \& Dermott 1999).  Indeed, studies of the
Earth's rotation provide evidence that its $Q$-value is approximately
constant over a wide range of frequency (Munk \& MacDonald 1960), even
though a variety of mechanisms may be responsible.  The tidal
$Q$-value may be predominantly determined by turbulent dissipation in
the shallow seas (Taylor 1920; Jeffreys 1920; see Munk \& MacDonald
1960 for a discussion).  In the case of the Moon, the observational
constraints on $Q$ are much weaker, but the synchronized spin of the
Moon suggests that the presence of an ocean is not necessary for tidal
dissipation.  In the context of binary stars, the assumption of a
constant lag time is usually adopted (Hut 1981; Eggleton, Kiseleva, \&
Hut 1998) as in Darwin's viscous model.

Evidence of tidal dissipation can also be found in gaseous giant
planets such as Jupiter and Saturn.  For example, the Laplace
resonance of Io, Europa, and Ganymede may be entered through the tidal
interaction of the moons with Jupiter (Goldreich 1965; Peale, Cassen,
\& Reynolds 1979; Lin \& Papaloizou 1979b).  In order to maintain this
resonant configuration despite the dissipation inside Io, tidal
dissipation within Jupiter must continually induce angular momentum
transfer from its spin to Io's orbit. The magnitude of the $Q$-value
for Jupiter, inferred from Io's dissipation rate, is in the range
$6\times 10^4-2\times 10^6$ (Yoder \& Peale 1981).  These values are
similar to those inferred, for solar-type stars, from the
circularization of close binaries as a function of their age and
semi-major axis (Mathieu 1994; Terquem et al. 1998).  Using these
values of $Q$, the orbital evolution of a planetary or a stellar
companion of a main-sequence star can be estimated by extrapolation,
under the assumption of either a constant lag angle (Goldreich \&
Soter 1966), a constant lag time (Hut 1981; Eggleton et al.  1998), or
an intermediate approach (Mardling \& Lin 2002).

In the extended convective envelopes of gaseous giant planets and
low-mass stars, turbulence can lead to dissipation of the motion that
results from the continual adjustment of the equilibrium tide.
However, the turbulent viscosity estimated from the mixing-length
theory ought to be reduced by a frequency-dependent factor owing to
the fact that the convective turnover timescale is usually much longer
than the period of the tidal forcing.  Estimating this reduction in
the efficiency of the turbulent viscosity, Zahn (1977, 1989) analyzed
the dissipation of equilibrium tides in low-mass stars and computed an
efficiency of angular momentum transfer for a non-rotating star that
matches the observationally inferred circularization timescale of
solar-type binary stars (Mathieu 1994).  However, with a different
prescription for the efficiency reduction factor (Goldreich \& Keeley
1977; Goodman \& Oh 1997), the derived rate of angular momentum
transfer falls short of that required by nearly two orders of
magnitude (Terquem et al. 1998).  Based on a similar approach, which
gives $Q\approx5\times10^{13}$ from the dissipation of the equilibrium
tide in Jupiter, Goldreich \& Nicholson (1977) suggested that the
tidal interaction between the Galilean satellites and Jupiter cannot
drive any significant orbital evolution.  For the short-period
extrasolar planets, such a high $Q$-value would imply circularization
times considerably longer than the ages of their host stars.

\subsection{Dynamical tides}

The problem of tidal dissipation is even more acute for high-mass
stars in close binaries because they have extended radiative envelopes
that may be expected to be free from turbulence.  Yet their orbits are
indeed circularized despite their short lifespans (Primini, Rappaport,
\& Joss 1977).  In these systems, the tidal perturbation of the
companion can induce the resonant excitation of low-frequency g-mode
oscillations in the radiative region, which carry both energy and
angular momentum fluxes (Cowling 1941; Zahn 1970).  This wavelike,
dynamical tide exists in addition to the equilibrium tide and provides
an alternative avenue for tidal dissipation. The g~modes are primarily
excited in the radiative envelope close to the convective core where
the Brunt--V\"ais\"al\"a frequency is comparable to the tidal forcing
frequency and the wavelengths of the gravity waves are sufficiently
long to couple well with the tidal potential (Goldreich \& Nicholson
1989). Using asymptotic methods in the limit that the forcing
frequency is small compared to the dynamical frequency of the star,
Zahn demonstrated the existence of resonant modes that amplify the
dynamical response of the star at particular frequencies.  This local
analysis was generalized to modest frequencies in a numerical analysis
by Savonije \& Papaloizou (1983, 1984).  Despite the absence of
convectively driven turbulence in the radiative envelope, the g-mode
oscillations can be dissipated through radiative damping on the
wavelength scale.  The radiative loss in the stellar interior is
generally weak so that the waves can propagate in an approximately
adiabatic manner to the stellar surface where they are dissipated and
the angular momentum they carry is deposited.  This concept gives rise
to the interesting suggestion that the stars become synchronized from
the outside in (Goldreich \& Nicholson 1989).  As the dissipation
layer first establishes a synchronized rotation, it presents a barrier
to the outwardly propagating waves in the form of a corotation
resonance at which the group velocity of the waves vanishes.  Just
below the corotation resonance, waves are stalled with large amplitude
and dissipated.  This process continues, initially inducing
differential rotation but eventually leading to the global
synchronization of the stellar spin.  This theoretical model is in
agreement with the observed spin and orbital evolution of high-mass
close binary stars (Giuricin, Mardirossian, \& Mezzetti 1984a, 1984b).

In the interior of a solar-type star, where a radiative core is
surrounded by a convective envelope, g-mode oscillations are also
excited in the radiative region and, inasmuch as they extend into the
outer convective region, they are dissipated by turbulent viscosity.
For some special forcing frequencies, these g-mode oscillations can
also attain a global resonance that strongly enhances the oscillation
amplitude in the radiative region and the rate of dissipation in the
convective region (Terquem et al. 1998). Nevertheless, the overall
synchronization process is limited by the less efficient tidal
dissipation that occurs in between resonant frequencies.  A
calibration with the observed circularization periods suggests that
the required viscosity is nearly two orders of magnitude larger than
that inferred from the standard mixing-length model for turbulent
convection and the Goldreich \& Keeley (1977) reduction factor.

Although g-mode excitation is a powerful process that drives the
dynamical tidal response in stars, it can be effective only in
radiative regions.  Since the envelopes of gaseous giant planets are
mostly convective, the relevance of g-mode oscillations is less well
established.  In a series of papers, Ioannou \& Lindzen (1993a, 1993b,
1994) considered the dynamical response of Jupiter to the tidal
perturbation of Io with a prescribed model for Jupiter's interior.
They showed that g~modes can be resonantly excited if the interior is
slightly subadiabatic to the extent that an appropriately tuned wave
cavity is created.  When these waves are transmitted into the
atmosphere and dissipated, they can produce an effective $Q$-value
comparable to that inferred from the supposed orbital evolution of Io.
However, current models suggest that convection may extend throughout
the entire envelope of Jupiter and is such an efficient heat-transport
process that Jupiter's envelope is adiabatically stratified and
neutrally buoyant to a very high degree (Guillot et al. 2003).  In
this limit, Ioannou \& Lindzen's mechanism does not work and only a
non-wavelike dynamical response to Io's tidal perturbation would
remain with a much reduced amplitude and $Q>10^{10}$.

In contrast to Jupiter, the surface layers of short-period extrasolar
planets may be stabilized against convection because they are
intensely heated by their host stars and may attain a radiative state.
If so, g-mode oscillations may be excited in the radiative layer just
above its interface with the planet's convective envelope and
dissipated through radiative or nonlinear damping (Lubow, Tout, \&
Livio 1997) as suggested for high-mass stars.  However, the one-sided
stellar irradiation of the surface of a synchronized planet induces a
large-scale, shallow circulation (Burkert et al. 2003), which does not
necessarily suppress convection and enhance the Brunt-V\"ais\"al\"a
frequency over an extended region in the atmosphere. Thus the
efficiency of dynamical tidal dissipation via g-mode oscillations
remains an outstanding issue.

In their calculation Ioannou \& Lindzen (1993a, 1993b) also considered
the effects of uniform rotation.  For the excitation of the dynamical
response, they adopted the so-called `traditional approximation' in
which only the radial component of the angular velocity is included in
the computation of the Coriolis force (Eckart 1960; Chapman \& Lindzen
1970; Unno et al. 1989).  This approach was justified in their model
on the basis that the dynamical response occurs primarily near the
atmosphere of Jupiter where the horizontal scale of the motion is
large and the radial velocity perturbation is relatively small.  It
enabled them to separate the radial and angular variables in the
governing linearized equations for the perturbations.  The meridional
structure of Jupiter's response due to Io's perturbation is then
determined by Laplace's tidal equation, the solutions of which are
Hough modes (Hough 1897, 1898) instead of spherical harmonics, while
the radial structure is determined by a set of ordinary differential
equations.  Using a local (WKB) approximation, they obtained a
dispersion relation describing a mixture of gravity, inertial, and
acoustic oscillations.  Although inertial waves can be excited in the
nearly adiabatic planetary interior, Ioannou \& Lindzen (1993b) did
not consider them relevant for tidal dissipation.  In fact the main
effect of rotation in their model is that tidal forcing by a single
solid harmonic projects on to many Hough modes, each with the
potential to resonate.  They noted that the traditional approximation
of neglecting the latitudinal component of the angular velocity is, in
fact, not appropriate in the planet's interior.

The effects of uniform rotation have been investigated in greater
depth by Savonije, Papaloizou, \& Alberts (1995) in the context of
high-mass binary stars, which consist of extended radiative envelopes
around small convective cores.  These authors were the first to
attempt a two-dimensional numerical solution of the linearized
equations governing tidal disturbances, including the full Coriolis
force.  By `switching off' terms deriving from the non-radial
component of the angular velocity, they were able to verify that the
traditional approximation can give a reasonably accurate measure of
the tidal torque on a highly stratified, slowly rotating star.  Like
Ioannou \& Lindzen (1993b), they found that the possibilities for
resonant excitation of g~modes are enhanced in a rotating star because
the Coriolis force couples spherical harmonics of different degrees.
Recently Savonije \& Witte (2002) applied the traditional
approximation to carry out a high-resolution study of the dynamical
tidal interaction between a uniformly rotating solar-type star and an
orbital companion including the effects of stellar evolution.

At the same time, Savonije et al. (1995) noted that the Coriolis force
introduces a rich spectrum of inertial modes in a certain frequency
range, which are not adequately represented by the traditional
approximation.  When the full Coriolis force is included, the response
of the star to tidal forcing in the frequency range of the inertial
modes contains large-amplitude, very short-wavelength components that
could not be resolved on the grid.  Using an improved
finite-difference numerical scheme including viscosity (Savonije \&
Papaloizou 1997) and an asymptotic perturbation analysis (Papaloizou
\& Savonije 1997), they later showed that inertial modes can be
resonantly excited in the adiabatic convective cores of high-mass
stars and can interact with the rotationally modified g~modes and
torsional r~modes in the radiative envelopes.  When these waves are
dissipated through radiative damping, the efficiency of the tidal
interaction is greatly enhanced.

\subsection{Inertial waves}

The study of inertial waves is said to have originated with Poincar\'e
(see Greenspan 1968).  Much attention has been devoted to the
oscillations of an incompressible fluid contained in a rotating
spherical, spheroidal, or cylindrical container.  When the Ekman
number is small, the Coriolis force provides the only significant
restoring force, away from viscous boundary layers, and the resulting
inertial waves have remarkable mathematical and physical properties.
The low-frequency oscillations of an adiabatically stratified star or
planet in uniform rotation are governed by a very similar problem
(Papaloizou \& Pringle 1981, 1982).  In each case the inertial waves
have frequencies (as observed in the rotating frame) no greater in
magnitude than twice the angular velocity of the fluid, and have a
rich spectrum that is dense or continuous in the absence of viscosity.

In the simplest problems, such as that of a full sphere of
incompressible fluid, an apparently complete set of inviscid inertial
wave solutions can be obtained analytically (Greenspan 1968; Zhang et
al. 2001).  However, the calculation of inertial waves in a spherical
annulus, or in a star or planet with an arbitrary density
stratification, requires two-dimensional numerical computations.
Moreover, since the inviscid Poincar\'e equation is spatially
hyperbolic in the relevant range of frequencies, the problem is
ill-posed unless an explicit viscosity is included.  Rieutord \&
Valdettaro (1997) calculated inertial waves in a spherical annulus,
noting the behavior of the frequencies, damping rates, and
eigenfunctions in the limit of small Ekman number.  Their results were
further elucidated by Rieutord, Georgeot, \& Valdettaro (2001), who
explained the important influence of the solid core.  In particular,
the viscous eigenfunctions are concentrated on characteristic rays of
the Poincar\'e equation, which typically converge towards limit cycles
as they reflect repeatedly from the inner and outer boundaries.  A
balance between the focusing of the wave energy along the converging
rays and a lateral viscous diffusion sets the characteristic width of
the eigenfunctions and determines the damping rates of the modes.

Dintrans \& Ouyed (2001) used similar methods to calculate inertial
waves in a slowly rotating polytrope at more modest Ekman numbers,
with potential application to Jovian seismology.  They considered a
pure inertial wave problem, eliminating acoustic and gravity waves by
adopting the anelastic approximation and using a neutrally buoyant
model.  Indeed, in contrast to either high-mass or solar-type stars in
close binaries, the entire envelope of a gaseous giant planet is
likely to be convective except for a shallow atmospheric layer.  In
contrast to the assumption adopted by Ioannou \& Lindzen (1993b),
efficient heat transport by convection prevents the propagation or
excitation of g-mode oscillations.  Inertial modes, however, can
propagate throughout the planetary envelope.

Jupiter's spin frequency is a significant fraction of its dynamical
frequency and is more than twice the orbital frequencies of the
Galilean satellites.  Short-period extrasolar planets probably formed
as rapid rotators similar to Jupiter and Saturn, although most of them
may have already established a state of synchronous rotation.  In each
case the tidal forcing occurs in the frequency range of inertial
waves, which will therefore constitute the natural and dominant
response of the planet.  Moreover, in this frequency range, the
Coriolis coupling between the radial and angular motions is strong and
the traditional approximation is inappropriate.  Therefore
two-dimensional computations are required.

\subsection{Plan of the paper}

In this paper, we revisit the issue of the excitation and dissipation
of dynamical tides within gaseous giant planets.  The response of the
planet to low-frequency tidal forcing separates naturally into an
equilibrium tide and a dynamical tide.  In the convective regions of
the planet the dynamical tide takes the form of inertial waves and we
consider a turbulent viscosity associated with convective eddies to
act on the tidal disturbance.  If the planet has an outer radiative
layer, outwardly propagating Hough waves are also excited and we
consider them to be damped in the atmosphere.  We therefore allow, in
principle, for three avenues of tidal dissipation: viscous dissipation
of the equilibrium tide, viscous dissipation of inertial waves, and
emission of Hough waves.  Our analysis permits the planet to rotate
differentially, although we focus on the case of uniform rotation in
our numerical calculations.

In Section~2, we briefly recapitulate the different components of the
tidal potential that are responsible for evolution towards
synchronization of the planet's spin and circularization of its orbit.
In Section~3, we carry out an analysis of low-frequency oscillation
modes in a rotating planet.  We formulate the perturbation equations
governing both the convective and radiative regions of the planet's
interior and analyze the matching conditions between these regions. In
Section~4, we consider the tidally forced disturbances, including the
equilibrium tide, the dynamical response in both regions, and the
matching across the interface.  We construct, in Section~5, two
numerical schemes to obtain global solutions of the forced
perturbation equations.  In Section~6 we present and discuss numerical
results appropriate for the internal structure of gaseous giant
planets.  In Section~7, we compare our results with those obtained in
previous investigations. We summarize our findings and discuss their
implications in Section~8.

\section{COMPONENTS OF THE TIDAL POTENTIAL}

\label{components}

We consider two bodies in a mutual Keplerian orbit, and adopt a
non-rotating coordinate system with origin at the center of body~1.
In our study body~1 will be a giant planet and body~2 its host star.
Let $\br$ be the position vector of an arbitrary point $P$ in body~1,
and let $\bR(t)$ be the position vector of the center of mass of
body~2.  When evaluating the tidal force acting within body~1 it is
usually adequate to treat body~2 as a point mass, i.e., to neglect its
rotational and tidal deformation.  The potential of body~2 experienced
at the point $P$, including the indirect potential resulting from the
acceleration of the origin of the coordinate system, is then
\begin{equation}
  -{{GM_2}\over{|\bR-\br|}}+{{GM_2}\over{R^3}}(\bR\cdot\br)=
  -{{GM_2}\over{R}}+
  {{GM_2}\over{2R^5}}\left[R^2r^2-3(\bR\cdot\br)^2\right]+O(r^3).
\end{equation}
The tidal potential is usually considered to be the non-trivial term
of lowest order in $r$,
\begin{equation}
  \Psi={{GM_2}\over{2R^5}}\left[R^2r^2-3(\bR\cdot\br)^2\right].
\end{equation}

Without loss of generality, let body~2 orbit in the $xy$-plane, so
that its Cartesian coordinates are
\begin{equation}
  \bR=R(\cos\varphi,\sin\varphi,0),
\end{equation}
while the Cartesian coordinates of the point $P$ are
\begin{equation}
  \br=r(\sin\theta\cos\phi,\sin\theta\sin\phi,\cos\theta),
\end{equation}
where $(r,\theta,\phi)$ are spherical polar coordinates.  Then
\begin{equation}
  \Psi={{GM_2}\over{2R^3}}r^2\left[1-3\sin^2\theta\cos^2(\phi-\varphi)\right].
\end{equation}

For a slightly eccentric orbit of semi-major axis $a$, eccentricity
$e$, and mean motion $n$, and with a convenient choice for the
longitude and time of pericenter passage, we have
\begin{equation}
  R=a(1-e\cos nt)+O(e^2),\qquad
  \varphi=nt+2e\sin nt+O(e^2).
\end{equation}
The tidal potential may then be expanded in powers of $e$ as
\begin{equation}
  \Psi=\Psi^{(0)}+\Psi^{(1)}+O(e^2),
\end{equation}
where
\begin{equation}
  \Psi^{(0)}={{GM_2}\over{4a^3}}r^2
  \left\{2-3\sin^2\theta\left[1+\cos(2\phi-2nt)\right]\right\},
\end{equation}
\begin{equation}
  \Psi^{(1)}={{3GM_2}\over{8a^3}}er^2\left\{4\cos nt+
  \sin^2\theta\left[\cos(2\phi-nt)-7\cos(2\phi-3nt)-6\cos nt\right]\right\}.
\end{equation}

Let
\begin{equation}
  \tilde P^m_\ell(\cos\theta)=
  \left[{{(2\ell+1)(\ell-m)!}\over{2(\ell+m)!}}\right]^{1/2}
  P^m_\ell(\cos\theta),
\end{equation}
where $0\le m\le\ell$, denote an associated Legendre polynomial
normalized such that
\begin{equation}
  \int_0^\pi\left[\tilde P^m_\ell(\cos\theta)\right]^2\sin\theta\,d\theta=1.
\end{equation}
The tidal potential correct to $O(e)$ can then be expanded in a series
of rigidly rotating solid harmonics of second degree,
\begin{equation}
  \Psi^{(0)}+\Psi^{(1)}={{GM_2}\over{a^3}}{\rm Re}\sum_{j=1}^5
  A_jr^2\tilde P^{m_j}_2(\cos\theta)e^{i(m_j\phi-\omega_jt)},
\end{equation}
where
\begin{eqnarray}
  &&m_1=0,\qquad\omega_1=0,\qquad A_1=\left({{1}\over{10}}\right)^{1/2},
  \nonumber\\
  &&m_2=2,\qquad\omega_2=2n,\qquad A_2=
  -\left({{3}\over{5}}\right)^{1/2},\nonumber\\
  &&m_3=0,\qquad\omega_3=n,\qquad A_3=
  3e\left({{1}\over{10}}\right)^{1/2},\nonumber\\
  &&m_4=2,\qquad\omega_4=n,\qquad A_4=
  e\left({{3}\over{20}}\right)^{1/2},\nonumber\\
  &&m_5=2,\qquad\omega_5=3n,\qquad A_5=
  -7e\left({{3}\over{20}}\right)^{1/2}.
\end{eqnarray}
Zahn (1977) gives an equivalent expansion carried to $O(e^2)$.

When body~1 is a differentially rotating planet with angular velocity
$\Omega(r,\theta)$, the effective frequency of component~$j$ as
experienced by the rotating fluid is the Doppler-shifted frequency
$\hat\omega_j=\omega_j-m_j\Omega$.  Component~1 can be neglected
because it is independent of time and causes only a hydrostatic
distortion of the planet but no tidal dissipation.  The effective
forcing frequencies of the remaining four components are
\begin{equation}
  \hat\omega_2=2(n-\Omega),\qquad
  \hat\omega_3=n,\qquad
  \hat\omega_4=n-2\Omega,\qquad
  \hat\omega_5=3n-2\Omega.
\end{equation}
This prescription assumes that the rotation axis of the planet is
perpendicular to its orbit.  Otherwise an additional component of the
tidal potential appears at first order in the obliquity, having
$m_6=1$ and $\omega_6=n$, and therefore
$\hat\omega_6=n-\Omega=\hat\omega_2/2$.

In the special case of a uniformly rotating planet, the forcing
frequency of component~$j$ lies in the spectrum of inertial waves if
$-2\le\hat\omega_j/\Omega\le2$.  In this frequency range, the Coriolis
coupling between the radial and angular motions is strong (Savonije et
al. 1995). For any planet being spun down by tidal dissipation towards
a synchronized state (i.e., $\Omega>n>0$), all four effective forcing
frequencies lie within the spectrum of inertial waves, as illustrated
in Fig.~1.  Once synchronism is achieved (i.e., $\Omega=n$),
component~2 ceases to cause tidal dissipation but, if the orbit
remains elliptical, the three eccentric components of the tidal
potential continue to provide forcing at $\hat\omega=\pm\Omega$,
clearly within the spectrum of inertial waves.

These considerations are directly relevant to the short-period
extrasolar planets.  All the known satellites of Jupiter and Saturn,
except the very distant, retrograde ones, also provide low-frequency
tidal forcing within the spectrum of inertial waves.  We therefore
suggest that the Coriolis force has a dominant influence on tidal
interactions in many cases of interest.  Except in work by Savonije
and co-workers on tides in uniformly rotating stars (Savonije et al.
1995; Savonije \& Papaloizou 1997; Papaloizou \& Savonije 1997; Witte
\& Savonije 1999a, 1999b, 2001, 2002; Savonije \& Witte 2002), the
Coriolis force has been unjustifiably neglected owing to the
mathematical, physical, and computational difficulties that it
introduces.

\section{LOW-FREQUENCY OSCILLATION MODES OF A ROTATING PLANET}

\subsection{Basic equations}

We consider a planet that is predominantly gaseous, but may contain a
solid core.  We initially consider the fluid regions to be ideal,
satisfying Euler's equation of motion,
\begin{equation}
  {{D\bu}\over{Dt}}=-{{1}\over{\rho}}\nabla p-\nabla\Phi,
\end{equation}
the equation of mass conservation,
\begin{equation}
  {{D\rho}\over{Dt}}=-\rho\nabla\cdot\bu,
\end{equation}
and the adiabatic condition,
\begin{equation}
  {{1}\over{\gamma}}{{D\ln p}\over{Dt}}-{{D\ln\rho}\over{Dt}}=0,
\end{equation}
where $\bu$ is the velocity, $\rho$ the density, $p$ the pressure,
$\gamma$ the adiabatic exponent, and
\begin{equation}
  {{D}\over{Dt}}={{\partial}\over{\partial t}}+\bu\cdot\nabla
\end{equation}
the Lagrangian time-derivative.  The gravitational potential satisfies
Poisson's equation,
\begin{equation}
  \nabla^2\Phi=4\pi G\rho.
\end{equation}

When expressed in spherical polar coordinates $(r,\theta,\phi)$, these
equations take the form
\begin{equation}
  {{Du_r}\over{Dt}}-{{u_\theta^2}\over{r}}-{{u_\phi^2}\over{r}}=
  -{{1}\over{\rho}}{{\partial p}\over{\partial r}}-
  {{\partial\Phi}\over{\partial r}},
  \label{dur}
\end{equation}
\begin{equation}
  {{Du_\theta}\over{Dt}}+{{u_ru_\theta}\over{r}}-
  {{u_\phi^2\cos\theta}\over{r\sin\theta}}=
  -{{1}\over{\rho r}}{{\partial p}\over{\partial\theta}}-
  {{1}\over{r}}{{\partial\Phi}\over{\partial\theta}},
\end{equation}
\begin{equation}
  {{Du_\phi}\over{Dt}}+{{u_ru_\phi}\over{r}}+
  {{u_\theta u_\phi\cos\theta}\over{r\sin\theta}}=
  -{{1}\over{\rho r\sin\theta}}{{\partial p}\over{\partial\phi}}-
  {{1}\over{r\sin\theta}}{{\partial\Phi}\over{\partial\phi}},
  \label{dup}
\end{equation}
\begin{equation}
  {{D\rho}\over{Dt}}=
  -\rho\left[{{1}\over{r^2}}{{\partial}\over{\partial r}}(r^2 u_r)+
  {{1}\over{r\sin\theta}}{{\partial}\over{\partial\theta}}(u_\theta\sin\theta)+
  {{1}\over{r\sin\theta}}{{\partial u_\phi}\over{\partial\phi}}\right],
\end{equation}
\begin{equation}
  {{1}\over{\gamma}}{{D\ln p}\over{Dt}}-{{D\ln\rho}\over{Dt}}=0,
\end{equation}
\begin{equation}
  {{1}\over{r^2}}{{\partial}\over{\partial r}}
  \left(r^2{{\partial\Phi}\over{\partial r}}\right)+
  {{1}\over{r^2\sin\theta}}{{\partial}\over{\partial\theta}}
  \left(\sin\theta{{\partial\Phi}\over{\partial\theta}}\right)+
  {{1}\over{r^2\sin^2\theta}}{{\partial^2\Phi}\over{\partial\phi^2}}=
  4\pi G\rho,
  \label{poisson}
\end{equation}
with
\begin{equation}
  {{D}\over{Dt}}={{\partial}\over{\partial t}}+
  u_r{{\partial}\over{\partial r}}+
  {{u_\theta}\over{r}}{{\partial}\over{\partial\theta}}+
  {{u_\phi}\over{r\sin\theta}}{{\partial}\over{\partial\phi}}.
\end{equation}

\subsection{Basic state}

The unperturbed state is a steady, axisymmetric planet which may
rotate differentially in its fluid regions.  We introduce the small
parameter $\epsilon$, being a characteristic value of the ratio of the
spin frequency to the dynamical frequency $(GM_1/R_1^3)^{1/2}$ of the
planet, which has mass $M_1$ and radius $R_1$.  Centrifugal
distortions of the planet are of fractional order $\epsilon^2$ and so
we may pose the expansions
\begin{eqnarray}
  \rho&=&\rho_0(r)+\epsilon^2\rho_2(r,\theta)+O(\epsilon^4),\nonumber\\
  p&=&p_0(r)+\epsilon^2p_2(r,\theta)+O(\epsilon^4),\nonumber\\
  \Phi&=&\Phi_0(r)+\epsilon^2\Phi_2(r,\theta)+O(\epsilon^4).
\end{eqnarray}
We neglect any meridional circulation and set $u_r=u_\theta=0$.  The
angular velocity is
\begin{equation}
  {{u_\phi}\over{r\sin\theta}}=\Omega=\epsilon\Omega_{(1)}(r,\theta).
\end{equation}

From equations (\ref{dur}) and (\ref{poisson}) at order $\epsilon^0$,
we obtain the standard equations of structure for a spherical planet.
We define the (inward) radial gravitational acceleration $g(r)$ and
the Brunt--V\"ais\"al\"a frequency $N(r)$ in the usual way by
\begin{equation}
  g={{d\Phi_0}\over{dr}},\qquad
  N^2={{1}\over{\rho_0}}{{dp_0}\over{dr}}
  \left({{d\ln\rho_0}\over{dr}}-{{1}\over{\gamma}}{{d\ln p_0}\over{dr}}\right).
\end{equation}
Equation (\ref{dup}) is satisfied identically.  In practice the
profile of differential rotation is determined by the meridional
circulation and by viscous, turbulent, or magnetic stresses that do
not feature in the equations as written above.  We will not require a
solution of the basic state at order $\epsilon^2$ or higher.

Our analysis is based on asymptotic expansions in the small parameter
$\epsilon^2$.  Jupiter is a relatively rapid rotator, having
$\epsilon^2\approx\Omega^2R_1^3/GM_1\approx0.08$.  This number
increases to about $0.14$ in the case of Saturn, but is greatly
reduced, to about $0.0016$, if Jupiter is spun down to a period of
three days.  This last case may be taken as representative of
short-period extrasolar planets once they achieve synchronous
rotation.

\subsection{Linearized equations}

\label{linearized}

We consider the linearized response to a tidal potential component of
the form
\begin{equation}
  {\rm Re}\left[\Psi(r,\theta)\,e^{i(m\phi-\omega t)}\right],
\end{equation}
where $m\in{\bf Z}$ is the azimuthal wavenumber and $\omega\in{\bf R}$
the forcing frequency.  As in Section~\ref{components}, we adopt the
convention that $m\ge0$.  The induced Eulerian velocity perturbation
is
\begin{equation}
  {\rm Re}\left[\bu'(r,\theta)\,e^{i(m\phi-\omega t)}\right],
\end{equation}
and similarly for other perturbed quantities.  The linearized
equations then take the form
\begin{equation}
  -i\hat\omega u_r'-2\Omega\sin\theta\,u_\phi'=
  -{{\partial W}\over{\partial r}}+
  {{1}\over{\rho^2}}\left(\rho'{{\partial p}\over{\partial r}}-
  p'{{\partial\rho}\over{\partial r}}\right),
  \label{u_r'}
\end{equation}
\begin{equation}
  -i\hat\omega u_\theta'-2\Omega\cos\theta\,u_\phi'=
  -{{1}\over{r}}{{\partial W}\over{\partial\theta}}+
  {{1}\over{\rho^2r}}\left(\rho'{{\partial p}\over{\partial\theta}}-
  p'{{\partial\rho}\over{\partial\theta}}\right),
  \label{u_t'}
\end{equation}
\begin{equation}
  -i\hat\omega u_\phi'+{{1}\over{r\sin\theta}}
  \left(u_r'{{\partial}\over{\partial r}}+
  {{u_\theta'}\over{r}}{{\partial}\over{\partial\theta}}\right)
  (\Omega r^2\sin^2\theta)=-{{imW}\over{r\sin\theta}},
\end{equation}
\begin{equation}
  -i\hat\omega\rho'+u_r'{{\partial\rho}\over{\partial r}}+
  {{u_\theta'}\over{r}}{{\partial\rho}\over{\partial\theta}}=
  -\rho\left[{{1}\over{r^2}}{{\partial}\over{\partial r}}(r^2 u_r')+
  {{1}\over{r\sin\theta}}{{\partial}\over{\partial\theta}}
  (u_\theta'\sin\theta)+
  {{im u_\phi'}\over{r\sin\theta}}\right],
\end{equation}
\begin{equation}
  -i\hat\omega\left({{p'}\over{\gamma p}}-{{\rho'}\over{\rho}}\right)+
  u_r'\left({{1}\over{\gamma}}{{\partial\ln p}\over{\partial r}}-
  {{\partial\ln\rho}\over{\partial r}}\right)+
  {{u_\theta'}\over{r}}\left({{1}\over{\gamma}}
  {{\partial\ln p}\over{\partial\theta}}-
  {{\partial\ln\rho}\over{\partial\theta}}\right)=0,
  \label{p'}
\end{equation}
\begin{equation}
  {{1}\over{r^2}}{{\partial}\over{\partial r}}
  \left(r^2{{\partial\Phi'}\over{\partial r}}\right)+
  {{1}\over{r^2\sin\theta}}{{\partial}\over{\partial\theta}}
  \left(\sin\theta{{\partial\Phi'}\over{\partial\theta}}\right)-
  {{m^2\Phi'}\over{r^2\sin^2\theta}}=4\pi G\rho',
\end{equation}
where
\begin{equation}
  \hat\omega=\omega-m\Omega
\end{equation}
is the Doppler-shifted forcing frequency, and
\begin{equation}
  W={{p'}\over{\rho}}+\Phi'+\Psi
\end{equation}
is a modified pressure perturbation.  In adiabatically stratified
regions, where $\nabla\ln p=\gamma\nabla\ln\rho$, the Eulerian
perturbations of pressure and density are related by
$p'/p=\gamma\rho'/\rho$, and the terms in parentheses in equations
(\ref{u_r'}) and (\ref{u_t'}) cancel.

\subsection{Free modes of oscillation}

We first consider the free modes of oscillation of the planet,
neglecting the tidal forcing.  We treat the convective regions of the
planet as being adiabatically stratified ($\nabla\ln
p=\gamma\nabla\ln\rho$), and the radiative regions as being stably
stratified such that the Brunt--V\"ais\"al\"a frequency is comparable
to the dynamical frequency.  Where a radiative region adjoins a
convective region, however, there is a transition region in which the
Brunt--V\"ais\"al\"a frequency goes to zero.  We obtain asymptotic
solutions for low-frequency oscillations in each region separately and
then match them together.  For clarity, we use accented symbols, e.g.,
$\check a$, $\hat a$, and $\bar a$, to distinguish solution components
in convective, radiative, and transition regions, respectively.

As explained in Section~\ref{components}, we are interested in
oscillation frequencies of order $\epsilon$, comparable to the angular
velocity of the planet but small compared to the dynamical frequency.
Accordingly, we set $\omega=\epsilon\omega_{(1)}$.  To achieve such a
low frequency in the presence of potentially large restoring forces,
the perturbations are highly constrained.  They should be almost
incompressible (or, more correctly, anelastic) in order to avoid a
large acoustic restoring force.  In radiative regions the
perturbations should also involve predominantly horizontal motions in
order to avoid a large buoyant restoring force.  In fact, the relevant
low-frequency modes are {\it inertial waves\/} in convective regions
and what we call {\it generalized Hough waves\/} in radiative regions.
In the solid core, if present, the rigidity is such that no
low-frequency modes are permissible.

\subsection{Convective regions}

\label{convective}

In convective regions we seek solutions of the unforced linearized
equations of the form
\begin{eqnarray}
  u_r'&\sim&\epsilon\check u_r'(r,\theta),\nonumber\\
  u_\theta'&\sim&\epsilon\check u_\theta'(r,\theta),\nonumber\\
  u_\phi'&\sim&\epsilon\check u_\phi'(r,\theta),\nonumber\\
  \rho'&\sim&\epsilon^2\check\rho'(r,\theta),\nonumber\\
  p'&\sim&\epsilon^2\check p'(r,\theta),\nonumber\\
  \Phi'&\sim&\epsilon^2\check\Phi'(r,\theta).
  \label{ordering_c}
\end{eqnarray}
The overall amplitude of the perturbations is arbitrary, but the
scaling adopted here is relevant to the tidally forced problem
considered in Section~\ref{tidal} below.  The relative scalings of the
perturbations are chosen to satisfy the linearized equations of
Section~\ref{linearized}, which reduce at leading order to
\begin{equation}
  -i\hat\omega_{(1)}\check u_r'-2\Omega_{(1)}\sin\theta\,\check u_\phi'=
  -{{\partial\check W}\over{\partial r}},
  \label{ur1'}
\end{equation}
\begin{equation}
  -i\hat\omega_{(1)}\check u_\theta'-2\Omega_{(1)}\cos\theta\,\check u_\phi'=
  -{{1}\over{r}}{{\partial\check W}\over{\partial\theta}},
\end{equation}
\begin{equation}
  -i\hat\omega_{(1)}\check u_\phi'+{{1}\over{r\sin\theta}}
  \left(\check u_r'{{\partial}\over{\partial r}}+
  {{\check u_\theta'}\over{r}}{{\partial}\over{\partial\theta}}\right)
  (\Omega_{(1)}r^2\sin^2\theta)=-{{im\check W}\over{r\sin\theta}},
\end{equation}
\begin{equation}
  {{1}\over{r^2\rho_0}}{{\partial}\over{\partial r}}(r^2\rho_0\check u_r')+
  {{1}\over{r\sin\theta}}{{\partial}\over{\partial\theta}}
  (\check u_\theta'\sin\theta)+{{im\check u_\phi'}\over{r\sin\theta}}=0,
  \label{rho2'}
\end{equation}
\begin{equation}
  {{\check p'}\over{p_0}}={{\gamma\check\rho'}\over{\rho_0}},
\end{equation}
\begin{equation}
  {{1}\over{r^2}}{{\partial}\over{\partial r}}
  \left(r^2{{\partial\check\Phi'}\over{\partial r}}\right)+
  {{1}\over{r^2\sin\theta}}{{\partial}\over{\partial\theta}}
  \left(\sin\theta{{\partial\check\Phi'}\over{\partial\theta}}\right)-
  {{m^2\check\Phi'}\over{r^2\sin^2\theta}}=4\pi G\check\rho',
\end{equation}
with
\begin{equation}
  \hat\omega_{(1)}=\omega_{(1)}-m\Omega_{(1)},
\end{equation}
and
\begin{equation}
  \check W={{\check p'}\over{\rho_0}}+\check\Phi'.
\end{equation}
Equations (\ref{ur1'})--(\ref{rho2'}) are decoupled from those that
follow, and, when combined with appropriate boundary conditions,
define the inertial wave problem.  We note that equation (\ref{rho2'})
is equivalent to the constraint $\nabla\cdot(\rho_0\bu')=0$ of the
anelastic approximation (Ogura \& Phillips 1962), which eliminates
acoustic waves from the problem.  Self-gravitation also plays no role
for inertial waves, because $\check\Phi'$ simply combines with $\check
p'$ in the combination $\check W$.

\subsection{Radiative regions}

\label{radiative}

In radiative regions of the planet, the oscillation frequency is small
compared to the Brunt--V\"ais\"al\"a frequency, implying that the
motions should be predominantly horizontal in addition to being
anelastic.  These constraints are satisfied when the radial wavelength
is short compared to the radius of the planet, and the perturbations
then have a WKB form in the radial direction.  Accordingly, we seek
solutions of the unforced linearized equations of the form
\begin{eqnarray}
  u_r'&\sim&\epsilon^2\hat u_r'(r,\theta)E(r),\nonumber\\
  u_\theta'&\sim&\epsilon\hat u_\theta'(r,\theta)E(r),\nonumber\\
  u_\phi'&\sim&\epsilon\hat u_\phi'(r,\theta)E(r),\nonumber\\
  \rho'&\sim&\epsilon\hat\rho'(r,\theta)E(r),\nonumber\\
  p'&\sim&\epsilon^2\hat p'(r,\theta)E(r),\nonumber\\
  \Phi'&\sim&\epsilon^2\check\Phi'(r,\theta)+\epsilon^3\hat\Phi'(r,\theta)E(r),
\end{eqnarray}
with
\begin{equation}
  E(r)=\epsilon^s\exp\left[{{i}\over{\epsilon}}\int k(r)\,dr\right].
\end{equation}
The exponent $s<0$ remains arbitrary until the solutions in adjacent
convective and radiative regions can be connected.  We show in
Section~\ref{matching} below that this procedure yields $s=-1/6$.
Note that the term $\epsilon^2\check\Phi'(r,\theta)$, which does not
have a WKB form, is the smooth continuation of the gravitational
potential perturbation associated with the inertial waves in
convective regions.

The linearized equations of Section~\ref{linearized} then reduce at
leading order to
\begin{equation}
  0=-ik\hat W-{{g\hat\rho'}\over{\rho_0}},
  \label{rad1}
\end{equation}
\begin{equation}
  -i\hat\omega_{(1)}\hat u_\theta'-2\Omega_{(1)}\cos\theta\,\hat u_\phi'=
  -{{1}\over{r}}{{\partial\hat W}\over{\partial\theta}},
\end{equation}
\begin{equation}
  -i\hat\omega_{(1)}\hat u_\phi'+{{\hat u_\theta'}\over{\sin\theta}}
  {{\partial}\over{\partial\theta}}(\Omega_{(1)}\sin^2\theta)=
  -{{im\hat W}\over{r\sin\theta}},
\end{equation}
\begin{equation}
  ik\hat u_r'+{{1}\over{r\sin\theta}}{{\partial}\over{\partial\theta}}
  (\hat u_\theta'\sin\theta)+{{im\hat u_\phi'}\over{r\sin\theta}}=0,
\end{equation}
\begin{equation}
  i\hat\omega_{(1)}{{\hat\rho'}\over{\rho_0}}+{{N^2}\over{g}}\hat u_r'=0,
\end{equation}
\begin{equation}
  -k^2\hat\Phi'=4\pi G\hat\rho',
\end{equation}
with
\begin{equation}
  \hat\omega_{(1)}=\omega_{(1)}-m\Omega_{(1)},
\end{equation}
and
\begin{equation}
  \hat W={{\hat p'}\over{\rho_0}}.
  \label{rad8}
\end{equation}
Since $\hat\Phi'$ appears only in Poisson's equation, self-gravitation
is irrelevant for waves of such short radial wavelength.  The
remaining equations may be combined into the form
\begin{equation}
  {\cal L}\hat W={{k^2r^2}\over{N^2}}\hat W,
  \label{lw}
\end{equation}
where ${\cal L}$ is a linear operator defined by
\begin{equation}
  {\cal L}w=-{{1}\over{\sin\theta}}\left({{\partial}\over{\partial\theta}}+
  {{2m\Omega_{(1)}\cos\theta}\over{\hat\omega_{(1)}\sin\theta}}\right)
  \left[{{\sin\theta}\over{D}}\left({{\partial}\over{\partial\theta}}-
  {{2m\Omega_{(1)}\cos\theta}\over{\hat\omega_{(1)}\sin\theta}}\right)w\right]+
  {{m^2w}\over{\hat\omega_{(1)}^2\sin^2\theta}},
  \label{call}
\end{equation}
with
\begin{equation}
  D=\hat\omega_{(1)}^2-4\Omega_{(1)}^2\cos^2\theta-
  2\Omega_{(1)}{{\partial\Omega_{(1)}}\over{\partial\theta}}
  \cos\theta\sin\theta.
\end{equation}

Let $\{\lambda_i\}$ and $\{w_i(\theta)\}$ be the eigenvalues and
eigenfunctions of ${\cal L}$, satisfying
\begin{equation}
  {\cal L}w_i=\lambda_iw_i.
\end{equation}
If the angular velocity $\Omega_{(1)}$ is independent of $r$, the
eigensolutions will be also.  For a general pattern of differential
rotation, however, the eigensolutions depend parametrically on $r$.
In either case, each eigensolution with $\lambda>0$ corresponds to a
freely propagating wave solution (which can be considered a
rotationally modified g~mode) satisfying the dispersion relation
\begin{equation}
  \lambda_i={{k^2r^2}\over{N^2}}.
  \label{dispersion}
\end{equation}
For a given wave frequency $\omega$ and azimuthal wavenumber $m$, this
relation determines the radial wavenumber $k$ at each $r$.  Evanescent
eigensolutions with $\lambda<0$ are also possible in a rotating
planet.

As given above, the solution represents a traveling wave with phase
velocity $\epsilon^2\omega_{(1)}/k$.  A standing wave is formed from a
superposition of two such solutions with opposite choices of the sign
of $k$.  Depending on how the wave is excited, and on whether it is
allowed to reflect from the boundaries of the radiative region, the
relevant solution may be a traveling wave, a standing wave or, in the
case of partial reflection, an intermediate solution.  In the case of
a traveling wave the sign of $k$ should be chosen with due regard to
the direction of the radial group velocity.

In the absence of dissipation, the waves carry conserved fluxes of
energy and angular momentum, owing to the invariance of the basic
state with respect to $t$ and $\phi$.  These conservation relations
can be derived from the theory of `wave action' using a Lagrangian
formulation of ideal fluid dynamics (e.g., Hayes 1970).  For a
traveling wave, the average flux of energy wave action through a
spherical surface can be shown to be
\begin{eqnarray}
  F&=&\pi\int_0^\pi{\rm Re}\left[\left({{\omega}\over{\hat\omega}}\right)
  p^{\prime*}u_r'\right]\,r^2\sin\theta\,d\theta\\
  \label{flux}
  &\sim&-\epsilon^{4+2s}\left({{\pi\rho_0r^2\omega_{(1)}}\over{N^2}}\right)
  {\rm Re}\left[k\right]\int_0^\pi|\hat W|^2\,\sin\theta\,d\theta,
  \label{flux_wkb}
\end{eqnarray}
and the flux of angular momentum wave action differs only by a factor
$m/\omega_{(1)}$.  Outwardly propagating waves, having positive radial
group velocity, are those with ${\rm sgn}(\hat\omega_{(1)}/k)<0$.
This property occurs because the frequency of g~modes decreases with
increasing radial wavenumber.  In the absence of dissipation the flux
$F$ should be independent of radius.

We note some properties of the operator ${\cal L}$.  Singular points
of the equation ${\cal L}w=\lambda w$ occur at the poles and wherever
$\hat\omega_{(1)}=0$ (`corotation resonance') or $D=0$ (`Lindblad
resonance'), although the latter is only an apparent singularity.
Close to the north pole, two independent solutions are $w\sim\theta^m$
and $w\sim\theta^{-m}$ (or $w\sim1$ and $w\sim\ln\theta$ in the case
$m=0$).  The condition that $w$ be bounded selects the regular
solution.  This, and a similar condition at the south pole, provide
the two boundary conditions for the eigenvalue problem.

In a differentially rotating planet, a corotation resonance can occur,
in principle, on any surface of constant angular velocity.  As noted
by Goldreich \& Nicholson (1989), this feature is expected to present
an absorbing barrier to propagating waves.  In this paper we disregard
the possibility of corotation resonances, but their influence merits
further investigation, which we defer to future work.

The operator ${\cal L}$ is self-adjoint in the sense that
\begin{equation}
  \int_0^\pi u^*({\cal L}v)\sin\theta\,d\theta=
  \left[\int_0^\pi v^*({\cal L}u)\sin\theta\,d\theta\right]^*
\end{equation}
for any two functions $u(\theta)$ and $v(\theta)$ satisfying the
regularity conditions at the poles.  The eigenvalues $\lambda$ are
therefore real and the eigenfunctions corresponding to distinct
eigenvalues are orthogonal in the sense that
\begin{equation}
  \int_0^\pi w_i^*w_j\sin\theta\,d\theta=0,\qquad i\ne j.
\end{equation}
Furthermore, if $w(\theta)$ is a
function satisfying the regularity conditions,
\begin{equation}
  \int_0^\pi w^*({\cal L}w)\sin\theta\,d\theta=
  \int_0^\pi\left[{{1}\over{D}}\left|\left({{\partial}\over{\partial\theta}}-
  {{2m\Omega_{(1)}\cos\theta}\over
  {\hat\omega_{(1)}\sin\theta}}\right)w\right|^2+
  {{m^2|w|^2}\over{\hat\omega_{(1)}^2\sin^2\theta}}\right]\sin\theta\,d\theta.
\end{equation}
This equation shows that the eigenvalues of ${\cal L}$ are all
positive when $D>0$ throughout $0<\theta<\pi$.  If $D$ changes sign in
$0<\theta<\pi$, there will be sequences of positive and negative
eigenvalues.

Later, we will need to know under what conditions the operator ${\cal
  L}$ can be inverted.  Provided that the eigenfunctions form a
complete set within the relevant function space, the inversion is
possible unless ${\cal L}$ has a null eigenvalue $\lambda=0$, in which
case the inversion can be performed only within the subspace of
functions orthogonal to the null eigenfunction.  In the case $m=0$,
the only possible null eigenfunction is $w_0={\rm constant}$, which
exists for all values of $\omega_{(1)}$.  In the case $m\ne0$, this
solution does not occur, but there may nevertheless be null
eigensolutions for isolated values of $\omega_{(1)}$.

In the case of a non-rotating planet, ${\cal L}$ reduces to a multiple
of the horizontal Laplacian operator and the eigensolutions are
\begin{equation}
  \lambda_i={{n(n+1)}\over{\omega_{(1)}^2}},\qquad
  w_i(\theta)=\tilde P^m_n(\cos\theta),\qquad
  n=m+i,\qquad i\ge0.
  \label{eigen}
\end{equation}
In the case of a uniformly rotating planet, or one in which the
angular velocity is independent of $\theta$, ${\cal L}$ reduces to
Laplace's tidal operator and the eigenfunctions are the Hough
functions (e.g., Chapman \& Lindzen 1970).  We note that, in the case
$m=0$, Laplace's tidal operator has null eigenfunctions when
$\hat\omega_{(1)}=-2m\Omega_{(1)}/n(n+1)$ for integers $n\ge m$.
These solutions are known variously as planetary waves, Rossby waves,
r~modes, or toroidal modes, and correspond to global modes (hence
$k\to0$) of a uniformly rotating radiative region (e.g., Papaloizou \&
Pringle 1978).

For a general pattern of differential rotation in which $\Omega$
depends on $\theta$, we call the eigensolutions `generalized Hough
modes'.  The general solution in radiative regions involves a linear
combination of generalized Hough waves.

\subsection{Matching between convective and radiative regions}

\label{matching}

In planets and stars, convective and radiative regions can be disposed
in a variety of configurations, and planets may also contain solid
cores.  We consider in detail the case in which a single radiative
envelope exists exterior to a single convective region, possibly
containing a solid core.  This is relevant to short-period extrasolar
planets and also to high-mass stars, although in the former case the
radiative envelope is shallow.  The same principles can be used to
construct solutions for alternative arrangements of convective and
radiative regions.

Near the boundary $r=r_{\rm b}$ between convective and radiative
regions, we assume that $N^2$ is zero on the convective side and rises
linearly from zero on the radiative side, so that
\begin{equation}
  {\cal D}={{dN^2}\over{dr}}\Bigg|_{r=r_{\rm b}+}>0.
\end{equation}
As $r\to r_{\rm b}$ from above, the wavelengths of all the generalized
Hough modes tend to infinity because $k\propto N$, and the
approximations adopted in Section~\ref{radiative} break down.  The
situation is related to the problem of Lindblad resonances in
differentially rotating discs (e.g., Goldreich \& Tremaine 1978; Lin
\& Papaloizou 1979a; Lubow \& Ogilvie 1998).  By analogy with that
problem, the characteristic radial extent of the transition region can
be identified as being of order $\epsilon^{2/3}$.  We therefore write
\begin{equation}
  r=r_{\rm b}+\epsilon^{2/3}x,
\end{equation}
where $x$ is an inner variable, of order unity within the transition
region.  We then have
\begin{equation}
  N^2=\epsilon^{2/3}{\cal D}x+O(\epsilon^{4/3}).
\end{equation}

We pose the following expansions for the perturbations within the
transition region:
\begin{eqnarray}
  u_r'&\sim&\epsilon^{5/3}\bar u_r'(x,\theta),\nonumber\\
  u_\theta'&\sim&\epsilon\bar u_\theta'(x,\theta),\nonumber\\
  u_\phi'&\sim&\epsilon\bar u_\phi'(x,\theta),\nonumber\\
  \rho'&\sim&\epsilon^{4/3}\bar\rho'(x,\theta),\nonumber\\
  p'&\sim&\epsilon^2\bar p'(x,\theta),\nonumber\\
  \Phi'&\sim&\epsilon^2\check\Phi'(r_{\rm b},\theta)+
  \epsilon^{8/3}\bar\Phi'(x,\theta).
  \label{ordering_t}
\end{eqnarray}
Note again that the term $\epsilon^2\check\Phi'(r_{\rm b},\theta)$ is
the smooth gravitational potential perturbation associated with the
inertial waves in convective regions, evaluated at the boundary.

The linearized equations of Section~\ref{linearized} then reduce to
\begin{equation}
  0=-{{\partial\bar W}\over{\partial x}}-{{g\bar\rho'}\over{\rho_0}},
\end{equation}
\begin{equation}
  -i\hat\omega_{(1)}\bar u_\theta'-2\Omega_{(1)}\cos\theta\,\bar u_\phi'=
  -{{1}\over{r}}{{\partial\bar W}\over{\partial\theta}},
\end{equation}
\begin{equation}
  -i\hat\omega_{(1)}\bar u_\phi'+{{\bar u_\theta'}\over{\sin\theta}}
  {{\partial}\over{\partial\theta}}(\Omega_{(1)}\sin^2\theta)=
  -{{im\bar W}\over{r\sin\theta}},
\end{equation}
\begin{equation}
  {{\partial\bar u_r'}\over{\partial x}}+
  {{1}\over{r\sin\theta}}{{\partial}\over{\partial\theta}}
  (\bar u_\theta'\sin\theta)+{{im\bar u_\phi'}\over{r\sin\theta}}=0,
\end{equation}
\begin{equation}
  i\hat\omega_{(1)}{{\bar\rho'}\over{\rho_0}}+
  {{{\cal D}x}\over{g}}\bar u_r'=0,
  \label{rho43}
\end{equation}
\begin{equation}
  {{\partial^2\bar\Phi'}\over{\partial x^2}}=4\pi G\bar\rho',
\end{equation}
with
\begin{equation}
  \bar W={{\bar p'}\over{\rho_0}}+\check\Phi'.
\end{equation}
In these equations all coefficients are to be considered independent
of $x$, except where it appears explicitly, and are to be evaluated at
$r=r_{\rm b}$.  Elimination of perturbations in favor of $\bar W$ now
yields
\begin{equation}
  {\cal L}\bar W+{{r^2}\over{\cal D}}{{\partial}\over{\partial x}}
  \left({{1}\over{x}}{{\partial\bar W}\over{\partial x}}\right)=0,
  \label{lw2}
\end{equation}
where, again, all coefficients are to be evaluated at $r=r_{\rm b}$.
This is the equivalent of equation (\ref{lw}) in the transition region.

Equation (\ref{lw2}) can be solved by separation of variables.  Let
$\kappa_i$ be the real solution of
\begin{equation}
  \kappa_i^3=-{{{\cal D}\lambda_i}\over{r^2}}.
\end{equation}
Then the general solution of equation (\ref{lw2}), regular at the
poles, is
\begin{equation}
  \bar W=\sum_i\bar W^{(i)}=
  \sum_ia_i\left[{\rm Ai}'(\kappa_ix)+s_ii\,{\rm Bi}'(\kappa_ix)\right]
  w_i(\theta),
\end{equation}
where $\{a_i\}$ and $\{s_i\}$ are arbitrary coefficients, and ${\rm
  Ai}'$ and ${\rm Bi}'$ denote the derivatives of the Airy functions
of the first and second kinds.  This solution must be matched to that
in the radiative region at $x\gg1$, and to the solution in the
convective region at $x=0$.

For generalized Hough modes with $\lambda>0$, $\kappa$ is negative and
the wave propagates in the radiative region $x>0$.  If the boundary
condition at the outer surface of the radiative region is such that
the waves are not reflected, the relevant solution is a traveling wave
with outwardly directed group velocity.  This is achieved by setting
either $s_i=\pm1$, because the asymptotic form of the Airy functions
is such that, for $\kappa<0$,
\begin{equation}
  {\rm Ai}'(\kappa x)\pm i\,{\rm Bi}'(\kappa x)\sim-\pi^{-1/2}(-\kappa x)^{1/4}
  \exp\left\{\mp i\left[{{2}\over{3}}(-\kappa x)^{3/2}+
  {{\pi}\over{4}}\right]\right\}
\end{equation}
as $x\to\infty$.  This matches on to the WKB form of the solution in
the radiative region, because the wavenumber can be identified as
\begin{equation}
  \mp\epsilon^{-2/3}{{d}\over{dx}}
  \left[{{2}\over{3}}(-\kappa x)^{3/2}+{{\pi}\over{4}}\right]\sim
  \mp\epsilon^{-1}{{\lambda^{1/2}N}\over{r}},
\end{equation}
in agreement with the WKB wavenumber $\epsilon^{-1}k$ given by
equation (\ref{dispersion}).  In fact the relevant solution has
$s_i={\rm sgn}(\hat\omega_{(1)})$.

For modes with $\lambda<0$, $\kappa$ is positive and the wave is
evanescent in the radiative region.  The relevant solution is then
$s_i=0$, because ${\rm Ai}'(\kappa x)$ decays exponentially as $\kappa
x\to\infty$, while ${\rm Bi}'(\kappa x)$ grows exponentially.

To match the solution in the transition region to that in the
convective region at $x=0$, we first compare the ordering schemes
(\ref{ordering_c}) and (\ref{ordering_t}), and find that $u_r'$ and
$\rho'$ are formally mismatched in order.  Now equation (\ref{rho43})
implies that $\bar\rho'\to0$ as $x\to0$, which resolves the mismatch
in $\rho'$.  The matching of $u_r'$ implies that the effective
boundary condition for inertial waves in the convective region is that
\begin{equation}
  \check u_r'(r_{\rm b},\theta)=0.
  \label{rigid}
\end{equation}
Matching of the other variables is achieved by requiring that $W$ be
continuous at $r=r_{\rm b}$, i.e.,
\begin{equation}
  \check W(r_{\rm b},\theta)=\bar W(0,\theta).
\end{equation}
Note that, if $W$ is continuous across the spherical boundary,
$u_\theta'$ and $u_\phi'$ are also continuous because they are related
to one another only by horizontal derivatives, and in exactly the same
way on each side of the boundary.  The matching condition
\begin{equation}
  \check W(r_{\rm b},\theta)=\sum_ia_i\left[{\rm Ai}'(0)+
  s_ii\,{\rm Bi}'(0)\right]w_i(\theta)
\end{equation}
determines the amplitude coefficients according to the orthogonal
projection
\begin{equation}
  a_i={{\int_0^\pi w_i^*\check W(r_{\rm b},\theta)\,\sin\theta\,d\theta}\over
  {\left[{\rm Ai}'(0)+s_ii\,{\rm Bi}'(0)\right]
  \int_0^\pi|w_i|^2\,\sin\theta\,d\theta}}.
  \label{ai}
\end{equation}

The physical interpretation of the matching conditions is that the
transition region appears as a rigid boundary for inertial waves in
the convective region because the steeply rising entropy gradient
suppresses radial motions.  The generalized Hough waves in the
transition and radiative regions are excited by the modified pressure
perturbation $W$ acting at the boundary $r=r_{\rm b}$.

The outward energy flux in generalized Hough waves is found by
evaluating equation (\ref{flux}) in the transition region at large
$x$, with the result
\begin{eqnarray}
  F&=&\epsilon^{11/3}\rho_0\omega_{(1)}\,{\rm sgn}(\hat\omega_{(1)})
  \left({{r^2}\over{{\cal D}}}\right)^{1/3}
  \sum_i\lambda_i^{2/3}|a_i|^2\int_0^\pi|w_i|^2\,\sin\theta\,d\theta\nonumber\\
  &=&\epsilon^{11/3}{{3^{2/3}}\over{4}}
  \left[\Gamma\left({{1}\over{3}}\right)\right]^2\rho_0
  \omega_{(1)}\,{\rm sgn}(\hat\omega_{(1)})
  \left({{r^2}\over{{\cal D}}}\right)^{1/3}
  \sum_i\lambda_i^{2/3}{{\left|\int_0^\pi w_i^*\check W(r_{\rm b},\theta)
  \,\sin\theta\,d\theta\right|^2}\over
  {\int_0^\pi|w_i|^2\,\sin\theta\,d\theta}},
\end{eqnarray}
where the sum includes only those modes for which $\lambda>0$, and all
quantities are to be evaluated at $r=r_{\rm b}$.  Comparison with
equation (\ref{flux_wkb}) indicates that $s=-1/6$, as anticipated
above.  A more detailed matching between the transition region and the
radiative region is not necessary for our purposes, but the constancy
of the energy flux could be used to determine how the amplitudes of
the Hough modes vary with radius throughout the radiative region.

It is likely that the waves will become nonlinear and be damped in the
outer layers of the planet.  In this case the emission of generalized
Hough waves at the boundary between convective and radiative regions
provides a dissipation mechanism for the inertial waves.  The rate of
energy loss in Hough modes is formally of order $\epsilon^{11/3}$,
while the energy content of the inertial modes is of order
$\epsilon^2$.  This route of dissipation must be compared with viscous
or turbulent damping of the inertial waves in the convective region
itself.

\subsection{Viscosity and inertial waves}

\label{viscosity}

The procedure suggested by the foregoing analysis is to solve the
inertial wave problem defined by equations (\ref{ur1'})--(\ref{rho2'})
in the convective region, using the rigid boundary condition
(\ref{rigid}) at the outer surface.  If the planet has a solid core, a
rigid boundary condition is also appropriate there.  In the absence of
a core, regularity conditions apply at $r=0$.  The amplitudes of the
generalized Hough modes in the radiative region can then be computed
from equation (\ref{ai}).

In a uniformly rotating planet, equations (\ref{ur1'})--(\ref{rho2'})
admit a dense or continuous spectrum of inertial wave solutions (e.g.,
Papaloizou \& Pringle 1982) occupying the frequency interval
$-2|\Omega_{(1)}|\le\hat\omega_{(1)}\le2|\Omega_{(1)}|$.  In a
differentially rotating planet, a similar phenomenon occurs, although
the extent of the spectrum is slightly different (e.g., Lin,
Papaloizou, \& Kley 1993).  This property is associated with the fact
that the system of equations is spatially hyperbolic, rather than
elliptic, for wave frequencies within the dense or continuous
spectrum.  For such frequencies, the mathematical problem requires the
imposition of Cauchy boundary conditions on an open surface, while
physical considerations lead to rigid (non-Cauchy) boundary conditions
on a closed surface.  The inertial wave problem is therefore
mathematically ill-posed in the absence of viscosity.

When studying the tidally forced problem, we are interested in
calculating the rate of dissipation of energy, or, equivalently, the
tidal torque exerted on the planet.  For effective forcing frequencies
within the range of the dense or continuous spectrum of inertial
waves, the possibility exists of resonantly exciting an infinite
number of modes.  We therefore introduce a small kinematic viscosity
$\nu$ in convective regions, partly to regularize the problem and
partly in an effort to model the effects of turbulent convection on
the wave motion.  In the presence of viscosity, the inertial wave
problem is mathematically well posed and the spectrum is discrete.  As
the viscosity tends to zero the spectrum becomes increasingly dense
(e.g., Dintrans \& Ouyed 2001) and the probability of resonant
excitation of wave modes increases.  It is therefore possible to
envisage a situation in which the dissipation rate would not vanish
linearly with $\nu$ as $\nu\to0$.  Indeed, it might even be
independent of $\nu$ as $\nu\to0$, as occurs in hydrodynamic
turbulence.

In the presence of viscosity it is necessary to introduce additional
boundary conditions for the inertial wave problem.  If the viscosity
is restricted to the convective region $r<r_{\rm b}$, then continuity
of the velocity and of the stress requires that the $r\theta$- and
$r\phi$-components of the viscous stress tensor vanish at $r=r_{\rm
  b}$.  The boundary conditions at $r=r_{\rm b}$ are therefore
\begin{equation}
  \check u_r'=
  {{\partial}\over{\partial r}}\left({{\check u_\theta'}\over{r}}\right)=
  {{\partial}\over{\partial r}}\left({{\check u_\phi'}\over{r}}\right)=0.
  \label{stress-free}
\end{equation}
If the planet has a solid core of radius $r_{\rm c}$, the no-slip
boundary conditions
\begin{equation}
  \check u_r'=\check u_\theta'=\check u_\phi'=0
  \label{no-slip}
\end{equation}
are appropriate at $r=r_{\rm c}$.

\section{TIDALLY FORCED OSCILLATIONS}

\label{tidal}

\subsection{The equilibrium tide}

\label{equilibrium}

The response of a body to tidal forcing is traditionally calculated
using the theory of the {\it equilibrium tide}, in which an initially
spherical body adjusts hydrostatically to the varying tidal potential
(e.g., Darwin 1880).  This approach is usually justified on the basis
that the tidal forcing frequency is much less than the dynamical
frequency of the body.  By introducing an arbitrary phase lag into the
response, usually parametrized by the quality factor $Q$, the theory
allows for a tidal torque to be exerted and energy to be dissipated.
The efficiency of known tidal interactions in the solar system places
constraints on the $Q$-values of the bodies involved (Goldreich \&
Soter 1966).

Such a theory may offer a reasonable description of the tidal response
of solid bodies, in which the dissipation factor $1/Q$ reflects the
imperfect elasticity of the medium.  However, it is quite
inappropriate for fluid bodies, in which the dissipation is of a
different nature and, moreover, a wavelike response, the {\it
  dynamical tide}, occurs at low forcing frequencies (Cowling 1941;
Zahn 1966).  Although, in principle, tidal dissipation in a fluid body
can still be parametrized by $Q$, the required $Q$-value will depend
on the tidal forcing frequency and on the properties of the body in
such a complicated way that the parametrization is of questionable
value.

In modern studies, the response of a fluid body to tidal forcing is
conveniently separated into the equilibrium tide, which represents a
large-scale, quasi-hydrostatic distortion of the body (without
introducing a phase lag), and the dynamical tide, which usually
constitutes a relatively small, mostly wavelike correction that is
critically important because of its contribution to the dissipation
rate.  The equilibrium tide associated with a tidal potential $\Psi$
is a Lagrangian displacement field $\bxi$ having the properties (e.g.,
Goldreich \& Nicholson 1989)
\begin{equation}
  \xi_r=-{{(\Phi_{(e)}'+\Psi)}\over{g}},\qquad
  \nabla\cdot\bxi=0.
\end{equation}
The Eulerian perturbations associated with the equilibrium tide are
\begin{equation}
  \bu'_{(e)}=\epsilon\left[-i\hat\omega_{(1)}\bxi-
  r\sin\theta(\bxi\cdot\nabla\Omega_{(1)})\be_\phi\right],
\end{equation}
\begin{equation}
  \rho'_{(e)}=-\xi_r{{d\rho_0}\over{dr}},\qquad
  p'_{(e)}=-\xi_r{{dp_0}\over{dr}},
\end{equation}
where $\Phi_{(e)}'$ and $\rho'_{(e)}$ are related by
\begin{equation}
  \nabla^2\Phi_{(e)}'=4\pi G\rho'_{(e)}.
\end{equation}
Note that $\nabla\cdot\bxi=0$ implies $\nabla\cdot\bu'_{(e)}=0$, each
implying that the equilibrium tide is incompressible.\footnote{In
  expressions such as $\nabla\cdot\bxi$ and $\nabla^2\Phi_{(e)}'$ the
  azimuthal part of the operator is intended to be included, even
  though a Fourier analysis in $\phi$ has been carried out.}

We consider a single component of the tidal potential in the form of a
spherical harmonic of degree $\ell\ge m$,
\begin{equation}
  \Psi=\tilde\Psi(r)\tilde P^m_\ell(\cos\theta).
\end{equation}
As noted in Section~\ref{components}, the more important components of
the tidal potential have $\ell=2$ and $m=0$, $1$, or $2$.  The
solution for the equilibrium tide is of the form
\begin{eqnarray}
  \xi_r&=&\tilde\xi_r(r)\tilde P^m_\ell(\cos\theta),\nonumber\\
  \xi_\theta&=&\tilde\xi_{\rm h}(r)
  {{d}\over{d\theta}}\tilde P^m_\ell(\cos\theta),\nonumber\\
  \xi_\phi&=&\tilde\xi_{\rm h}(r)
  {{im}\over{\sin\theta}}\tilde P^m_\ell(\cos\theta),\nonumber\\
  \rho_{\rm e}'&=&\tilde\rho_{\rm e}'(r)\tilde P^m_\ell(\cos\theta),\nonumber\\
  p_{\rm e}'&=&\tilde p_{\rm e}'(r)\tilde P^m_\ell(\cos\theta),\nonumber\\
  \Phi_{\rm e}'&=&\tilde\Phi_{\rm e}'(r)\tilde P^m_\ell(\cos\theta),
\end{eqnarray}
where the radial functions satisfy the ordinary differential equations
\begin{equation}
  \tilde\xi_r=-{{(\tilde\Phi_{\rm e}'+\tilde\Psi)}\over{g}},
\end{equation}
\begin{equation}
  {{1}\over{r^2}}{{d}\over{dr}}(r^2\tilde\xi_r)-
  {{\ell(\ell+1)}\over{r}}\tilde\xi_{\rm h}=0,
\end{equation}
\begin{equation}
  \tilde\rho_{\rm e}'=-\tilde\xi_r{{d\rho_0}\over{dr}},
\end{equation}
\begin{equation}
  \tilde p_{\rm e}'=-\tilde\xi_r{{dp_0}\over{dr}},
\end{equation}
\begin{equation}
  {{1}\over{r^2}}{{d}\over{dr}}
  \left(r^2{{d\tilde\Phi_{\rm e}'}\over{dr}}\right)-
  {{\ell(\ell+1)}\over{r^2}}\tilde\Phi_{\rm e}'=4\pi G\tilde\rho_{\rm e}'.
\end{equation}
The boundary conditions are that the solution should be regular at the
center and at the surface of the planet.  At the surface,
$\tilde\Phi_{\rm e}'$ should also match on to a decaying solid
harmonic $\propto r^{-(\ell+1)}$, with continuity in the function and
its first radial derivative.

Note that the equilibrium tide is not affected by rotation, and
involves only a single spherical harmonic.  This is not true of the
dynamical tide.

In both convective and radiative regions, the perturbations
$\rho'_{(e)}$, $p'_{(e)}$, and $\Phi'_{(e)}$ associated with the
equilibrium tide constitute the dominant part of the response to tidal
forcing, and it is convenient to subtract them out.  In radiative
regions it is also convenient to subtract out the radial velocity
perturbation $u'_{r(e)}$, which dominates over the wavelike part.

\subsection{Convective regions}

\label{convective_forced}

In convective regions the forced solution is of the form
\begin{eqnarray}
  u_r'&\sim&\epsilon\check u_r'(r,\theta),\nonumber\\
  u_\theta'&\sim&\epsilon\check u_\theta'(r,\theta),\nonumber\\
  u_\phi'&\sim&\epsilon\check u_\phi'(r,\theta),\nonumber\\
  \rho'-\rho'_{(e)}&\sim&\epsilon^2\check\rho'(r,\theta),\nonumber\\
  p'-p'_{(e)}&\sim&\epsilon^2\check p'(r,\theta),\nonumber\\
  \Phi'-\Phi'_{(e)}&\sim&\epsilon^2\check\Phi'(r,\theta),\nonumber\\
  W&\sim&\epsilon^2\check W(r,\theta).
\end{eqnarray}
The linearized equations of Section~\ref{linearized} reduce to
\begin{equation}
  -i\hat\omega_{(1)}\check u_r'-2\Omega_{(1)}\sin\theta\,\check u_\phi'=
  -{{\partial\check W}\over{\partial r}},
  \label{dt1}
\end{equation}
\begin{equation}
  -i\hat\omega_{(1)}\check u_\theta'-2\Omega_{(1)}\cos\theta\,\check u_\phi'=
  -{{1}\over{r}}{{\partial\check W}\over{\partial\theta}},
  \label{dt2}
\end{equation}
\begin{equation}
  -i\hat\omega_{(1)}\check u_\phi'+{{1}\over{r\sin\theta}}
  \left(\check u_r'{{\partial}\over{\partial r}}+
  {{\check u_\theta'}\over{r}}{{\partial}\over{\partial\theta}}\right)
  (\Omega_{(1)}r^2\sin^2\theta)=-{{im\check W}\over{r\sin\theta}},
  \label{dt3}
\end{equation}
\begin{equation}
  {{1}\over{r^2\rho_0}}{{\partial}\over{\partial r}}(r^2\rho_0\check u_r')+
  {{1}\over{r\sin\theta}}{{\partial}\over{\partial\theta}}
  (\check u_\theta'\sin\theta)+{{im\check u_\phi'}\over{r\sin\theta}}=
  -i\hat\omega_{(1)}\xi_r{{d\ln\rho_0}\over{dr}}.
  \label{dt4}
\end{equation}

Equations (\ref{dt1})--(\ref{dt4}) are identical to
(\ref{ur1'})--(\ref{rho2'}), except for what is effectively a forcing
term on the right-hand side of equation (\ref{dt4}).  The unusual
positioning of the forcing term in the equation of mass conservation
reflects the fact that the dynamical tide is not forced directly by
the tidal potential.  Indeed, the tidal forcing is balanced in the
equation of motion by the pressure and potential perturbations
associated with the equilibrium tide.  However, the time-dependence of
the associated density perturbation provides an unbalanced term in the
equation of mass conservation, which appears as the driving term for
$\bu'$ in the equations as written above.  In fact, the equilibrium
tide $\bu'_{(e)}$ satisfies equation (\ref{dt4}) exactly, but fails to
satisfy the equation of motion because of its contribution to the
inertial terms on the left-hand side.  Therefore the dynamical tide is
forced very indirectly, through these inertial terms.

Note that the dynamical tide $\bu'-\bu'_{(e)}$ in convective regions
is anelastic, like the free oscillations studied in
Section~\ref{convective}, while the equilibrium tide is
incompressible.  The total tide is therefore neither anelastic nor
incompressible, but satisfies
$\nabla\cdot(\rho_0\bu')=\bu'_{(e)}\cdot\nabla\rho_0$, which is
equivalent to equation (\ref{dt4}).  In a similar way, although
self-gravitation is unimportant for the dynamical tide, which
therefore satisfies the Cowling approximation, it cannot be neglected
for the equilibrium tide.

\subsection{Radiative regions}

In radiative regions the forced solution is of the form
\begin{eqnarray}
  u_r'-u_{r(e)}'&\sim&\epsilon^2\hat u_r'(r,\theta)E(r),\nonumber\\
  u_\theta'&\sim&\epsilon\hat u_\theta'(r,\theta)E(r),\nonumber\\
  u_\phi'&\sim&\epsilon\hat u_\phi'(r,\theta)E(r),\nonumber\\
  \rho'-\rho'_{(e)}&\sim&\epsilon\hat\rho'(r,\theta)E(r),\nonumber\\
  p'-p'_{(e)}&\sim&\epsilon^2\hat p'(r,\theta)E(r),\nonumber\\
  \Phi'-\Phi'_{(e)}&\sim&\epsilon^2\check\Phi'(r,\theta)+
  \epsilon^3\hat\Phi'(r,\theta)E(r),\nonumber\\
  W&\sim&\epsilon^2\hat W(r,\theta)E(r),
\end{eqnarray}
with $E(r)$ as in Section~\ref{radiative}.  The linearized equations
then reduce exactly to equations (\ref{rad1})--(\ref{rad8}).  This
implies that the dynamical tide is not directly forced in radiative
regions.

\subsection{Matching between convective and radiative regions}

\label{matching_forced}

In a transition region of the kind considered in
Section~\ref{matching} the forced solution is of the form
\begin{eqnarray}
  u_r'-u_{r(e)}'&\sim&\epsilon^{5/3}\bar u_r'(x,\theta),\nonumber\\
  u_\theta'&\sim&\epsilon\bar u_\theta'(x,\theta),\nonumber\\
  u_\phi'&\sim&\epsilon\bar u_\phi'(x,\theta),\nonumber\\
  \rho'-\rho'_{(e)}&\sim&\epsilon^{4/3}\bar\rho'(x,\theta),\nonumber\\
  p'-p'_{(e)}&\sim&\epsilon^2\bar p'(x,\theta),\nonumber\\
  \Phi'-\Phi'_{(e)}&\sim&\epsilon^2\check\Phi'(r_{\rm b},\theta)+
  \epsilon^{8/3}\bar\Phi'(x,\theta),\nonumber\\
  W&\sim&\epsilon^2\bar W(x,\theta).
\end{eqnarray}
The linearized equations reduce to
\begin{equation}
  0=-{{\partial\bar W}\over{\partial x}}-{{g\bar\rho'}\over{\rho_0}},
  \label{w2_forced}
\end{equation}
\begin{equation}
  -i\hat\omega_{(1)}\bar u_\theta'-2\Omega_{(1)}\cos\theta\,\bar u_\phi'=
  -{{1}\over{r}}{{\partial\bar W}\over{\partial\theta}},
  \label{ut1_forced}
\end{equation}
\begin{equation}
  -i\hat\omega_{(1)}\bar u_\phi'-i\hat\omega_{(1)}\xi_r{{\sin\theta}\over{r}}
  {{\partial}\over{\partial r}}(r^2\Omega_{(1)})+
  {{\bar u_\theta'}\over{\sin\theta}}
  {{\partial}\over{\partial\theta}}(\Omega_{(1)}\sin^2\theta)=
  -{{im\bar W}\over{r\sin\theta}},
\end{equation}
\begin{equation}
  -{{1}\over{r^2}}{{\partial}\over{\partial r}}(i\hat\omega_{(1)}r^2\xi_r)+
  {{\partial\bar u_r'}\over{\partial x}}+
  {{1}\over{r\sin\theta}}{{\partial}\over{\partial\theta}}
  (\bar u_\theta'\sin\theta)+{{im\bar u_\phi'}\over{r\sin\theta}}=0,
  \label{divu_forced}
\end{equation}
\begin{equation}
  i\hat\omega_{(1)}{{\bar\rho'}\over{\rho_0}}+
  {{{\cal D}x}\over{g}}\bar u_r'=0,
  \label{rho43_forced}
\end{equation}
\begin{equation}
  {{\partial^2\bar\Phi'}\over{\partial x^2}}=4\pi G\bar\rho',
\end{equation}
with
\begin{equation}
  \bar W={{\bar p'}\over{\rho_0}}-{{p'_{(e)}\rho_2}\over{\rho_0^2}}+
  \check\Phi'.
\end{equation}
As in Section~\ref{matching}, all coefficients in equations
(\ref{w2_forced})--(\ref{rho43_forced}) are to be considered
independent of $x$, except where it appears explicitly, and are to be
evaluated at $r=r_{\rm b}$.  The equations now contain forcing terms
(those involving $\xi_r$) that are also independent of $x$.  To deal
with these, it is convenient to split the solution into a `particular
solution', which is independent of $x$ and satisfies equations
(\ref{w2_forced})--(\ref{rho43_forced}) as written, and a
`complementary function', which depends on $x$ and satisfies equations
(\ref{w2_forced})--(\ref{rho43_forced}) with the forcing terms
omitted.  Thus
\begin{equation}
  \bar W(x,\theta)=\bar W^{(p)}(\theta)+\bar W^{(c)}(x,\theta),
\end{equation}
and similarly for $\bar u_\theta'$ and $\bar u_\phi'$, while $\bar
u_r'$ and $\bar\rho'$ do not appear in the particular solution.  The
ordinary differential equations defining the particular solution can
be combined, if desired, in the form
\begin{equation}
  {\cal L}\bar W^{(p)}={\cal F},
  \label{lw2p}
\end{equation}
where
\begin{equation}
  {\cal F}=
  {{1}\over{\hat\omega_{(1)}}}{{\partial}\over{\partial r}}
  (\hat\omega_{(1)}r^2\xi_r)+
  {{1}\over{\hat\omega_{(1)}\sin\theta}}{{\partial}\over{\partial\theta}}
  \left[{{2\hat\omega_{(1)}\Omega_{(1)}\sin^2\theta\cos\theta}\over{D}}
  \xi_r{{\partial}\over{\partial r}}(r^2\Omega_{(1)})\right]+
  {{m\hat\omega_{(1)}}\over{D}}\xi_r{{\partial}\over{\partial r}}
  (r^2\Omega_{(1)})
\end{equation}
is a forcing combination independent of $x$.

For the complementary function, elimination of perturbations in favor
of $\bar W^{(c)}$ again yields
\begin{equation}
  {\cal L}\bar W^{(c)}+{{r^2}\over{\cal D}}{{\partial}\over{\partial x}}
  \left({{1}\over{x}}{{\partial\bar W^{(c)}}\over{\partial x}}\right)=0,
\end{equation}
with the solution
\begin{equation}
  \bar W^{(c)}=\sum_ia_i\left[{\rm Ai}'(\kappa_ix)+
  s_ii\,{\rm Bi}'(\kappa_ix)\right]w_i(\theta),
\end{equation}
where $s_i={\rm sgn}(\hat\omega_{(1)})$ for propagating modes with
$\lambda_i>0$, and $s_i=0$ for evanescent modes with $\lambda_i<0$.
The wavelike part of the solution dominates the particular solution at
large $x$.

Matching $W$ to the solution in the convective region at $x=0$
determines the amplitude coefficients according to
\begin{equation}
  a_i={{\int_0^\pi w_i^*\left[\check W(r_{\rm b},\theta)-\bar W^{(p)}
  \right]\,\sin\theta\,d\theta}\over
  {\left[{\rm Ai}'(0)+s_ii\,{\rm Bi}'(0)\right]
  \int_0^\pi|w_i|^2\,\sin\theta\,d\theta}}.
\end{equation}
The outward energy flux in generalized Hough waves is then
\begin{equation}
  F=\epsilon^{11/3}{{3^{2/3}}\over{4}}
  \left[\Gamma\left({{1}\over{3}}\right)\right]^2\rho_0
  \omega_{(1)}\,{\rm sgn}(\hat\omega_{(1)})
  \left({{r^2}\over{{\cal D}}}\right)^{1/3}
  \sum_i\lambda_i^{2/3}{{\left|\int_0^\pi w_i^*
  \left[\check W(r_{\rm b},\theta)-\bar W^{(p)}\right]
  \,\sin\theta\,d\theta\right|^2}\over
  {\int_0^\pi|w_i|^2\,\sin\theta\,d\theta}},
  \label{flux_forced}
\end{equation}
where, again, the sum includes only those modes for which $\lambda>0$,
and all quantities are to be evaluated at $r=r_{\rm b}$.

The physical interpretation of this analysis is that the particular
solution $W^{(p)}$ represents the bulk response of the transition
region to residual tidal forcing (i.e., the failure of the equilibrium
tide to provide an exact solution to the problem), while the wavelike
complementary function is its response to the pressure of the inertial
waves acting at the boundary $r=r_{\rm b}$, as occurs in the case of
free oscillations (Section~\ref{matching}).

As discussed in Section~\ref{radiative}, the inversion of the operator
${\cal L}$ to solve equation (\ref{lw2p}) for $\bar W^{(p)}$ requires
careful interpretation in cases in which ${\cal L}$ has a null
eigenvalue.  In the case $m=0$, it is easy to show that ${\cal F}$ is
orthogonal to the null eigenfunction $w_0={\rm constant}$.  The
solution $\bar W^{(p)}$ therefore exists but is non-unique, as any
multiple of $w_0$ can be added to it.  This ambiguity affects only the
amplitude of this non-propagating mode, however, and not the energy
flux in propagating Hough modes.  In the case $m\ne0$, if ${\cal L}$
has a null eigenvalue for an isolated value of $\omega_{(1)}$, and if
${\cal F}$ has a non-zero projection on to the null eigenfunction, the
corresponding mode will be resonantly excited and the present analysis
breaks down (formally, $W^{(p)}\to\infty$).  This phenomenon, called
`toroidal mode resonance' by Papaloizou \& Savonije (1997), is
discussed further in Section~\ref{solutions}.

The matching of $u_r'$ implies that the boundary condition for tidally
forced inertial waves in the convective region is
\begin{equation}
  \check u_r'-\epsilon^{-1}u'_{r(e)}=0\qquad\hbox{at}\quad r=r_{\rm b},
  \label{rigid_forced}
\end{equation}
The transition region therefore appears as a rigid boundary for the
dynamical tide in the convective region, while the equilibrium tide
continues smoothly across the boundary.  In the presence of an
effective viscosity confined to the convective region, the equivalents of
the boundary conditions (\ref{stress-free}) and (\ref{no-slip}) are
then
\begin{equation}
  \check u_r'-\epsilon^{-1}u'_{r(e)}=
  {{\partial}\over{\partial r}}\left({{\check u_\theta'}\over{r}}\right)=
  {{\partial}\over{\partial r}}\left({{\check u_\phi'}\over{r}}\right)=0\qquad
  \hbox{at}\quad r=r_{\rm b},
  \label{stress-free_forced}
\end{equation}
\begin{equation}
  \check u_r'-\epsilon^{-1}u'_{r(e)}=
  \check u_\theta'-\epsilon^{-1}u'_{\theta(e)}=
  \check u_\phi'-\epsilon^{-1}u'_{\phi(e)}=0\qquad\hbox{at}\quad r=r_{\rm c}.
  \label{no-slip_forced}
\end{equation}
The assumption underlying the boundary condition at $r=r_{\rm c}$ is
that the core is distorted exactly according to the equilibrium tide,
and so the dynamical tide satisfies a no-slip condition.  This
approximation neglects the influence of the rigidity of the core in
modifying its tidal distortion, which is however only a small error
for typical cores of several Earth masses.

\section{NUMERICAL ANALYSIS FOR UNIFORM OR `SHELLULAR' ROTATION}

\label{solutions}

\subsection{Planet model and equilibrium tide}

In the remainder of this paper we construct detailed numerical
solutions for a simple model of a planet.  Where possible, we omit the
ordering parameter $\epsilon$ and the ordering subscripts that were
needed in deriving the equations in previous sections.

Our numerical method is suited to the case of `shellular' rotation in
which the angular velocity $\Omega(r)$ is independent of latitude
(Zahn 1992).  Of course, this prescription includes the simple case of
uniform rotation, which will be the main focus of our initial
investigation.  We consider the planet to be fully convective except
for a small solid core, $r<r_{\rm c}$, and a shallow radiative
envelope, $r_{\rm b}<r<R_1$.  We model the planet as a polytrope of
index $1$, so that the density is
\begin{equation}
  \rho=\left({{\pi M_1}\over{4R_1^3}}\right){{\sin kr}\over{kr}},
\end{equation}
where $k$ is related to the radius $R_1$ of the planet by $kR_1=\pi$.  The
gravitational acceleration is
\begin{equation}
  g={{GM_1}\over{\pi r^2}}(\sin kr-kr\cos kr).
\end{equation}
While it is common practice to model fully convective giant planets as
polytropes of index $1$, we neglect any modifications of this
structure within the convective region associated with the presence of
a core and a radiative envelope.

The tidal potential is of the form
\begin{equation}
  \tilde\Psi={{GM_2}\over{a^3}}Ar^2,
\end{equation}
where $A$ is a dimensionless constant, as in Section~\ref{components}.
The equations for the equilibrium tide (Section~\ref{equilibrium})
imply that
\begin{equation}
  \nabla^2(\Phi'_{\rm e}+\Psi)={{4\pi G}\over{g}}
  {{d\rho}\over{dr}}(\Phi'_{\rm e}+\Psi).
  \label{phie}
\end{equation}
For a polytrope of index $1$ this simplifies to the Helmholtz equation
\begin{equation}
  (\nabla^2+k^2)(\Phi'_{\rm e}+\Psi)=0,
\end{equation}
with the solution (regular at $r=0$)
\begin{equation}
  \tilde\Phi'_{\rm e}+\tilde\Psi={{GM_2}\over{a^3}}
  BR_1^{5/2}r^{-1/2}J_{5/2}(kr),
\end{equation}
where $B$ is a dimensionless constant to be determined.  At $r=R_1$,
$\tilde\Phi_{\rm e}'$ must match smoothly on to a decaying solid
harmonic $\propto r^{-(\ell+1)}$, and therefore
\begin{equation}
  {{d\ln|\tilde\Phi_{\rm e}'|}\over{d\ln r}}=-3\qquad
  \hbox{at}\quad r=R_1.
\end{equation}
This condition determines the ratio $B/A=5/\sqrt{2}$.

The detailed solution for the equilibrium tide is
\begin{equation}
  \tilde\Phi'_{\rm e}={{GM_2}\over{a^3}}A\left\{{{5R_1^2}\over{k^3r^3}}
  \left[(3-k^2r^2)\sin kr-3kr\cos kr\right]-r^2\right\},
\end{equation}
\begin{equation}
  \tilde\xi_r=-\left({{M_2}\over{M_1}}\right)\left({{R_1}\over{a}}\right)^3
  {{5A}\over{k^2r}}
  \left[{{(3-k^2r^2)\sin kr-3kr\cos kr}\over{\sin kr-kr\cos kr}}\right],  
\end{equation}
\begin{equation}
  \tilde\xi_{\rm h}={{1}\over{6r}}{{d}\over{dr}}(r^2\tilde\xi_r).
\end{equation}
We note that, if a less idealized density profile is adopted, the
equilibrium tide must be determined by solving the ordinary
differential equation (\ref{phie}) numerically.

\subsection{Projection of the equations on to spherical harmonics}

The basic equations of Section~\ref{convective_forced} governing
tidally forced inertial waves in the convective region are
\begin{equation}
  -i\hat\omega u_r-2\Omega\sin\theta\,u_\phi=
  -{{\partial W}\over{\partial r}}+{{F_r}\over{\rho}},
  \label{tide1}
\end{equation}
\begin{equation}
  -i\hat\omega u_\theta-2\Omega\cos\theta\,u_\phi=
  -{{1}\over{r}}{{\partial W}\over{\partial\theta}}+{{F_\theta}\over{\rho}},
  \label{tide2}
\end{equation}
\begin{equation}
  -i\hat\omega u_\phi+
  \left(2\Omega+{{d\Omega}\over{d\ln r}}\right)\sin\theta\,u_r+
  2\Omega\cos\theta\,u_\theta=
  -{{imW}\over{r\sin\theta}}+{{F_\phi}\over{\rho}},
  \label{tide3}
\end{equation}
\begin{equation}
  {{1}\over{r^2\rho}}{{\partial}\over{\partial r}}(r^2\rho u_r)+
  {{1}\over{r\sin\theta}}{{\partial}\over{\partial\theta}}(u_\theta\sin\theta)+
  {{imu_\phi}\over{r\sin\theta}}=-i\hat\omega\xi_r{{d\ln\rho}\over{dr}},
  \label{tide4}
\end{equation}
where we specialize to the case of shellular rotation and omit all
unnecessary subscripts and superscripts.  Included here, as explained
in Section~\ref{viscosity}, is a viscous force with components
\begin{eqnarray}
  F_r&=&{{1}\over{r^2\sin\theta}}
  \left[{{\partial}\over{\partial r}}(r^2\sin\theta\,T_{rr})+
  {{\partial}\over{\partial\theta}}(r\sin\theta\,T_{r\theta})+
  imrT_{r\phi}\right]-
  {{T_{\theta\theta}}\over{r}}-{{T_{\phi\phi}}\over{r}},\nonumber\\
  F_\theta&=&{{1}\over{r^3\sin\theta}}
  \left[{{\partial}\over{\partial r}}(r^3\sin\theta\,T_{r\theta})+
  {{\partial}\over{\partial\theta}}(r^2\sin\theta\,T_{\theta\theta})+
  imr^2T_{\theta\phi}\right]-
  {{T_{\phi\phi}\cot\theta}\over{r}},\nonumber\\
  F_\phi&=&{{1}\over{r^3\sin^2\theta}}
  \left[{{\partial}\over{\partial r}}(r^3\sin^2\theta\,T_{r\phi})+
  {{\partial}\over{\partial\theta}}(r^2\sin^2\theta\,T_{\theta\phi})+
  imr^2\sin\theta\,T_{\phi\phi}\right],
\end{eqnarray}
where
\begin{equation}
  {\bf T}=2\mu{\bf S}+\mu_{\rm b}(\nabla\cdot\bu){\bf1}
\end{equation}
is the Navier--Stokes viscous stress tensor, with
\begin{equation}
  {\bf S}={{1}\over{2}}\left[\nabla\bu+(\nabla\bu)^{\rm T}\right]-
  {{1}\over{3}}(\nabla\cdot\bu){\bf1}
\end{equation}
being the traceless shear tensor, $\mu(r)$ the (dynamic) shear
viscosity and $\mu_{\rm b}(r)$ the bulk viscosity.  The components of
${\bf T}$ are
\begin{eqnarray}
  T_{rr}&=&2\mu\left({{\partial u_r}\over{\partial r}}\right)+
  (\mu_{\rm b}-{\textstyle{{2}\over{3}}}\mu)\nabla\cdot\bu,\nonumber\\
  T_{r\theta}&=&\mu\left[r{{\partial}\over{\partial r}}
  \left({{u_\theta}\over{r}}\right)+
  {{1}\over{r}}{{\partial u_r}\over{\partial\theta}}\right],\nonumber\\
  T_{r\phi}&=&\mu\left[r{{\partial}\over{\partial r}}
  \left({{u_\phi}\over{r}}\right)+
  {{imu_r}\over{r\sin\theta}}\right],
  \nonumber\\
  T_{\theta\theta}&=&2\mu\left({{1}\over{r}}
  {{\partial u_\theta}\over{\partial\theta}}+
  {{u_r}\over{r}}\right)+
  (\mu_{\rm b}-{\textstyle{{2}\over{3}}}\mu)\nabla\cdot\bu,\nonumber\\
  T_{\theta\phi}&=&\mu\left[{{\sin\theta}\over{r}}
  {{\partial}\over{\partial\theta}}
  \left({{u_\phi}\over{\sin\theta}}\right)+
  {{imu_\theta}\over{r\sin\theta}}\right],
  \nonumber\\
  T_{\phi\phi}&=&2\mu\left({{imu_\phi}\over{r\sin\theta}}+
  {{u_r}\over{r}}+{{u_\theta\cot\theta}\over{r}}\right)+
  (\mu_{\rm b}-{\textstyle{{2}\over{3}}}\mu)\nabla\cdot\bu,
\end{eqnarray}
with
\begin{equation}
  \nabla\cdot\bu={{1}\over{r^2}}{{\partial}\over{\partial r}}(r^2u_r)+
  {{1}\over{r\sin\theta}}{{\partial}\over{\partial\theta}}
  (u_\theta\sin\theta)+{{imu_\phi}\over{r\sin\theta}}.
\end{equation}

Following the basic idea of Zahn (1966), we decompose the velocity
into `spheroidal' and `toroidal' parts and then project all variables
on to spherical harmonics such that
\begin{eqnarray}
  u_r&=&\sum a_n(r)\tilde P^m_n(\cos\theta),\nonumber\\
  u_\theta&=&r\sum\left[b_n(r){{d}\over{d\theta}}+
  c_n(r){{im}\over{\sin\theta}}\right]\tilde P^m_n(\cos\theta),\nonumber\\
  u_\phi&=&r\sum\left[b_n(r){{im}\over{\sin\theta}}-
  c_n(r){{d}\over{d\theta}}\right]\tilde P^m_n(\cos\theta),\nonumber\\
  W&=&\sum id_n(r)\tilde P^m_n(\cos\theta).
  \label{projection}
\end{eqnarray}
where the summations are over integers $n\ge m$.  Even if the tidal
forcing and equilibrium tide involve only one spherical harmonic, the
Coriolis force couples spherical harmonics of different degrees.  Note
that $c_n$ is the amplitude of the `toroidal' part of the velocity,
which is not present in the equilibrium tide but is induced by
rotation and is incompressible.

To project equations (\ref{tide1})--(\ref{tide4}) on to spherical
harmonics, we require the recurrence relations
\begin{equation}
  \cos\theta\,\tilde P^m_n(\cos\theta)=
  q_{n+1}\tilde P^m_{n+1}(\cos\theta)+q_n\tilde P^m_{n-1}(\cos\theta),
\end{equation}
\begin{equation}
  \sin\theta{{d}\over{d\theta}}\tilde P^m_n(\cos\theta)=
  nq_{n+1}\tilde P^m_{n+1}(\cos\theta)-(n+1)q_n\tilde P^m_{n-1}(\cos\theta),
\end{equation}
for {\it normalized\/} associated Legendre polynomials, where we
define the coefficients
\begin{equation}
  q_n=\left({{n^2-m^2}\over{4n^2-1}}\right)^{1/2},
\end{equation}
and it is to be understood that $\tilde P^m_n(\cos\theta)=0$ for
$n<m$.  To deal with the angular components of the equation of motion,
equations (\ref{tide2}) and (\ref{tide3}), we take combinations that
correspond to `divergence' and `curl':
\begin{eqnarray}
  \lefteqn{{{1}\over{r\sin\theta}}{{\partial}\over{\partial\theta}}
  \left[\sin\theta(-i\hat\omega u_\theta-2\Omega\cos\theta\,u_\phi)\right]}
  &\nonumber\\
  &&+{{im}\over{r\sin\theta}}\left[-i\hat\omega u_\phi+
  \left(2\Omega+{{d\Omega}\over{d\ln r}}\right)\sin\theta\,u_r+
  2\Omega\cos\theta\,u_\theta\right]\nonumber\\
  &&=-{{1}\over{r^2}}\left[{{1}\over{\sin\theta}}
  {{\partial}\over{\partial\theta}}
  \left(\sin\theta{{\partial W}\over{\partial\theta}}\right)-
  {{m^2W}\over{\sin^2\theta}}\right]+
  {{1}\over{r\sin\theta}}{{\partial}\over{\partial\theta}}
  \left({{F_\theta}\over{\rho}}\right)+
  {{im}\over{r\sin\theta}}\left({{F_\phi}\over{\rho}}\right),
  \label{div}
\end{eqnarray}
\begin{eqnarray}
  \lefteqn{{{1}\over{r\sin\theta}}{{\partial}\over{\partial\theta}}
  \left\{\sin\theta\left[-i\hat\omega u_\phi+
  \left(2\Omega+{{d\Omega}\over{d\ln r}}\right)\sin\theta\,u_r+
  2\Omega\cos\theta\,u_\theta\right]\right\}}&\nonumber\\
  &&-{{im}\over{r\sin\theta}}(-i\hat\omega u_\theta-2\Omega\cos\theta\,u_\phi)
  ={{1}\over{r\sin\theta}}{{\partial}\over{\partial\theta}}
  \left({{F_\phi}\over{\rho}}\right)-
  {{im}\over{r\sin\theta}}\left({{F_\theta}\over{\rho}}\right).
  \label{curl}
\end{eqnarray}

Equation (\ref{tide1}) then gives
\begin{eqnarray}
  \lefteqn{-i\hat\omega a_n-2im\Omega rb_n+
  2\Omega r\left[(n-1)q_nc_{n-1}-(n+2)q_{n+1}c_{n+1}\right]}&\nonumber\\
  &&=-i{{dd_n}\over{dr}}+{{1}\over{\rho}}{{d}\over{dr}}
  \left[2\mu{{da_n}\over{dr}}+(\mu_{\rm b}-
  {\textstyle{{2}\over{3}}}\mu)\Delta_n\right]+
  {{4\mu}\over{\rho}}{{d}\over{dr}}\left({{a_n}\over{r}}\right)
  \qquad\qquad\qquad\qquad\nonumber\\
  &&\qquad-n(n+1){{\mu}\over{\rho}}\left[{{a_n}\over{r^2}}+
  r^2{{d}\over{dr}}\left({{b_n}\over{r^2}}\right)\right],
  \label{eq1}
\end{eqnarray}
where
\begin{equation}
  \Delta_n={{1}\over{r^2}}{{d}\over{dr}}(r^2a_n)-n(n+1)b_n.
\end{equation}
Equations (\ref{div}) and (\ref{curl}) give
\begin{eqnarray}
  \lefteqn{-n(n+1)i\hat\omega b_n-
  {{im}\over{r}}\left(2\Omega+{{d\Omega}\over{d\ln r}}\right)a_n-
  2im\Omega b_n}&\nonumber\\
  &&+2\Omega\left[(n-1)(n+1)q_nc_{n-1}+n(n+2)q_{n+1}c_{n+1}\right]=
  -n(n+1){{id_n}\over{r^2}}\nonumber\\
  &&+{{n(n+1)}\over{\rho}}\left\{{{1}\over{r^6}}{{d}\over{dr}}(\mu r^4a_n)+
  (\mu_{\rm b}-{\textstyle{{2}\over{3}}}\mu){{\Delta_n}\over{r^2}}+
  {{1}\over{r^4}}{{d}\over{dr}}\left(\mu r^4{{db_n}\over{dr}}\right)-
  2[n(n+1)-1]{{\mu b_n}\over{r^2}}\right\}\nonumber\\
  \label{eq2}
\end{eqnarray}
and
\begin{eqnarray}
  \lefteqn{-n(n+1)i\hat\omega c_n+{{1}\over{r}}
  \left(2\Omega+{{d\Omega}\over{d\ln r}}\right)\left[
  (n+1)q_na_{n-1}-nq_{n+1}a_{n+1}\right]}&\nonumber\\
  &&-2\Omega\left[(n-1)(n+1)q_nb_{n-1}+
  n(n+2)q_{n+1}b_{n+1}\right]-2im\Omega c_n\nonumber\\
  &&={{n(n+1)}\over{\rho}}\left\{
  {{1}\over{r^4}}{{d}\over{dr}}\left(\mu r^4{{dc_n}\over{dr}}\right)-
  [n(n+1)-2]{{\mu c_n}\over{r^2}}\right\},
  \label{eq3}
\end{eqnarray}
respectively.  Finally, equation (\ref{tide4}) gives
\begin{equation}
  {{1}\over{r^2\rho}}{{d}\over{dr}}(r^2\rho a_n)-n(n+1)b_n=
  -i\hat\omega\tilde\xi_r{{d\ln\rho}\over{dr}}\delta_{n\ell},
  \label{eq4}
\end{equation}
where $\delta$ is the Kronecker delta function.

The couplings are such that, if $\ell=m=2$, the quantities $a_n$,
$b_n$, and $d_n$ are non-zero only for even values $n\ge2$, while
$c_n$ is non-zero only for odd values $n\ge3$.  If $\ell=2$ and $m=1$,
the quantity $c_1$ is also non-zero.  Finally, if $\ell=2$ and $m=0$,
the quantity $d_0$ is also non-zero, but $a_0$ and $b_0$ can be taken
to vanish, and equations (\ref{eq2}) and (\ref{eq4}) are not required
in the case $n=0$.

The boundary conditions (\ref{stress-free_forced}) and
(\ref{no-slip_forced}) translate into
\begin{equation}
  a_n+i\hat\omega\tilde\xi_r\delta_{n\ell}=
  {{a_n}\over{r^2}}+{{db_n}\over{dr}}={{dc_n}\over{dr}}=0
  \qquad\hbox{at}\quad r=r_{\rm b},
\end{equation}
\begin{equation}
  a_n+i\hat\omega\tilde\xi_r\delta_{n\ell}=
  rb_n+i\hat\omega\tilde\xi_{\rm h}\delta_{n\ell}=
  rc_n=0
  \qquad\hbox{at}\quad r=r_{\rm c}.
\end{equation}

The viscous dissipation rate per unit volume is
\begin{equation}
  2\mu{\bf S}^2+\mu_{\rm b}(\nabla\cdot\bu)^2,
\end{equation}
and the time-averaged tidal dissipation rate in the convective region
can be evaluated as
\begin{equation}
  D_{\rm visc}=\pi\int_{r_{\rm c}}^{r_{\rm b}}\sum_nD_n\,r^2\,dr,
\end{equation}
where
\begin{eqnarray}
  \lefteqn{D_n=\mu\Bigg[3\left|{{da_n}\over{dr}}-{{\Delta_n}\over{3}}\right|^2+
  n(n+1)\left(\left|{{a_n}\over{r}}+r{{db_n}\over{dr}}\right|^2+
  \left|r{{dc_n}\over{dr}}\right|^2\right)}&\nonumber\\
  &&\qquad\qquad
  +(n-1)n(n+1)(n+2)\left(|b_n|^2+|c_n|^2\right)\Bigg]+
  \mu_{\rm b}|\Delta_n|^2.
\end{eqnarray}

\subsection{Method of solution}

We truncate the system of equations at even order $L$, so that
$a_n=b_n=c_n=d_n=0$ for $n>L$, and thereby obtain a large but finite
number of linear, inhomogeneous ordinary differential equations
constituting a two-point boundary-value problem.  The order of the
system is in fact $3L$.

\subsubsection{Principal method}

Our principal method of solution makes use of a Chebyshev
pseudospectral approach (e.g., Boyd 2001) as used by, e.g., Rieutord
et al. (2001) in their study of free inertial waves in an
incompressible fluid contained in a spherical annulus.  This method is
well suited to the problem as it provides spectral accuracy, superior
to any finite-difference method with a similar number of radial nodes,
and naturally supplies additional resolution in the boundary layer
just outside the solid core.

We introduce the Chebyshev coordinate $x$ such that $-1<x<1$ and
\begin{equation}
  r=\left({{1-x}\over{2}}\right)r_{\rm c}+
  \left({{1+x}\over{2}}\right)r_{\rm b},
\end{equation}
and reorganize the differential equations by defining a new set of
dependent variables
\begin{equation}
  X_i=c_{2i-1},\qquad
  Y_i=a_{2i},\qquad
  Z_i={{db_{2i}}\over{dr}},\qquad
  1\le i\le{{L}\over{2}},
\end{equation}
using equation (\ref{eq4}) to substitute for $b_n$ wherever it
appears.  For each value of $i$ we then obtain an equation involving
the quantities
\begin{equation}
  {{d^2X_i}\over{dx^2}},\quad
  {{dX_i}\over{dx}},\quad
  X_i,\quad
  {{dY_i}\over{dx}},\quad
  Y_i,\quad
  {{dY_{i-1}}\over{dx}},\quad
  Y_{i-1}
\end{equation}
from equation (\ref{eq3}).  Differentiating equation (\ref{eq4}) with
respect to $r$, we obtain an equation involving the quantities
\begin{equation}
  {{d^2Y_i}\over{dx^2}},\quad
  {{dY_i}\over{dx}},\quad
  Y_i,\quad
  Z_i.
\end{equation}
Finally, eliminating $d_n$ between equations (\ref{eq1}) and
(\ref{eq2}), we obtain an equation involving the quantities
\begin{equation}
  {{dX_i}\over{dx}},\quad
  X_i,\quad
  {{d^2Y_i}\over{dx^2}},\quad
  {{dY_i}\over{dx}},\quad
  Y_i,\quad
  {{d^2Z_i}\over{dx^2}},\quad
  {{dZ_i}\over{dx}},\quad
  Z_i,\quad
  {{dX_{i+1}}\over{dx}},\quad
  X_{i+1}.
\end{equation}
The functions $X_i$, $Y_i$, and $Z_i$, for the appropriate values of
$i$, are represented by their values at the Gauss--Lobatto collocation
nodes
\begin{equation}
  x_j=\cos\left({{j\pi}\over{N}}\right),\qquad 0\le j\le N,
\end{equation}
where $N$ is the Chebyshev truncation order.  Each differential
equation is then represented at each of the internal collocation
nodes, using the first and second Chebyshev collocation derivative
matrices.  At the boundary nodes, the boundary conditions are
represented instead, again using the Chebyshev collocation derivative
where needed.

The result of this procedure is a block-tridiagonal system of linear,
inhomogeneous algebraic equations.  Each block is of dimension
$3(N+1)$ and relates to the functions $c_{n-1}$, $a_n$, and $b_n$ for
a single value of $n$.  There are $L$ blocks along the diagonal, and
blocks immediately above and below the diagonal, which originate from
the Coriolis terms in the equation of motion.  The blocks are dense
owing to the structure of the Chebyshev collocation derivative matrix.
We solve the system by a standard procedure for block-tridiagonal
matrices, which involves $O(LN^3)$ operations and has a memory
requirement $O(LN^2)$, although this can be reduced to $O(N^2)$ if
ample temporary storage is available on disk.

\subsubsection{Alternative method}

Our alternative method is based on a direct integration of the
differential equations using a fifth-order Runge--Kutta method with
adaptive stepsize control.  This method does not work when the
viscosity is very small, because the equations are then very stiff,
but we use it to verify the results obtained by the principal method
in an accessible parameter regime.

We recast the system as a set of $3L$ first-order differential
equations for the functions
\begin{equation}
  a_{2i},\quad
  b_{2i},\quad
  c_{2i-1},\quad
  d_{2i},\quad
  \nu r^4{{db_{2i}}\over{dr}},\quad
  \nu r^4{{dc_{2i-1}}\over{dr}},\qquad
  1\le i\le{{L}\over{2}},
\end{equation}
where $\nu=\mu/\rho$ is the kinematic viscosity.  To eliminate
$da_n/dr$ and $d^2a_n/dr^2$ from the equations, we use equation
(\ref{eq4}) to express these quantities in terms of $b_n$ and
$db_n/dr$.  To start the integration at $r=r_{\rm b}$ we must guess
$3L/2$ initial values in addition to those specified by the boundary
conditions, and there are $3L/2$ corresponding boundary conditions to
satisfy at $r=r_{\rm c}$.  As the problem is linear, a conventional
shooting method is not required.  Instead, we integrate the equations
once from $r_{\rm c}$ to $r_{\rm b}$ using null initial values, to
obtain a `particular solution'.  We then integrate the equations
$3L/2$ times, omitting the forcing terms, setting a different one of
the initial values to unity each time (the others being null).  This
approach generates a set of `complementary functions' that spans the
space of admissible initial values.  The desired solution is then the
particular solution plus a linear combination of the complementary
functions; the coefficients of the complementary functions are found
by imposing the boundary conditions at $r_{\rm b}$, requiring the
inversion of a square matrix of dimension $3L/2$.

\subsection{Computation of the Hough functions}

\label{hough}

In order to determine the amplitudes with which the Hough waves are
excited in the radiative region, according to equation
(\ref{flux_forced}), the numerical solution at the boundary $r=r_{\rm
  b}$ must be projected on to the Hough functions.  In the case of
shellular rotation, the operator ${\cal L}$ given by equation
(\ref{call}) reduces to Laplace's tidal operator, the eigenfunctions
of which have often been expanded in associated Legendre polynomials
(e.g., Longuet-Higgins 1968 and references therein).  We make use of
the identity
\begin{equation}
  \tilde P^m_n(\cos\theta)=\hat\omega_{(1)}^2{\cal L}\left[
  A_n\tilde P^m_n(\cos\theta)+B_{n+1}\tilde P^m_{n+2}(\cos\theta)+
  B_{n-1}\tilde P^m_{n-2}(\cos\theta)\right],
  \label{identity}
\end{equation}
valid for $n\ge2$, where
\begin{equation}
  A_n={{1}\over{n^2(n+1)^2}}\left\{n(n+1)+fm-
  f^2\left[{{(n-1)^2(n+1)^2q_n^2}\over{(n-1)n+fm}}+
  {{n^2(n+2)^2q_{n+1}^2}\over{(n+1)(n+2)+fm}}\right]\right\},
\end{equation}
\begin{equation}
  B_n=-{{f^2q_nq_{n+1}}\over{n(n+1)+fm}},
\end{equation}
and $f=2\Omega_{(1)}/\hat\omega_{(1)}$.  Equation (\ref{identity}) is readily
verified by direct substitution.  It becomes indeterminate in the case
$m\ne0$ when $f=-n(n+1)/m$ for integers $n\ge m$, owing to the
existence of a null eigenfunction of ${\cal L}$ corresponding to a
toroidal mode, as discussed in Section~\ref{radiative}.

To solve the eigenvalue problem ${\cal L}w=\lambda w$ we consider the
equivalent problem $(\hat\omega_{(1)}^2{\cal L})^{-1}w=\Lambda w$,
where $\Lambda=(\hat\omega_{(1)}^2\lambda)^{-1}$.  In cases when
${\cal L}$ has a null eigenvalue, we consider the restriction of this
equation to the subspace of functions orthogonal to the null
eigenfunction.  Expanding $w$ in normalized associated Legendre
polynomials,
\begin{equation}
  w=\sum e_n\tilde P^m_n(\cos\theta),
  \label{wsum}
\end{equation}
where the sum is over even integers $n\ge 2$, we obtain from equation
(\ref{identity})
\begin{equation}
  A_ne_n+B_{n-1}e_{n-2}+B_{n+1}e_{n+2}=\Lambda e_n.
\end{equation}
We truncate at the same order $n=N$ as for the solution in the
convective region, and thereby obtain an eigenvalue problem involving
a tridiagonal, real symmetric matrix,
\begin{equation}
  \left[\matrix{
    A_2    &B_3    &0      &0      &\cdots &0      &0\cr
    B_3    &A_4    &B_5    &0      &\cdots &0      &0\cr
    0      &B_5    &A_6    &B_7    &\cdots &0      &0\cr
    0      &0      &B_7    &A_8    &\cdots &0      &0\cr
    \cdots &\cdots &\cdots &\cdots &\cdots &\cdots &\cdots\cr
    0      &0      &0      &0      &\cdots &A_{N-2}&B_{N-1}\cr
    0      &0      &0      &0      &\cdots &B_{N-1}&A_N\cr}\right]
  \left[\matrix{
  e_2\cr e_4\cr e_6\cr e_8\cr \cdots\cr e_{N-2}\cr e_N\cr}\right]=
  \Lambda\left[\matrix{
  e_2\cr e_4\cr e_6\cr e_8\cr \cdots\cr e_{N-2}\cr e_N\cr}\right].
  \label{hough_matrix}
\end{equation}
It is clear from this representation that, in the case of a
non-rotating planet ($f=0$), the matrix is diagonal and the
eigenvalues are the diagonal elements $\left[n(n+1)\right]^{-1}$, in
agreement with equation (\ref{eigen}).  More generally, the
eigensolutions of equation (\ref{hough_matrix}) are readily obtained
by standard numerical methods.  The inner product in equation
(\ref{flux_forced}) is easily computed in the Legendre polynomial
representation, because
\begin{equation}
  \int_0^\pi\left[\sum_na_n\tilde P^m_n(\cos\theta)\right]^*
  \left[\sum_{n'}b_{n'}\tilde P^m_{n'}(\cos\theta)\right]\sin\theta\,d\theta=
  \sum_na_n^*b_n.
\end{equation}

To determine the particular solution in the transition region, rather
than working with equation (\ref{lw2p}), we project equations
(\ref{ut1_forced})--(\ref{divu_forced}) for the particular solution on
to spherical harmonics in the manner of equation (\ref{projection}),
leading to (in an obvious notation)
\begin{eqnarray}
  \lefteqn{-n(n+1)i\hat\omega b_n^{(p)}-
  {{im}\over{r}}\left(2\Omega+{{d\Omega}\over{d\ln r}}\right)\tilde a_n-
  2im\Omega b_n^{(p)}}&\nonumber\\
  &&+2\Omega\left[(n-1)(n+1)q_nc_{n-1}^{(p)}+
  n(n+2)q_{n+1}c_{n+1}^{(p)}\right]=-n(n+1){{id_n^{(p)}}\over{r^2}},
\end{eqnarray}
\begin{eqnarray}
  \lefteqn{-n(n+1)i\hat\omega c_n^{(p)}+
  {{1}\over{r}}\left(2\Omega+{{d\Omega}\over{d\ln r}}\right)\left[
  (n+1)q_n\tilde a_{n-1}-nq_{n+1}\tilde a_{n+1}\right]}&\nonumber\\
  &&-2\Omega\left[(n-1)(n+1)q_nb_{n-1}^{(p)}+
  n(n+2)q_{n+1}b_{n+1}^{(p)}\right]-2im\Omega c_n^{(p)}=0,
  \label{cnp}
\end{eqnarray}
\begin{equation}
  {{1}\over{r^2}}{{d}\over{dr}}(r^2\tilde a_n)-n(n+1)b_n^{(p)}=0,
\end{equation}
where $\tilde a_n=-i\hat\omega\tilde\xi_r\delta_{n\ell}$.  It is then
straightforward to solve algebraically for the coefficients
$d_n^{(p)}$ defining the desired function $\bar W^{(p)}(\theta)$.
When $\ell=2$, the only non-zero coefficients are $\tilde a_2$,
$b_2^{(p)}$, $c_1^{(p)}$ (when $m<2$), $c_3^{(p)}$, $d_2^{(p)}$, and
$d_4^{(p)}$.  In the course of obtaining this solution, a singularity
arises if the coefficient of $c_n^{(p)}$ in equation (\ref{cnp})
vanishes.  This gives rise to a `toroidal mode resonance' as discussed
in Section~\ref{matching_forced}.  The only such resonances for
$\ell=2$ tidal forcing in a planet with shellular rotation occur at
$\hat\omega=-\Omega/3$ (when $m=2$), or at $\hat\omega=-\Omega$ and
$\hat\omega=-\Omega/6$ (when $m=1$).

\section{NUMERICAL RESULTS}

\subsection{Model parameters}

We present results for a uniformly rotating planet with a solid core
of fractional radius $r_{\rm c}/R_1=0.2$ and a convective--radiative
boundary of fractional radius $r_{\rm b}/R_1=0.9$.  These values are
suggested by the models of short-period extrasolar planets constructed
by Bodenheimer, Lin, \& Mardling (2001).

The kinematic shear viscosity $\nu$ is assumed to be uniform and is
parametrized by the dimensionless Ekman number
\begin{equation}
  Ek={{\nu}\over{2\Omega R_1^2}},
\end{equation}
while the bulk viscosity, which we found to be unimportant, is set to
zero.  Current models (Guillot et al. 2003) suggest that the
microscopic viscosity in the interior of Jupiter is of the order of
$10^{-2}\,{\rm cm}^2\,{\rm s}^{-1}$, corresponding to
$Ek\sim10^{-18}$.  The eddy viscosity associated with turbulent
convection is much larger, corresponding to $Ek\sim10^{-7}$.  However,
the characteristic timescale of the eddies is a year or longer, making
the convection inefficient at damping tidal disturbances with typical
periods of a few days.  Therefore the estimate $Ek\sim10^{-7}$ should
probably be reduced by several orders of magnitude, perhaps using the
prescription of Goldreich \& Keeley (1977).  In this initial study, we
make no attempt to model the detailed variations of the effective
viscosity, but investigate the systematic behavior of the solutions as
the uniform kinematic viscosity is reduced into the regime $Ek\ll1$.

Almost all the results reported here are obtained by the principal
method, using a standard numerical resolution of $L=N=100$, for which
each calculation takes approximately 30~seconds on a modern
workstation.  As discussed below, this resolution allows numerically
converged results to be obtained at Ekman numbers at least as small as
$10^{-7}$.  We have also carried out a small number of calculations
for isolated forcing frequencies at a very high resolution of
$L=N=500$, each taking several hours on a workstation, but allowing
access to yet smaller Ekman numbers.

We have computed the tidal dissipation rate by viscous dissipation in
the convective region and via the emission of Hough waves at the
convective--radiative boundary for forcing frequencies in the range
$-3\Omega<\hat\omega<3\Omega$, comfortably spanning the spectrum of
inertial waves.  Only the most important case of $m=2$ forcing was
considered.  To investigate the effect of a small but non-zero
viscosity, the calculations were carried out for $Ek=10^{-4}$,
$10^{-5}$, $10^{-6}$, and $10^{-7}$.

In the case of the uniformly rotating planet, the calculations were
also repeated omitting certain terms in the governing equations, in
order to test the validity of earlier and more tractable approaches to
the problem.  In the {\it traditional approximation\/} (e.g., Chapman
\& Lindzen 1970) the latitudinal component of the angular velocity is
neglected in computing the Coriolis force.  This amounts to neglecting
the $\Omega\sin\theta$ terms in equations (\ref{tide1}) and
(\ref{tide3}), while the Hough modes are unaffected.  The traditional
approximation, where valid, is useful because it allows the linearized
equations to be solved by separation of variables, although we do not
make use of that property here.  In the {\it no-Coriolis
  approximation\/} the Coriolis force is completely neglected and the
Hough modes reduce to spherical harmonics.  This is equivalent to
considering a non-rotating planet except that the Doppler shift of the
forcing frequency is taken into account.

\subsection{Dimensionless dissipation rates}

It is convenient to express the tidal dissipation rate in a
dimensionless form.  The natural unit for viscous dissipation is
\begin{equation}
  U_{\rm visc}=\left({{M_2}\over{M_1}}\right)^2\left({{R_1}\over{a}}\right)^6
  |A|^2M_1R_1^2|\Omega|^3.
\end{equation}
Use of this unit means that the dimensionless viscous dissipation rate
$D_{\rm visc}/U_{\rm visc}$ depends only on $\hat\omega/\Omega$ and
$Ek$ (in principle also on $r_{\rm c}/R_1$ and $r_{\rm b}/R_1$ if
these are varied).  We note that $D_{\rm visc}$ captures the viscous
dissipation of the equilibrium tide in addition to the damping of any
inertial waves that are excited as a dynamical tide, and $D_{\rm
  visc}$ is therefore non-zero even under the no-Coriolis
approximation.

To calculate the dissipation rate via Hough waves, we note that the
waves carry both energy and angular momentum fluxes, their ratio being
the angular pattern speed $\omega/m$.  When the waves damp, their
angular momentum is deposited locally in the planet.  Part of the
energy of the waves is then required in order to change the spin of
the planet, while the remainder is dissipated locally.  A simple
calculation shows that the dissipation rate is
\begin{equation}
  D_{\rm Hough}=\left({{\hat\omega}\over{\omega}}\right)F,
\end{equation}
where $F$ is the energy flux in equation (\ref{flux_forced}) (factors
of $\epsilon$ and ordering subscripts being omitted).

The dissipation rate via the emission of Hough waves is naturally expressed
in terms of the unit
\begin{equation}
  U_{\rm Hough}=f_{\rm Hough}U_{\rm visc},
  \label{u_hough}
\end{equation}
where
\begin{equation}
  f_{\rm Hough}={{\rho(r_{\rm b})}\over{\rho(0)}}
  \left({{\Omega^2}\over{r_{\rm b}{\cal D}}}\right)^{1/3}.
\end{equation}
is a model-dependent, dimensionless quantity.  The first factor in
$f_{\rm Hough}$, being the ratio of the density at the
convective--radiative boundary to the central density of the
polytropic model, is highly uncertain, being possibly in the range
$10^{-4}-10^{-3}$ for short-period extrasolar planets, according to
the models of spherically irradiated and tidally heated planets by
Bodenheimer et al. (2001).  The second factor involves the quantity
${\cal D}$, which is also very uncertain but is raised only to a weak
power.  For planets that are distant from their host stars, such as
Jupiter, $f_{\rm Hough}$ is probably so small that tidal dissipation
via Hough waves is entirely negligible.

According to the conventional theory of the equilibrium tide, in which
the efficiency of tidal dissipation is parametrized by a quality
factor $Q$ (e.g., Goldreich \& Soter 1966), the tidal dissipation rate
in a body of negligible rigidity would be
\begin{equation}
  \left({{5k_2}\over{4Q}}\right){{GM_2^2R_1^5}\over{a^6}}|A|^2|\hat\omega|=
  {{1}\over{f_QQ}}\left|{{\hat\omega}\over{\Omega}}\right|U_{\rm visc},
\end{equation}
where $k_2$ is the second-order Love number, and
\begin{equation}
  f_Q=\left({{4}\over{5k_2}}\right){{\Omega^2R_1^3}\over{GM_1}}
\end{equation}
is a dimensionless number of order $\epsilon^2$.  We can therefore
express the numerically determined tidal dissipation rates in terms of
effective $Q$-values given by
\begin{equation}
  Q_{\rm visc}^{-1}=f_Q\left({{D_{\rm visc}}\over{U_{\rm visc}}}\right)
  \left|{{\Omega}\over{\hat\omega}}\right|,\qquad
  Q_{\rm Hough}^{-1}=f_Qf_{\rm Hough}
  \left({{D_{\rm Hough}}\over{U_{\rm Hough}}}\right)
  \left|{{\Omega}\over{\hat\omega}}\right|.
  \label{qeff}
\end{equation}
For Jupiter, using the volumetric radius for $R_1$ and the value
$k_2\approx0.38$ from Gavrilov \& Zharkov (1977), we obtain
$f_Q\approx0.18$; when Jupiter is spun down to a period of three days,
illustrative of a short-period extrasolar planet, we obtain
$f_Q\approx0.0033$.

We note that Goldreich \& Soter (1966) assume that $k_2=3/2$, as for a
homogeneous body, and therefore the effective $Q$-values appropriate
for use with their formulae are approximately four times larger than
the values we quote below.

\subsection{Standard model and variations}

Fig.~2 shows the viscous dissipation rate as a function of forcing
frequency, for the standard model.  The four panels correspond to
Ekman numbers $10^{-4}$, $10^{-5}$, $10^{-6}$, and $10^{-7}$.  The
quantity plotted is the base-10 logarithm of $(D_{\rm visc}/U_{\rm
  visc})|\Omega/\hat\omega|$, which is directly related to the
effective $Q$-value.  The dotted line in panel (d) corresponds to a
fiducial value of $Q_{\rm visc}=10^5$ for Jupiter, or approximately
$50$ times higher for a short-period extrasolar planet.  Fig.~3 shows
the dissipation rate via Hough waves for the same model.  Although the
values obtained are much larger, they must be scaled down by the small
and uncertain factor $f_{\rm Hough}$ before they can be compared with
Fig.~2.  Figs~4--7 show the equivalent results obtained when either
the no-Coriolis approximation or the traditional approximation is
adopted.

It is clear from Fig.~2 that the results become more complicated, and
the computations more challenging, as the viscosity is reduced.  The
numerical convergence of the solutions is examined in two resolution
studies in Figs~8 and~9.  The first of these (Fig.~8) is carried out
at a forcing frequency $\hat\omega=\Omega$ within the spectrum of
inertial waves but not associated with an obvious resonant feature.
The figure shows that higher resolution is required at lower Ekman
numbers, and that the dissipation rate can be either underestimated or
overestimated if the resolution is insufficient.  The standard
resolution $L=N=100$ clearly gives numerically converged results for
all the Ekman numbers considered in Fig.~2.  The second resolution
study (Fig.~9) is carried out at a forcing frequency
$\hat\omega=0.489\,\Omega$, which lies near the peak of the largest
resonant feature seen in Fig.~2.  In this case the standard resolution
is just adequate to capture the peak dissipation rate at $Ek=10^{-7}$
with a small error.

Fig.~10 shows, as a dotted line, an expanded view of part of panel (d)
of Fig.~2, the graph of viscous dissipation rate for $Ek=10^{-7}$.
The solid line, for comparison, comes from a calculation at
$Ek=10^{-8}$, for which the increased numerical resolution $L=N=200$
was necessary.

Finally, Fig.~11 illustrates the spatial structure of the velocity
field induced by tidal forcing in the convective region.  This
calculation, at $Ek=10^{-9}$, is performed at the very high resolution
of $L=N=500$, and the forcing frequency $\hat\omega=-1.1181\,\Omega$
is chosen to be close to the peak of a large resonance.

\subsection{The Coriolis effect and the traditional approximation}

We begin by comparing the results for the uniformly rotating planet in
the full model (Figs~2 and~3) with those obtained under the
no-Coriolis approximation (Figs~4 and~5).  It is clear that, when the
Coriolis force is included, a number of resonant features are
superimposed on the otherwise smooth variation of the viscous
dissipation rate with forcing frequency.  As expected, the resonances
are restricted to the interval $-2\le\hat\omega/\Omega\le2$
corresponding to the spectrum of inertial waves in the convective
region.  For $Ek=10^{-4}$ the inertial waves show up as two resonant
peaks in the dissipation rate.  However, as $Ek$ is reduced further, a
host of resonances come into play.  This trend is consistent with the
idea that the spectrum of inertial waves is dense or continuous in the
absence of viscosity, while it consists of a discrete set of damped
modes in the presence of viscosity.  When $Ek=10^{-4}$ the discrete
nature of the modes is clearly evident, but when $Ek=10^{-7}$ the
individual resonances can barely be discerned, except for the most
prominent examples, and a resonant response is almost guaranteed for
any forcing frequency within the spectrum of inertial waves.  Under
the no-Coriolis approximation the convective regions of the planet
have no wavelike response and no dynamical tide is excited.  The slow,
large-scale motion associated with the equilibrium tide is damped by
viscosity, leading to a dissipation rate that vanishes linearly with
the viscosity.  The same dependence does not occur in the full model
for frequencies within the spectrum of inertial waves.  This behaviour
reflects the fact that the characteristic spatial scale of the tidally
forced inertial waves diminishes as the viscosity is reduced.  Rather
as occurs in hydrodynamic turbulence, the dissipation rate then has a
non-trivial finite limit as the viscosity tends to zero.

The dissipation rate via Hough waves is differently normalized, owing
to the model-dependent prefactor $f_{\rm Hough}$.  If this factor is
of the order of $10^{-4}-10^{-3}$, as may occur in short-period
extrasolar planets, the Hough dissipation rate could dominate over the
viscous dissipation rate.  A comparison of Figs~2 and~3 indicates that
the Hough dissipation rate responds to the inertial-mode resonances
when the full Coriolis effect is used.  This result reflects the fact
that the Hough waves are excited partly by the pressure of the
inertial waves at the convective--radiative boundary.  These resonant
features become stronger and sharper as the Ekman number is reduced,
but otherwise the Hough dissipation rate is essentially independent of
the viscosity.  The feature at $\hat\omega=-\Omega/3$ is the toroidal
mode resonance mentioned in Section~\ref{hough}.  In the neighborhood
of this forcing frequency, a global toroidal mode is excited in the
radiative region, contrary to the assumptions of this paper, and the
Hough dissipation rate cannot be trusted.  The no-Coriolis
approximation fails to reproduce any of these features but, generally,
significantly overestimates the Hough dissipation rate.  This
discrepancy may be attributable to the fact that, depending on their
frequency, the Hough modes are confined by the Coriolis force to
propagate only in certain ranges of latitude, unlike the g~modes in a
non-rotating planet.

Now comparing Figs~2 and~3 with Figs~6 and~7, we find that, within the
inertial region, the traditional approximation gives a totally
inaccurate representation of the tidal dissipation rate.  While the
traditional approximation does allow for a resonant response in the
convective region, the details are quite wrong and the viscous
dissipation rate is generally overestimated by about two orders of
magnitude.  This difference may occur because of a artificial
decoupling between the angular and radial structures of the response
that occurs when the traditional approximation is employed.  This
decoupling implies that, when the conditions for resonance are met, a
global amplification of the response occurs.  We conclude that, within
the spectrum of inertial waves, the traditional approximation is, if
anything, even worse than the no-Coriolis approximation, and should
therefore never be used for inertial waves in gaseous giant planets.

\subsection{The low-viscosity limit}

Returning to the full model, we see clearly from the expanded view in
Fig.~10 that the viscous dissipation rate does not simply vanish
linearly with the viscosity.  Instead, as the Ekman number is reduced
(in this case by a factor of $10$) the resonant peaks become more
numerous, sharper, and taller.  The frequency-averaged viscous
dissipation rate for the interval shown in Fig.~10 is very nearly {\it
  independent of the viscosity}, decreasing by approximately four per
cent when the Ekman number is reduced by a factor of $10$.  In fact,
the frequency sampling in Fig.~10 is such that the resonant peaks for
the case $Ek=10^{-8}$ are not perfectly captured, and this inaccuracy
may be sufficient to account for the four per cent reduction.

This finding suggests what may be a very important result: although
the effective $Q$-value derived from the dissipation of inertial waves
at very low Ekman number may be an exceedingly complicated function of
the forcing frequency, a robust average $Q$-value may be obtained that
is asymptotically independent of the viscosity as $Ek\to0$.  In
Appendix~A we present a toy model that may explain this effect.  The
toy model, based on a simplified Cartesian geometry, has the advantage
of possessing an analytical solution from which it can be shown that
the frequency-averaged dissipation rate is exactly independent of the
viscosity.  We suggest that this property may be typical of situations
in which forcing is applied to systems that exhibit a dense or
continuous spectrum in the absence of dissipation.

It is conceivable that the dissipation rate, averaged locally in
frequency, may be essentially independent of the details of the
dissipation mechanism, whether it is due to a Navier--Stokes
viscosity, damping by turbulent convection, or nonlinear wave damping.
We are far from having demonstrated such a result.  Nevertheless, our
results suggest that a $Q$-value of the order of $10^5$ can be
obtained for Jupiter or Saturn through this mechanism.  The value for
a synchronized short-period extrasolar planet would be some $50$ times
greater owing to its slower rotation.

Fig.~11 illustrates the spatial structure of the tidal response in the
convective region.  In line with work by Rieutord et al. (2001) on
free inertial waves in an incompressible fluid contained in a
spherical annulus, we find that the disturbance is localized near
rays, which are the characteristics of the spatially hyperbolic
equations governing inertial waves, and are seen to reflect many times
from the inner and outer boundaries.  In a uniformly rotating planet
the rays are straight and a simple relation exists between the wave
frequency and the angle made by the rays with the vertical.  A mild
near-singularity occurs when the rays approach the axis of rotation.
The overall wave pattern does not appear as neat as that in the
calculations by Rieutord et al. (2001), perhaps because the density is
non-uniform, but more probably because of an interference effect
between the equilibrium and dynamical tides.

\subsection{Effects of differential rotation}

We have also made a very preliminary investigation of the possible
effects of differential rotation.  Unfortunately, without a reliable
theory of how differential rotation is generated by convection and
other processes in giant planets and stars, the choice of angular
velocity profile is unconstrained.  We considered the case of a linear
profile $\Omega(r)$ of angular velocity in the convective region.
There are clear similarities between the graphs of viscous dissipation
rate versus forcing frequency (which we omit here) and those shown in
Fig.~2 for a uniformly rotating planet.  Even with a differential
rotation of $20$ per cent, however, the resonant response is
substantially richer and stronger.  The tidal response also looks
qualitatively similar to that shown in Fig.~11, although the rays are
curved in a differentially rotating planet.

We emphasize that we have not considered the possible effects of
corotation resonances.  While such resonances in the radiative region
may absorb outgoing Hough waves, as noted by Goldreich \& Nicholson
(1989), there may also be a direct tidal torque and energy dissipation
associated with tidal forcing at the corotation resonance, as occurs
in differentially rotating disks (e.g., Goldreich \& Tremaine 1980).
There is also the possibility of an angular velocity that depends on
latitude.  The numerical method used in this paper is not well suited
to such a situation, unless $\Omega$ depends in a sufficiently simple
way on $\theta$ that the couplings between different
spherical-harmonic components of the velocity field are strongly
restricted.  Therefore there remains much to be investigated with
regard to tides in differentially rotating planets and stars.

\section{COMPARISON WITH PREVIOUS WORK}

The previous work most closely related to the present paper is that by
Savonije \& Papaloizou (1997) and Papaloizou \& Savonije (1997).  The
former paper presents a numerical analysis of the linearized response
of a uniformly rotating high-mass star to $\ell=m=2$ tidal forcing.
Unlike the planet considered in our calculations, the star is
predominantly radiative and has only a small convective core.  In
comparison with our study, Savonije \& Papaloizou (1997) take a more
direct numerical approach to the full linearized equations, including
non-adiabatic effects and a small viscosity, but retaining only terms
of first order in the angular velocity.  Perhaps because the
convective core is small, they find that main effects of rotation
occur through the resonant excitation of Hough modes in the radiative
region, rather than inertial modes in the convective region.  Indeed,
they find a number of toroidal mode resonances for negative forcing
frequencies, whereas precisely one such resonance is allowed in our
analysis.  Our calculation is more dedicated to the exacting task of
resolving the tidally forced inertial waves in the convective region
and examining more carefully the limit of small viscosity.  The
reduction of the basic equations that we carry out allows us to solve
the resulting equations much more efficiently, and the numerical
method that we employ is particularly well suited to this problem.

The accompanying paper by Papaloizou \& Savonije (1997) provides an
asymptotic analysis of low-frequency tidal forcing in the same type of
star.  There is much in common between their concisely expressed
analysis and the present paper.  We have attempted to give a fuller
exposition of the problem, formally justifying the approximations
involved and allowing for the important possibility of differential
rotation.  We have thereby generalized the derivation of reduced
equations for inertial modes and Hough modes.  Furthermore, our
analysis should be useful for future direct numerical simulations of
tides, as it clarifies the role of the anelastic and Cowling
approximations.  These approximations are satisfied by the dynamical
tide but not by the equilibrium tide (in Papaloizou \& Savonije 1997
the effect of self-gravitation on the equilibrium tide is neglected).

Related work for non-rotating stars and planets has also been
published by Goldreich \& Nicholson (1989), Lubow et al. (1997), and
Goodman \& Dickson (1998), who arrive at asymptotic expressions for
the tidal dissipation rate through the emission of g~modes at a
convective--radiative interface.  There is a close formal relation
between these expressions and our equation (\ref{flux_forced}).
However, the expression can only be evaluated, in the case of a
rotating star or planet, if the inertial wave problem in the
convective region has been solved numerically.  In this connection it
would be interesting to re-examine the problem of tidal forcing in
rotating solar-type stars with convective envelopes (cf. Terquem et
al. 1998; Goodman \& Dickson 1998) taking into account either uniform
or differential rotation.  In the latter case the angular velocity
profile of the solar convective envelope, inferred from helioseismic
studies, could be used as a guide.

\section{SUMMARY AND CONCLUSIONS}

In this paper we have revisited the classical problem of the response
of a giant planet to low-frequency tidal forcing.  Much of our general
analysis of this problem applies equally to tidal forcing in stars.
The novel feature of our approach is that we take into account the
slow and possibly non-uniform rotation of the planet.  (Here `slow'
rotation means that we include the full Coriolis force but not the
centrifugal distortion of the planet.)  Convective regions of the
planet then support low-frequency inertial waves with an intricate
spatial structure and a rich frequency spectrum, while radiative
regions support (generalized) Hough modes, some of which may be
regarded as g~modes of high radial order, modified by the
(differential) rotation.  We have argued that, in many cases of
interest, such as that of an extrasolar planet that orbits close to
its host star, the effective tidal forcing frequency lies within the
spectrum of inertial waves in the convective regions of the planet,
and a resonant response can be expected.  As a result, the rate of
tidal dissipation may be greatly enhanced relative to a model that
neglects the rotation of the planet.  The dissipation occurs both
through the viscous or turbulent dissipation of the inertial waves in
the convective region and through the emission of generalized Hough
waves that propagate through the radiative envelope towards the
surface, where they presumably damp.  Enhancement of the tidal
dissipation rate implies a more rapid synchronization of the planet's
spin with its orbit, a faster circularization of the orbit, and a more
intense heating of the planet, which may lead in turn to inflation and
even Roche-lobe overflow (Gu et al. 2003).

We have presented a systematic asymptotic analysis of the linearized
response of a differentially rotating planet to tidal forcing with a
frequency that is small compared to the dynamical frequency of the
planet.  The response separates naturally into an equilibrium tide,
which represents a large-scale, quasi-hydrostatic distortion of the
planet in the imposed tidal potential, and a dynamical tide, which
constitutes a mostly wavelike correction.  We obtain the reduced
system of equations governing the dynamical tide in convective and
radiative regions separately, and explain the asymptotic matching
procedure between the two solutions.

In convective regions, which we model as being adiabatically
stratified, the dynamical tide takes the form of an indirectly forced
inertial wave confined in a spherical annulus.  The reduced equations
are intrinsically two-dimensional and require a numerical solution.
Moreover, for effective forcing frequencies within the dense or
continuous spectrum of free inertial waves, the problem is
mathematically ill-posed unless viscosity is included.  A question of
particular interest is how the dissipation rate varies with the
viscosity in the limit that the viscosity tends to zero.

In radiative regions, the dynamical tide involves predominantly
horizontal motions and takes the form of a wave (or evanescent
disturbance) with a short radial wavelength.  A separation of
variables is possible, and we find that the angular structure of the
wave is governed by a generalization of Laplace's tidal operator.  We
call the resulting solutions generalized Hough modes and discuss some
mathematical properties of the operator concerned.

Our analysis clarifies the role in the tidal problem of certain well
known approximations to the equations of fluid dynamics.  We find that
the so-called traditional approximation, in which the latitudinal
component of the angular velocity is neglected, is not applicable to
inertial waves in convective regions and should never be used for this
purpose.  In fact, it gives highly misleading results for the tidal
dissipation rate, perhaps worse than neglecting the Coriolis force
altogether.  On the positive side, we argue that the dynamical tide
satisfies both the anelastic and Cowling approximations, which
eliminate acoustic waves and self-gravitation, respectively, from the
problem.  However, the equilibrium tide satisfies neither of these
approximations.  These considerations are important not only for
analytical or semi-analytical studies but also for direct numerical
simulations of tidal forcing.  Such simulations will be needed in
order to determine properly how the dynamical tide interacts with
turbulent convection, as well as the role of nonlinearity in the tidal
problem.  It is highly convenient and appropriate to simulate
convection with a numerical method based on the anelastic and Cowling
approximations.  Our analysis indicates that the dynamical tide can
also be captured within this approach, and shows how the subtle,
indirect forcing of the dynamical tide by the equilibrium tide can be
achieved through the inertial terms in the equation of motion.

We have presented the results of full numerical calculations of the
tidal response for an idealized model of a giant planet that is
predominantly convective but also contains a solid core and has a thin
radiative envelope.  The numerical method is suited to the case of
`shellular' rotation in which the angular velocity is independent of
latitude, and we have focused mainly on the case of uniform rotation.
High-resolution calculations, using a pseudospectral method, are
required to access the physically interesting regime of small Ekman
number (small viscosity) and reveal the intricate spatial structure of
the inertial waves while properly resolving the dissipative
structures.  We have calculated the tidal dissipation rate, both by
viscous dissipation in the convective region and via the emission of
Hough waves, as a function of the tidal forcing frequency, for the
important case of a tidal potential proportional to the $\ell=m=2$
solid harmonic, and for forcing frequencies that span the spectrum of
inertial waves.  We find that the viscous dissipation rate is strongly
enhanced, relative to a calculation in which the Coriolis force is
neglected, in a number of inertial-mode resonant peaks that become
more numerous, sharper, and taller as the Ekman number is reduced.  As
a result, the viscous dissipation rate does not vanish linearly with
the viscosity.  Depending on the physical conditions at the
convective--radiative boundary, the Hough dissipation rate may
possibly exceed the viscous dissipation rate, and also responds to the
inertial-mode resonances.  A single `toroidal mode resonance' is also
possible, in which residual forcing excites a large-scale mode
(related to a Rossby wave), rather than a short-wavelength response,
in the radiative region.  When the planet rotates differentially, the
inertial-mode resonances are yet stronger and more numerous, and the
viscous dissipation rate is further enhanced.

Examination of the spatial structure of the tidal response in the
convective region shows, in line with work by Rieutord et al. (2001)
on free inertial waves in an incompressible fluid contained in a
spherical annulus, that the disturbance is localized near rays, which
are the characteristics of the spatially hyperbolic equations
governing inertial waves, and are seen to reflect many times from the
inner and outer boundaries.  The rays are straight or curved depending
on whether the planet rotates uniformly or differentially.

The dissipation rate associated with inertial waves at very low Ekman
number is, in principle, a highly erratic function of the forcing
frequency (Fig.~2).  As we will discuss in future work, when realistic
scenarios for tidal evolution are considered, there are various
reasons why it may be more appropriate to apply a smoothed version of
this `fractal' curve rather than taking every resonant peak literally.
We have presented both numerical evidence (Fig.~10), and an analytical
demonstration based on a toy model (Appendix~A), that the
frequency-averaged tidal dissipation rate associated with inertial
waves may be asymptotically independent of the viscosity in the limit
of small Ekman number.  If correct, this result is most important
because the Ekman number based on the microscopic viscosity in the
interior of Jupiter is exceptionally small and beyond the range of any
numerical calculation.  Even the eddy viscosity based on turbulent
convection gives rise to a very small Ekman number, which should
probably be further reduced owing to the mismatch between the typical
tidal forcing frequency and the characteristic timescale of the
convective motion.

The resulting tidal dissipation rates are not adequately represented
by a constant $Q$-value as is commonly adopted in parametrized models.
One reason for this is that the dissipation rate associated with
inertial waves scales naturally with the spin frequency of the planet
in a way that is not captured in the constant-$Q$ model.
Nevertheless, the effective $Q$-values obtained are of the order of
$10^5$ for Jupiter or Saturn, or some $50$ times greater for a
synchronized short-period extrasolar planet owing to its slower
rotation.

These values are probably adequate to explain the historical evolution
and current state of the Galilean satellites (e.g., Peale 1999).  We
consider the alternative model of Ioannou \& Lindzen (1993b)
improbable as it requires the interior of Jupiter to be
subadiabatically stratified, in contradiction of current models
(Guillot et al. 2003) which suggest that it has a minuscule
superadiabatic gradient owing to the high efficiency with which
convection can transport the required heat flux.

In the case of short-period extrasolar planets, a $Q$-value of the
order of $10^7$ following synchronization is probably sufficient to
explain the circularization of their orbits (cf. Marcy et al.  1997).
The additional route of dissipation via the emission of Hough waves in
an outer radiative layer may play a role here, although this mechanism
is subject to significant uncertainties.  The dissipation of inertial
waves in the convective region provides a deep source of heating that
can inflate the planet, as may have been observed in the case of the
transiting planet HD~209458~b (Brown et al. 2001).

In conclusion, we have shown that it is both important and feasible,
although numerically challenging, to include the effects of rotation
when studying the response of a planet or star to tidal forcing.  Our
general analysis lays the ground work for future numerical studies
including more realistic interior models and, possibly, the effects of
differential rotation or nonlinearity.  In our preliminary numerical
investigation we have demonstrated that a robust enhancement of tidal
dissipation results from the inclusion of rotational effects, which
can account for the rapid evolution of tidally interacting systems.
There remains much of interest to be explored in this problem.

\acknowledgments

GIO acknowledges the hospitality of UCO/Lick Observatory, where this
work was initiated, and also the support of the Royal Society through
a University Research Fellowship.  This work is supported in part by
NASA through NAG5-13177 and NSF through NSF-AST-9987417.

\appendix

\section{DISSIPATION OF INERTIAL WAVES IN A TOY MODEL}

In this appendix we present a toy model that explains some aspects of
the viscous dissipation of forced inertial waves.  Although the model
does not capture many of the subtleties of the tidal-forcing problem,
it has the advantage of possessing an analytical solution.

We consider an incompressible fluid of uniform density $\rho$ and
kinematic viscosity $\nu$.  The fluid initially rotates with uniform
angular velocity $\bOmega$, with a pressure gradient balancing the
centrifugal force and any gravitational force.

A body force $\bff$ per unit mass excites small disturbances
satisfying the linearized equations (written in the rotating frame)
\begin{equation}
  {{\partial\bu}\over{\partial t}}+2\bOmega\times\bu=
  -{{1}\over{\rho}}\nabla p+\nu\nabla^2\bu+\bff,
\end{equation}
\begin{equation}
  \nabla\cdot\bu=0,
\end{equation}
where $\bu$ and $p$ are the Eulerian perturbations of velocity and
pressure.

The fluid is contained in a square box, $0<x<L$ and $0<z<L$, referred
to Cartesian coordinates $(x,y,z)$ such that $\bOmega=\Omega\,\be_z$.
For the purposes of illustration, we assume that all quantities are
independent of $y$, and adopt the boundary conditions
\begin{equation}
  u_x=u_y={{\partial u_z}\over{\partial x}}=0
\end{equation}
on $x=0,L$ and
\begin{equation}
  {{\partial u_x}\over{\partial z}}={{\partial u_y}\over{\partial z}}=u_z=0
\end{equation}
on $z=0,L$.  Apart from the first condition on $u_y$, these can be
interpreted as meaning that the boundaries are rigid and stress-free.

We adopt units of mass, length, and time such that $\rho=1$, $L=\pi$
and $2\Omega=1$.  For a periodic forcing, all quantities may be
assumed to have the form
\begin{eqnarray}
  &&u_x={\rm Re}\left[u(x,z)\,e^{-i\omega t}\right],\qquad
  u_y={\rm Re}\left[v(x,z)\,e^{-i\omega t}\right],\nonumber\\
  &&u_z={\rm Re}\left[w(x,z)\,e^{-i\omega t}\right],\qquad
  p={\rm Re}\left[\psi(x,z)\,e^{-i\omega t}\right],\nonumber\\
  &&f_x={\rm Re}\left[f(x,z)\,e^{-i\omega t}\right],\qquad
  f_z={\rm Re}\left[h(x,z)\,e^{-i\omega t}\right].
\end{eqnarray}
We assume for simplicity that $f_y=0$.

The linearized equations then reduce to
\begin{eqnarray}
  -i\omega u-v&=&-\partial_x\psi+\nu(\partial_x^2+\partial_z^2)u+f,\\
  -i\omega v+u&=&\nu(\partial_x^2+\partial_z^2)v,\\
  -i\omega w&=&-\partial_z\psi+\nu(\partial_x^2+\partial_z^2)w+h,\\
  \partial_xu+\partial_zw&=&0,
\end{eqnarray}
subject to
\begin{equation}
  u=v=\partial_xw=0\qquad\hbox{at}\quad x=0,\pi
\end{equation}
and
\begin{equation}
  \partial_zu=\partial_zv=w=0\qquad\hbox{at}\quad z=0,\pi.
\end{equation}

For free modes of oscillation ($f=h=0$), solutions exist of the form
\begin{eqnarray}
  &&u=u_{mn}\sin mx\cos nz,\qquad
  v=v_{mn}\sin mx\cos nz,\nonumber\\
  &&w=w_{mn}\cos mx\sin nz,\qquad
  \psi=\psi_{mn}\cos mx\cos nz,
\end{eqnarray}
with $m$ and $n$ non-negative integers, not both equal to zero.  The
dispersion relation of these viscously damped inertial modes is
\begin{equation}
  \tilde\omega^2={{n^2}\over{m^2+n^2}},
\end{equation}
where
\begin{equation}
  \tilde\omega=\omega+i\nu(m^2+n^2).
\end{equation}
In the absence of viscosity, the eigenfrequencies are real and dense
in the interval $-1\le\omega\le1$.  Viscous damping causes the
eigenfrequencies to move below the real axis, increasingly so for
modes of higher `order'.

We return to the forced problem where $\omega$ is real and prescribed.
The force components $f$ and $h$ can be expanded in Fourier series
(the sums are over non-negative integers $m,n$)
\begin{eqnarray}
  f&=&\sum f_{mn}\sin mx\cos nz,\nonumber\\
  h&=&\sum h_{mn}\cos mx\sin nz.
\end{eqnarray}
The solution to the forced problem is then
\begin{eqnarray}
  u&=&\sum u_{mn}\sin mx\cos nz,\nonumber\\
  v&=&\sum v_{mn}\sin mx\cos nz,\nonumber\\
  w&=&\sum w_{mn}\cos mx\sin nz,\nonumber\\
  \psi&=&\sum \psi_{mn}\cos mx\cos nz,
\end{eqnarray}
with
\begin{equation}
  u_{mn}={{i\tilde\omega n(nf_{mn}-mh_{mn})}\over
  {(m^2+n^2)\tilde\omega^2-n^2}},
\end{equation}
\begin{equation}
  v_{mn}={{n(nf_{mn}-mh_{mn})}\over{(m^2+n^2)\tilde\omega^2-n^2}},
\end{equation}
\begin{equation}
  w_{mn}=-{{i\tilde\omega m(nf_{mn}-mh_{mn})}\over
  {(m^2+n^2)\tilde\omega^2-n^2}},
\end{equation}
\begin{equation}
  \psi_{mn}={{n(1-\tilde\omega^2)h_{mn}-m\tilde\omega^2f_{mn}}
  \over{(m^2+n^2)\tilde\omega^2-n^2}}.
\end{equation}

Of particular interest is the average rate of viscous dissipation per
unit volume,
\begin{equation}
  D={{1}\over{\pi^2}}\int_0^\pi\int_0^\pi\langle-\bu\cdot\nu\nabla^2\bu\rangle
  \,dx\,dz.
\end{equation}
This evaluates to
\begin{equation}
  D={{\nu}\over{8}}\sum(m^2+n^2)\left(|u_{mn}^2|+|v_{mn}^2|+|w_{mn}^2|\right),
\end{equation}
and can then be related to the forcing in the form
\begin{equation}
  D={{\nu}\over{8}}\sum{{(m^2+n^2)|\tilde\omega^2|+n^2}
  \over{|(m^2+n^2)\tilde\omega^2-n^2|^2}}\,(m^2+n^2)|nf_{mn}-mh_{mn}|^2.
  \label{D}
\end{equation}

The variation of $D$ with $\omega$ and $\nu$ is very complicated.  As
$\nu\to0$ for fixed $\omega$, $D$ eventually tends linearly to zero,
unless $\omega$ is one of the eigenvalues of the inviscid problem
(which form a dense set of measure zero), in which case $D$ tends to
infinity.  However, an `average' dissipation rate associated with the
spectrum of inertial waves can be identified by integrating $D$ over
all positive (or negative) frequencies, with the simple result
\begin{equation}
  \int_0^\infty D\,d\omega={{\pi}\over{16}}\sum
  {{|nf_{mn}-mh_{mn}|^2}\over{m^2+n^2}},
\end{equation}
which is {\it exactly independent of the viscosity}.

Fig.~12 illustrates how $D$ varies with $\omega$ for several values of
$\nu$, for a simple forcing of the form $f_{mn}=1/mn^2$, $h_{mn}=0$.
(In fact, for generic forcing, $f_{mn}=O(1/mn^2)$ for large $m,n$.)
As seen in the tidal-forcing problem, the dissipation is enhanced in
the neighborhood of inertial-mode resonances in the interval
$-1\le\omega\le1$.  As $\nu$ is reduced the resonant peaks become more
numerous, sharper, and taller, but the integrated dissipation rate
remains exactly the same.  In fact, the contribution of each mode to
the integrated dissipation rate is independent of the viscosity, and
the strongest resonances, which are with modes of lowest `order',
retain their identity and importance as $\nu$ is reduced.

\newpage

\begin{figure}
  \plotone{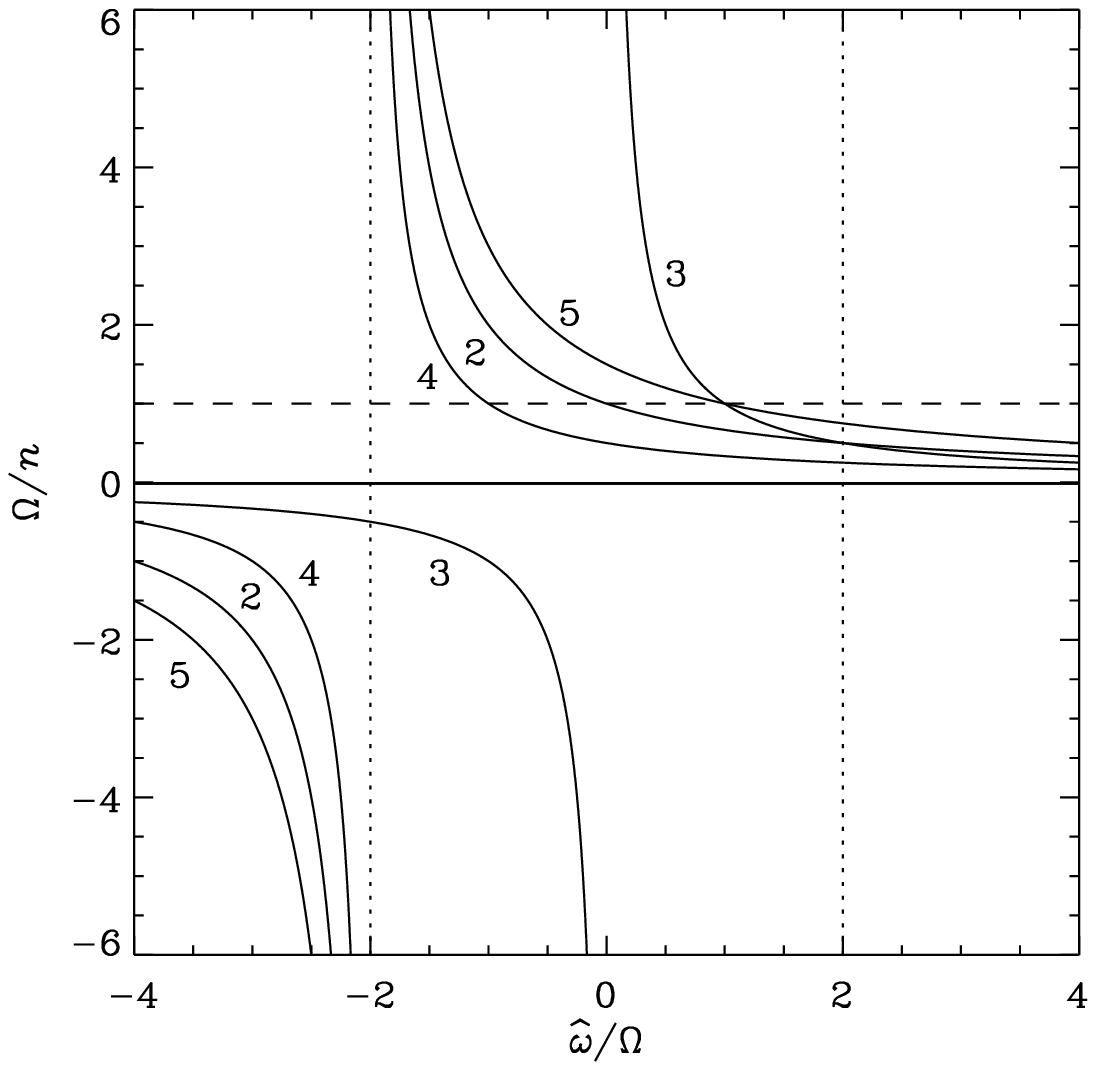} \figcaption{Tidal forcing frequencies and the
    spectrum of inertial waves in a uniformly rotating planet.  The
    ratio of the (spin) angular velocity of the planet to the mean
    motion of the orbit is plotted against the ratio of the effective
    forcing frequency to the angular velocity of the planet, for
    components 2--5 of the tidal potential.  The dotted lines indicate
    the extent of the spectrum of inertial waves.  A short-period
    extrasolar planet being spun down towards the synchronized state
    descends the tracks from near the top of the diagram to the dashed
    line.}
\end{figure}

\newpage

\begin{figure}
  \plotone{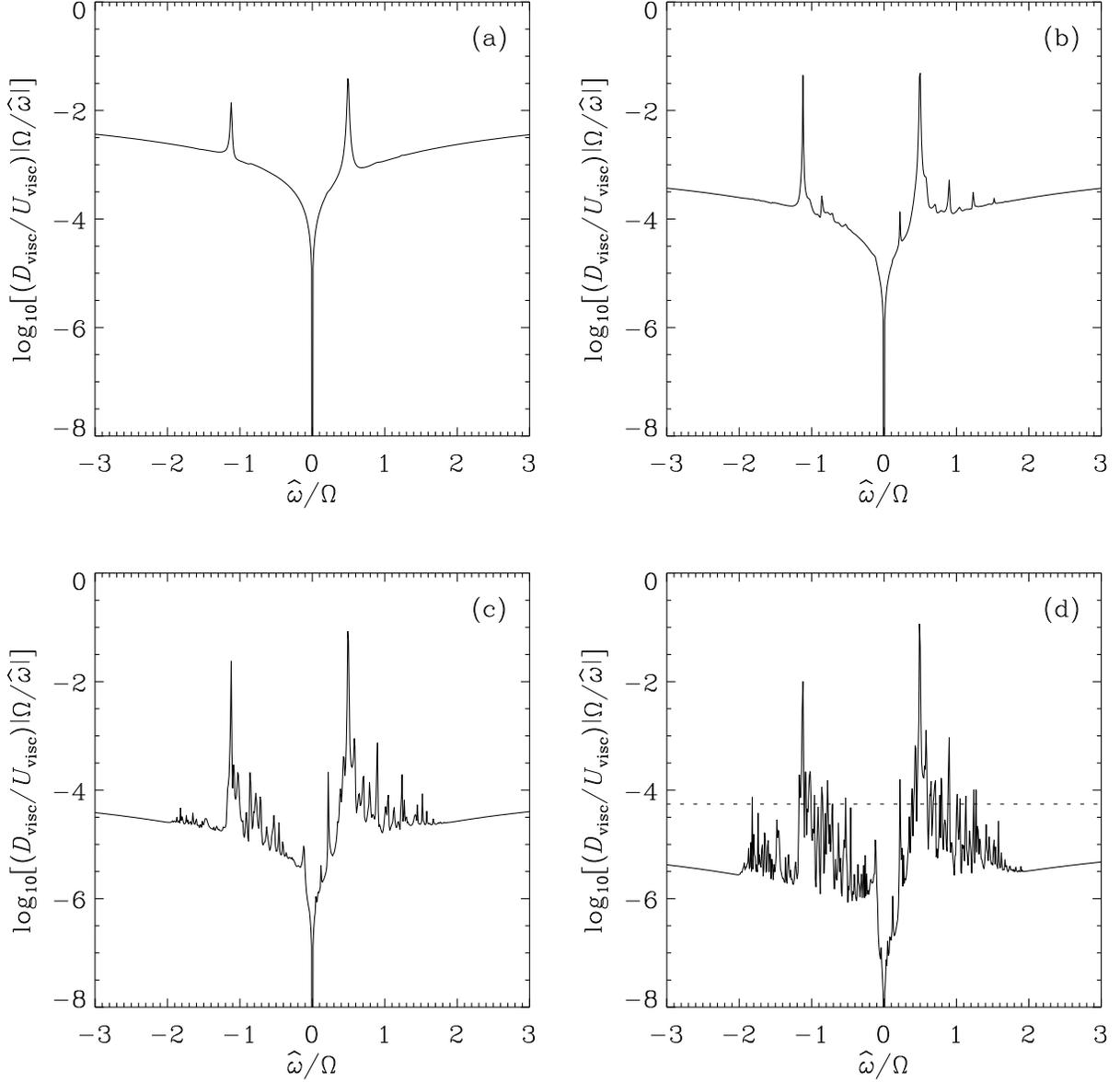} \figcaption{Tidal dissipation rate by viscosity
    in a uniformly rotating planet, according to the full model.  The
    dissipation rate is scaled for comparison with equation
    (\ref{qeff}).  Panels (a), (b), (c), and (d) are for Ekman numbers
    $10^{-4}$, $10^{-5}$, $10^{-6}$, and $10^{-7}$, respectively.  The
    spectrum of inertial waves corresponds to the interval $[-2,2]$.
    The dotted line in panel (d) indicates the dissipation rate in
    Jupiter corresponding to an effective quality factor $Q_{\rm
      visc}=10^5$.}
\end{figure}

\begin{figure}
  \plotone{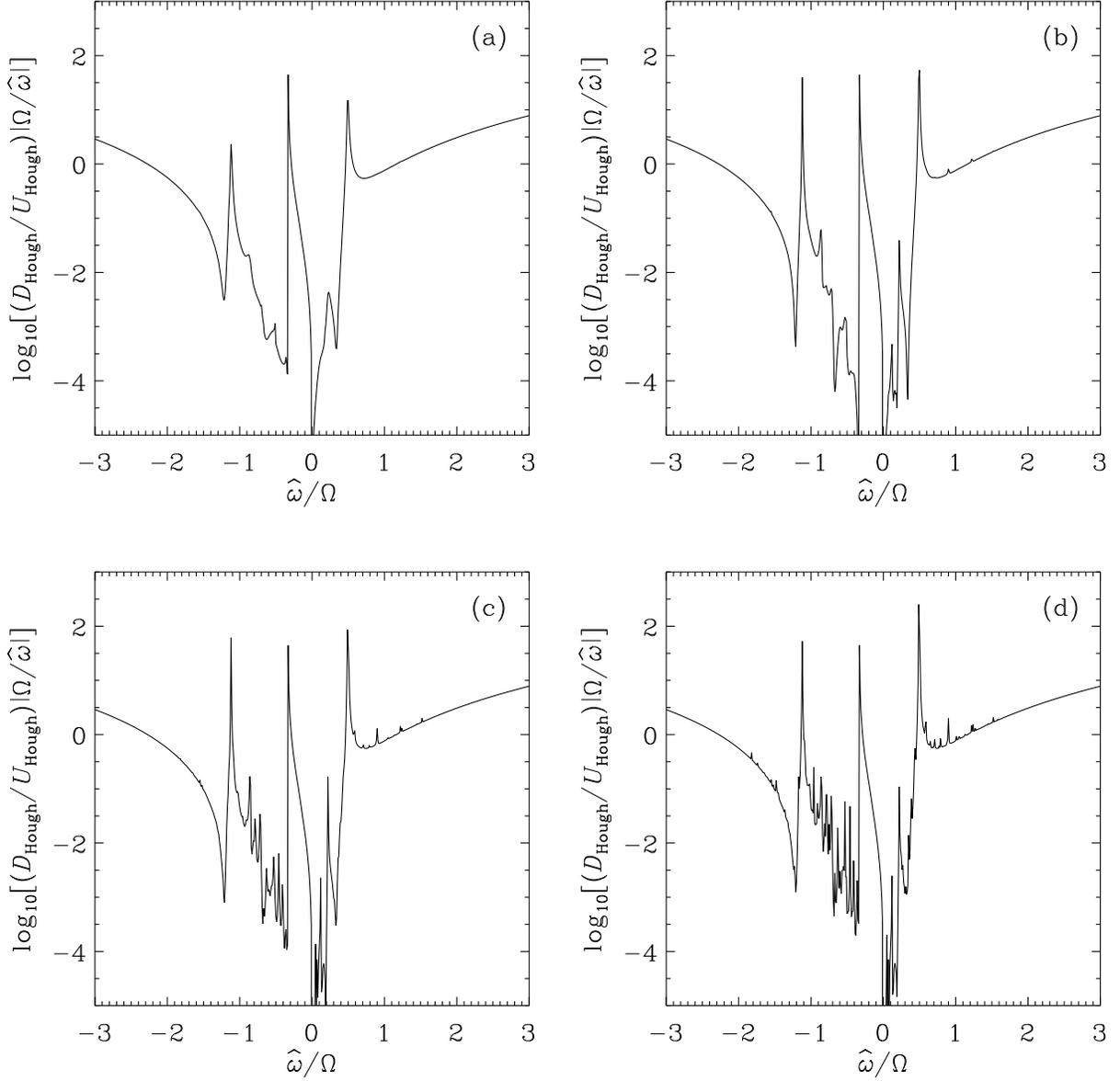} \figcaption{Tidal dissipation rate via Hough
    waves in a uniformly rotating planet, according to the full model.
    The dissipation rate is scaled for comparison with equation
    (\ref{qeff}).  Panels (a), (b), (c), and (d) are for Ekman numbers
    $10^{-4}$, $10^{-5}$, $10^{-6}$, and $10^{-7}$, respectively.}
\end{figure}

\newpage

\begin{figure}
  \plotone{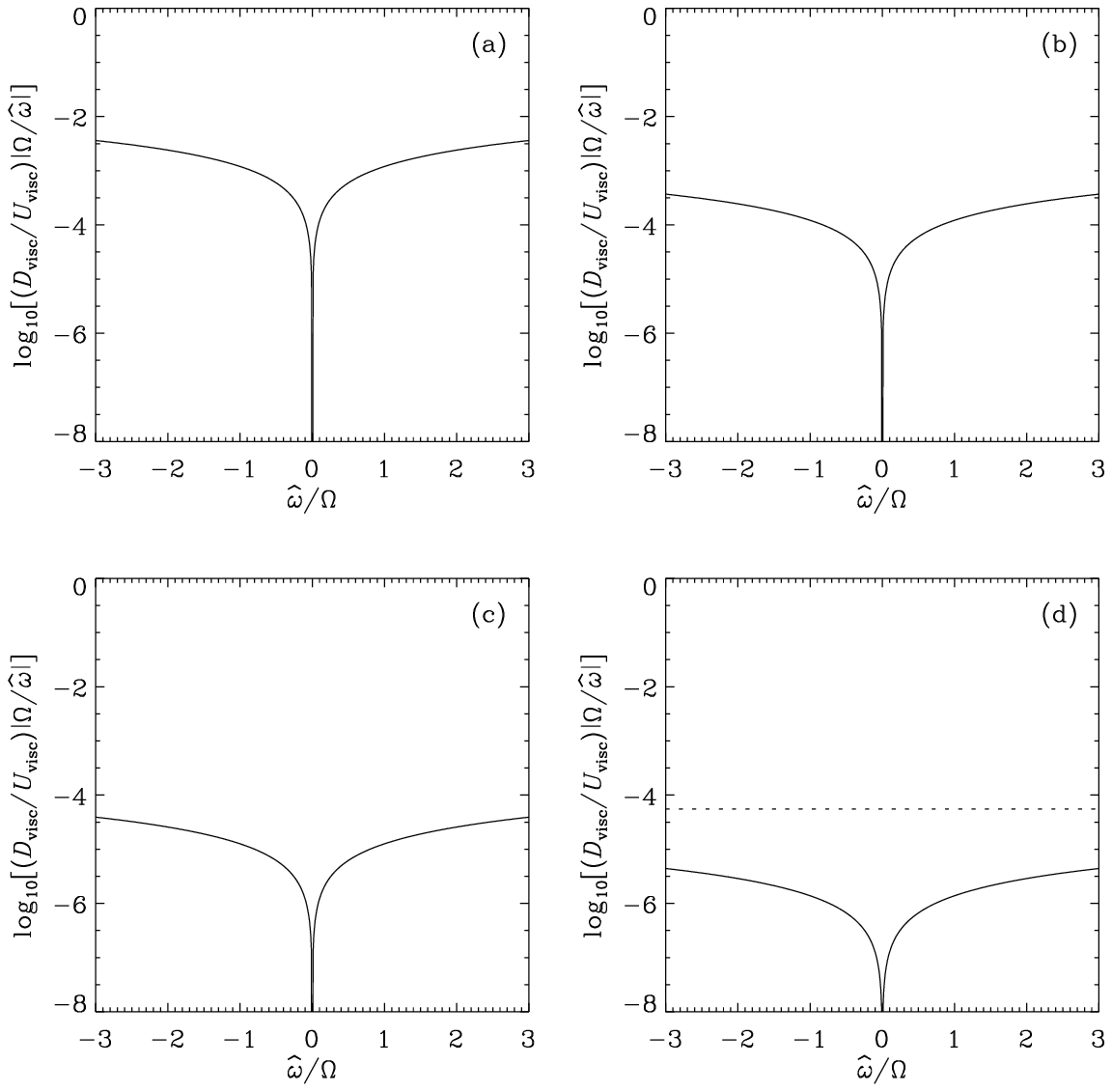} \figcaption{Tidal dissipation rate by viscosity
    in a uniformly rotating planet, according to the no-Coriolis
    approximation.  The parameters are as in Fig.~2.}
\end{figure}

\newpage

\begin{figure}
  \plotone{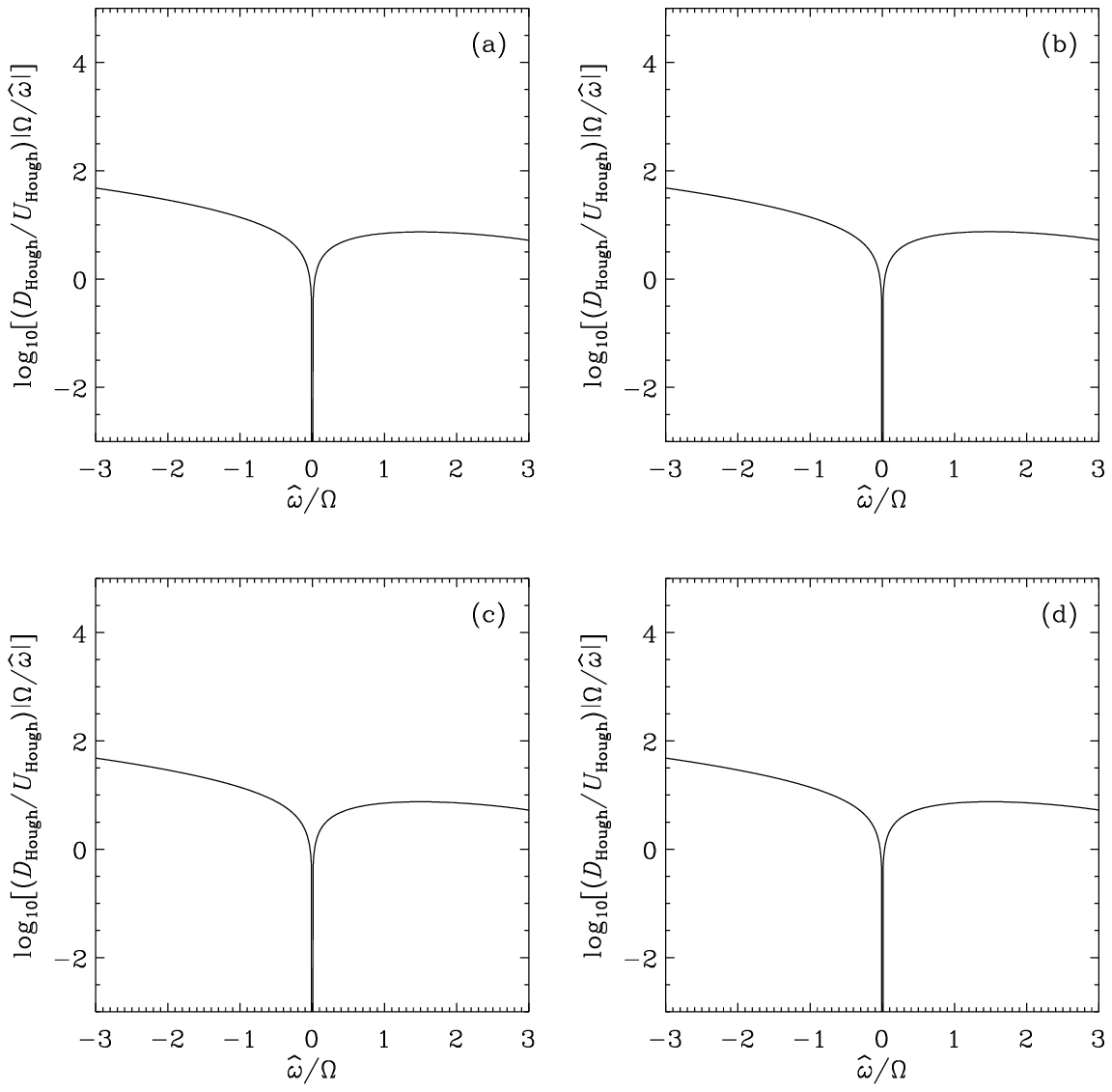} \figcaption{Tidal dissipation rate via Hough
    waves in a uniformly rotating planet, according to the no-Coriolis
    approximation.  The parameters are as in Fig.~3.}
\end{figure}

\newpage

\begin{figure}
  \plotone{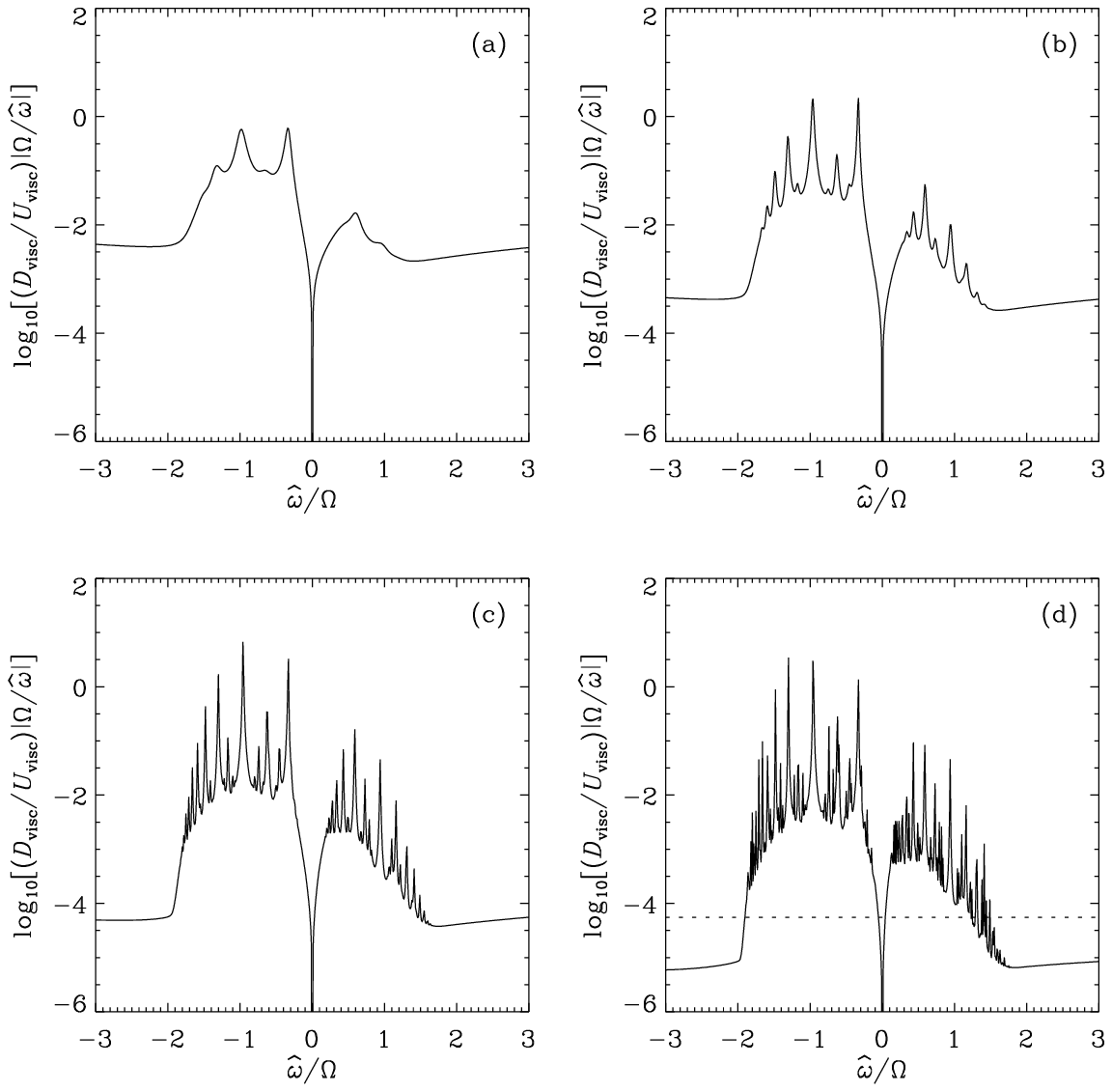} \figcaption{Tidal dissipation rate by viscosity
    in a uniformly rotating planet, according to the traditional
    approximation.  The parameters are as in Fig.~2.}
\end{figure}

\newpage

\begin{figure}
  \plotone{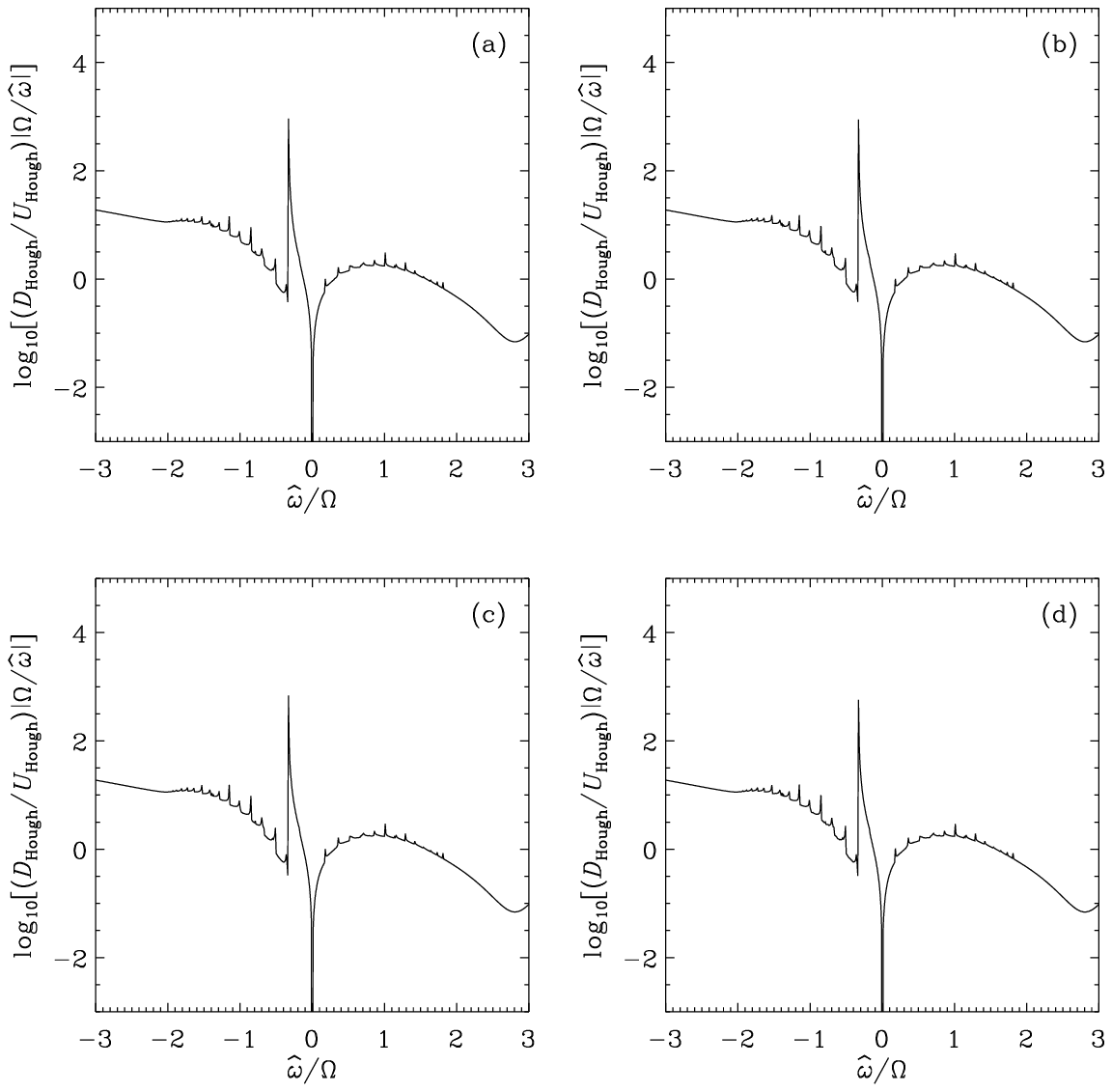} \figcaption{Tidal dissipation rate via Hough
    waves in a uniformly rotating planet, according to the traditional
    approximation.  The parameters are as in Fig.~3.}
\end{figure}

\newpage

\begin{figure}
  \plotone{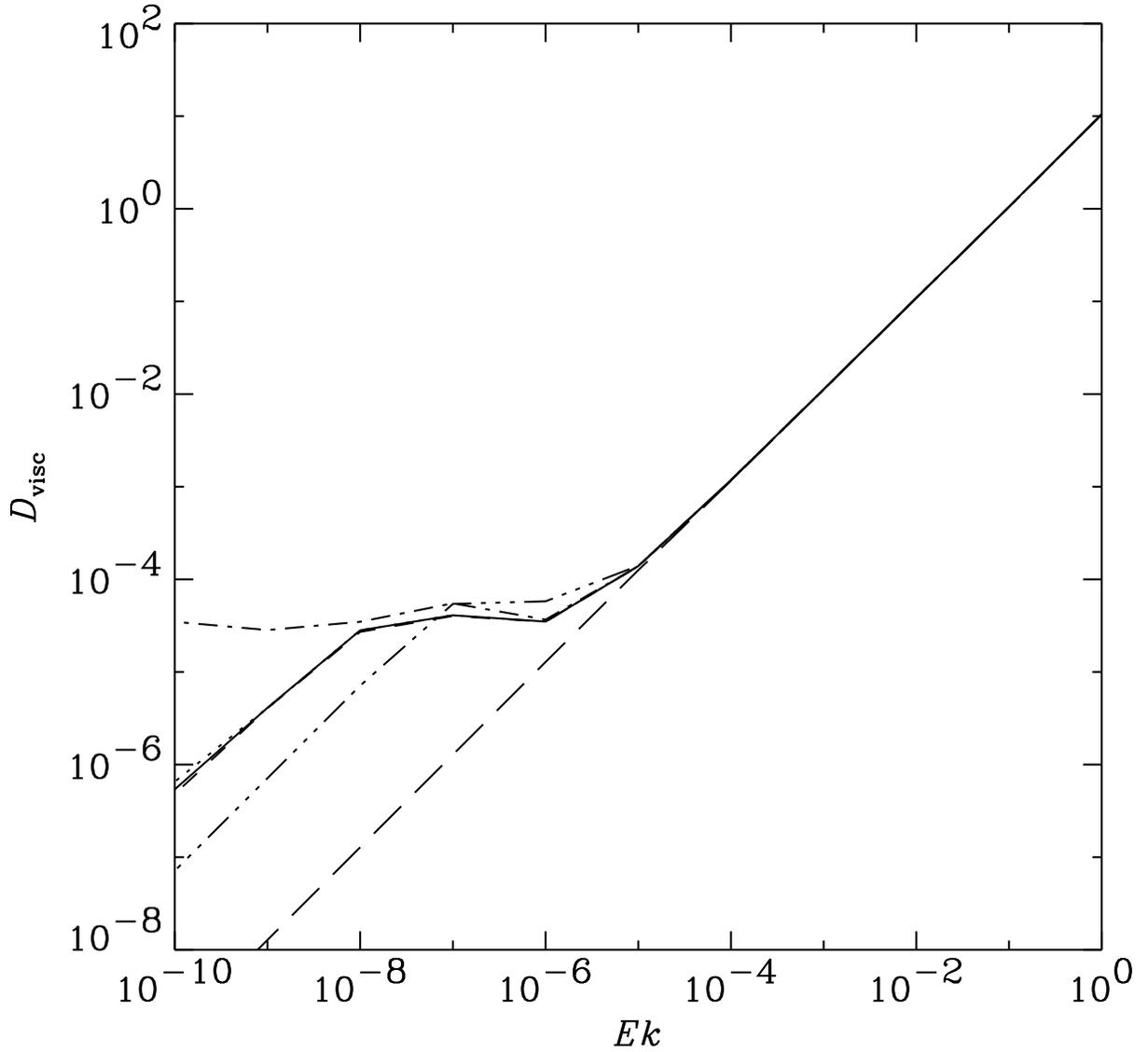} \figcaption{Resolution study away from resonance.
    The viscous dissipation rate for a forcing frequency
    $\hat\omega=\Omega$, not associated with an obvious resonant
    feature, is plotted versus the Ekman number for numerical
    resolutions $L=N=500$ (solid line), $200$ (dotted line), $100$
    (dashed line), $50$ (dash-dotted line), $20$ (dash-triple-dotted
    line), and $10$ (long-dashed line).  The three highest resolutions
    give results that are almost indistinguishable by eye.  The
    alternative numerical method reproduces the curve for $Ek>10^{-3}$
    to very high accuracy, but cannot be used for smaller $Ek$.}
\end{figure}

\newpage

\begin{figure}
  \plotone{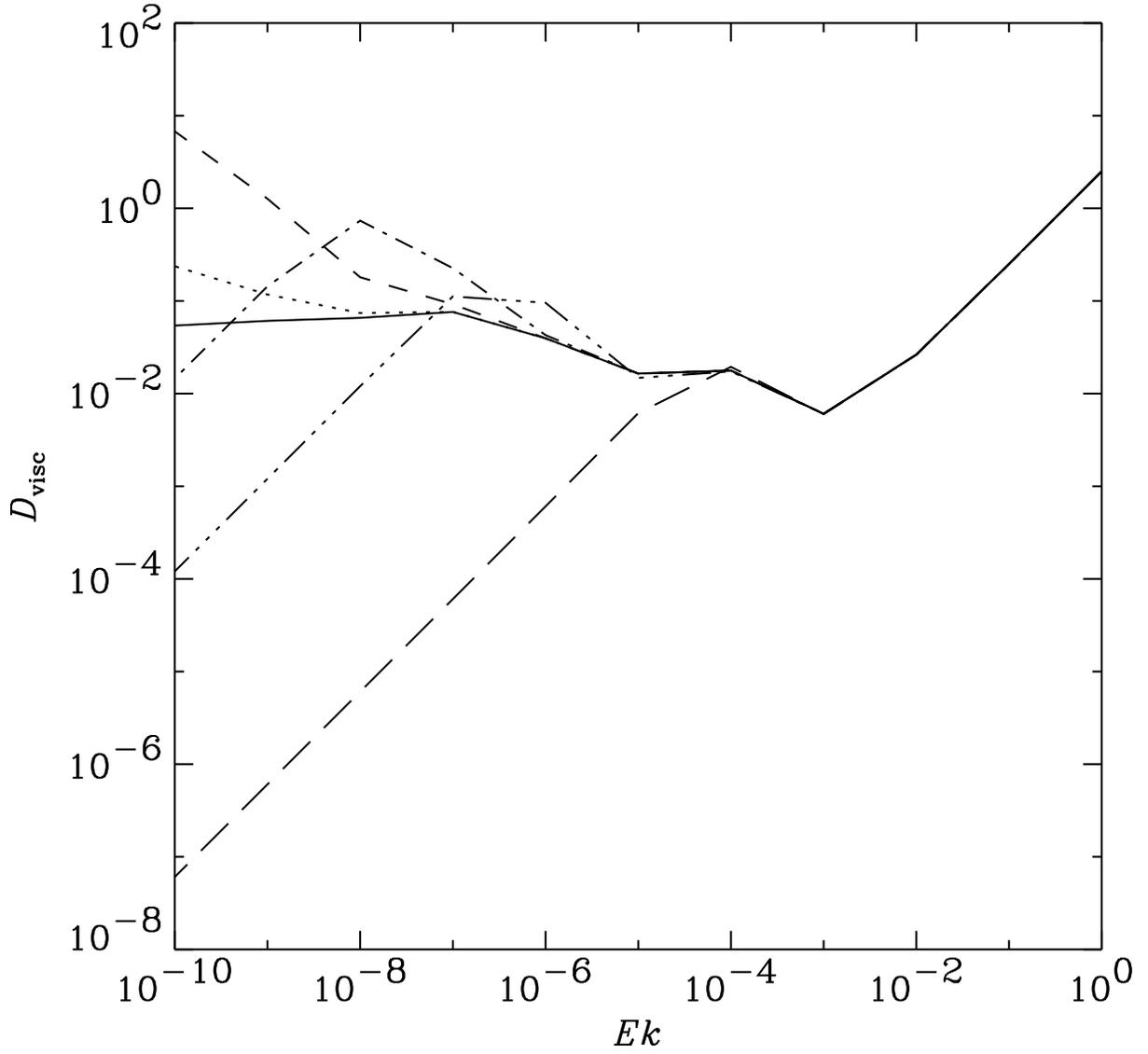} \figcaption{Resolution study on resonance.  As
    for Fig.~8, but for a forcing frequency
    $\hat\omega=0.489\,\Omega$, associated with the largest resonant
    feature seen in Fig.~2.}
\end{figure}

\newpage

\begin{figure}
  \plotone{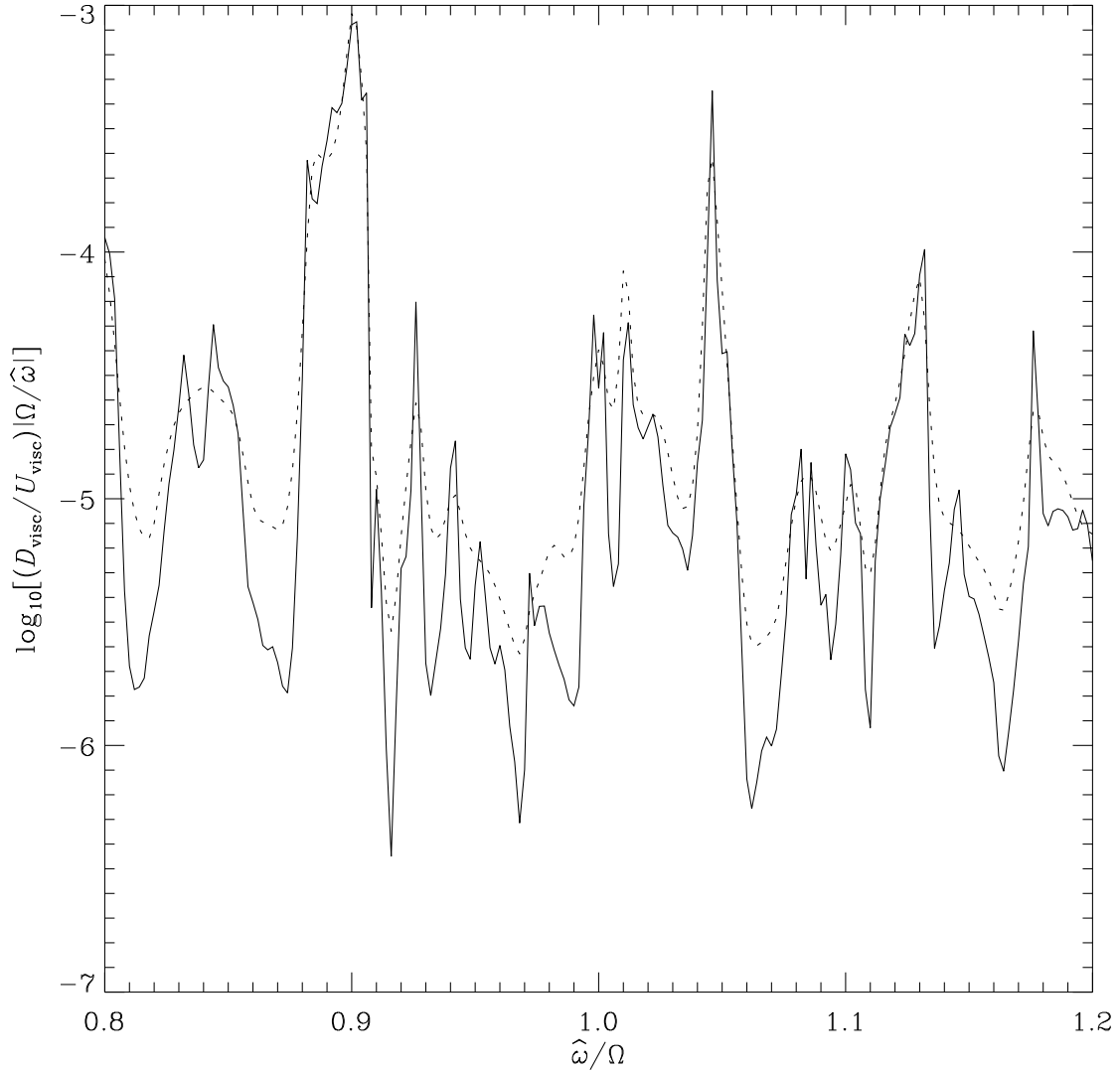} \figcaption{Expanded view of the graph of
    viscous dissipation rate in the full model, for forcing
    frequencies close to $\hat\omega=\Omega$.  The dotted line
    corresponds to $Ek=10^{-7}$ and is equivalent to panel (d) of
    Fig.~2.  The solid line corresponds to $Ek=10^{-8}$ and is
    calculated at a higher resolution of $L=N=200$.  The peaks of the
    curve are not perfectly captured at the frequency sampling of
    $\Delta\hat\omega/\Omega=0.002$.  Although the viscosities differ
    by a factor of ten, the frequency-averaged dissipation rates over
    this interval differ by at most a few per cent.}
\end{figure}

\newpage

\begin{figure}
  \plotone{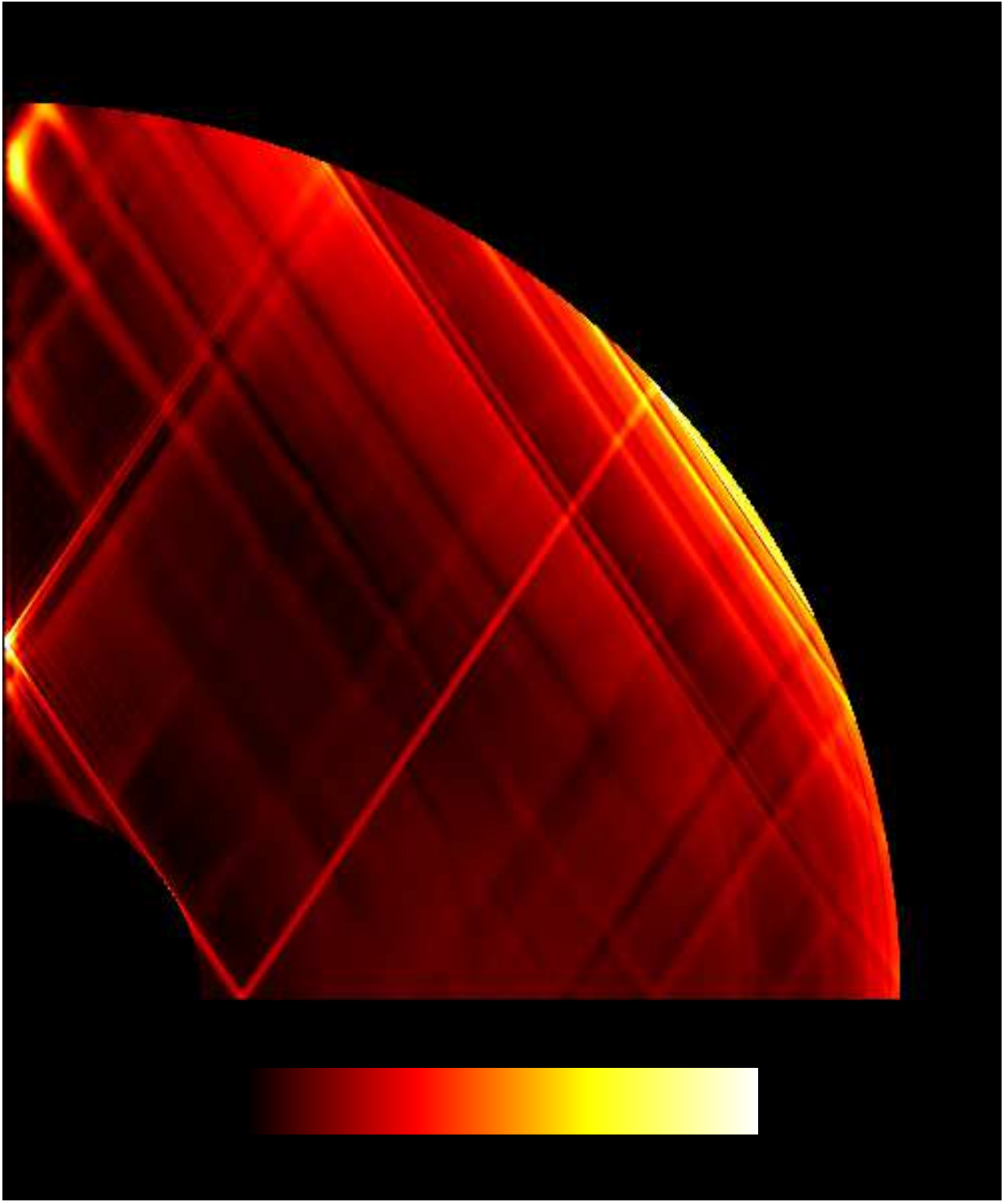} \figcaption{High-resolution calculation
    ($L=N=500$, $Ek=10^{-9}$) of the tidal response of a uniformly
    rotating planet.  The root-mean-square velocity of the total tide
    (equilibrium plus dynamical) is plotted in a meridional slice
    through the convective region only.  The velocity scale is linear,
    black representing zero.  The forcing frequency is chosen to be
    near the peak of an inertial-mode resonance.}
\end{figure}

\newpage

\begin{figure}
  \plotone{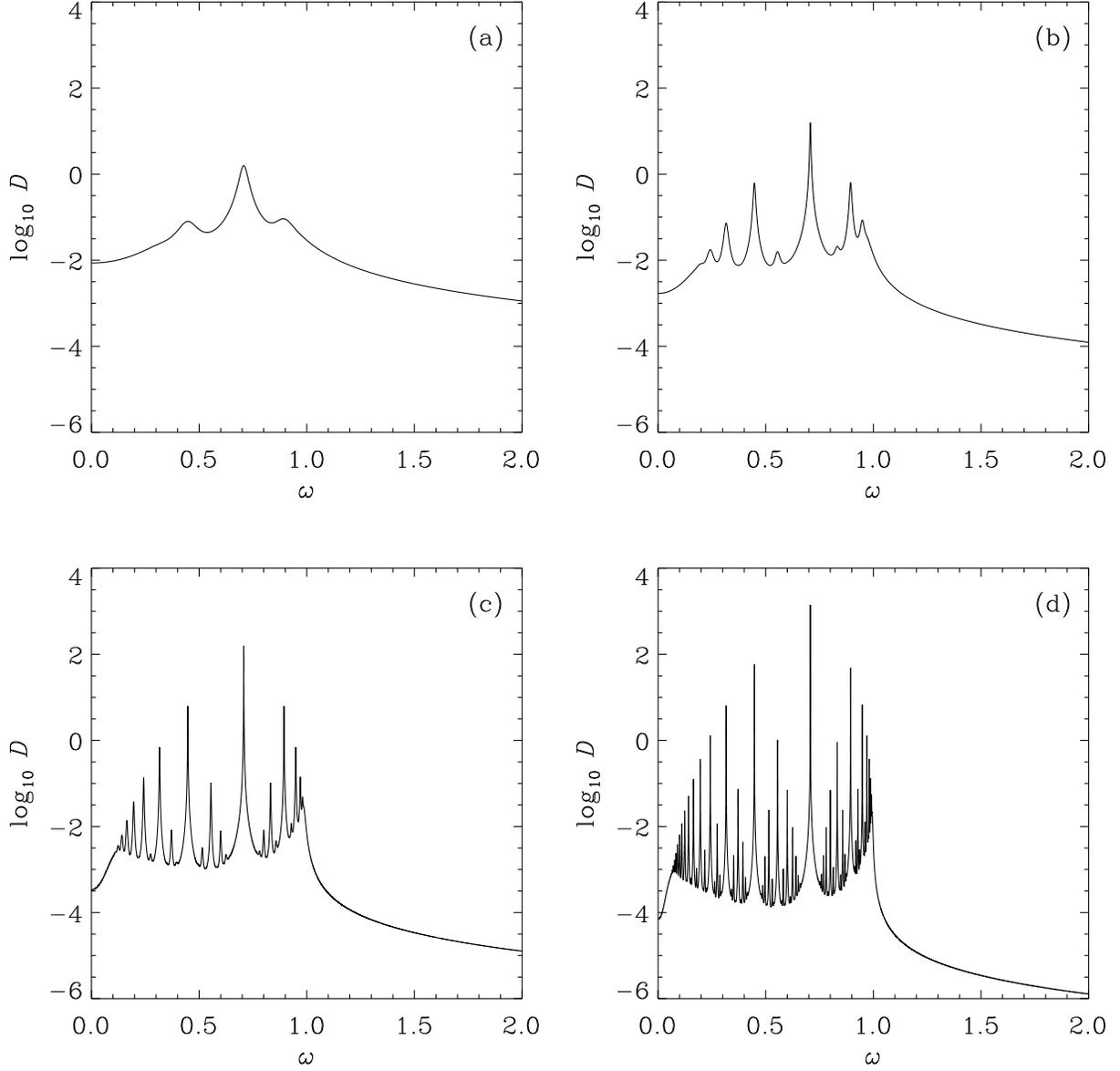} \figcaption{Variation of the viscous dissipation
    rate with forcing frequency, in the toy model for forced inertial
    waves (Appendix~A). Panels (a), (b), (c), and (d) are for
    dimensionless viscosities $\nu=10^{-2}$, $10^{-3}$, $10^{-4}$, and
    $10^{-5}$, respectively.  The spectrum of inertial waves
    corresponds to $-1\le\omega\le1$, and the graphs are symmetric
    about $\omega=0$.}
\end{figure}

\end{document}